\documentclass[
    aps,
    prresearch,
    reprint,
    superscriptaddress,
    amsmath,
    amssymb,
    longbibliography
]{revtex4-2}

\usepackage[T1]{fontenc}
\usepackage[utf8]{inputenc}

\usepackage{amsfonts}
\usepackage{bm}
\usepackage{physics}

\usepackage{graphicx}
\usepackage{booktabs}
\usepackage{multirow}

\usepackage{tensor-network}

\usepackage{xcolor}
\usepackage[
    colorlinks=true,
    linkcolor=blue,
    citecolor=blue,
    urlcolor=blue
]{hyperref}

\setcitestyle{numbers,square,comma,sort&compress}

\begin{document}

\title{Quantum Algorithms for Computational Fluid Dynamics}

\author{Mario Guillaume Cecile}
\email{mario.cecile@uni-hamburg.de}
\affiliation{Institute for Quantum Physics, University of Hamburg, Luruper Chaussee 149, 22761 Hamburg, Germany.}

\author{Nis-Luca van Hülst}
\affiliation{Institute for Quantum Physics, University of Hamburg, Luruper Chaussee 149, 22761 Hamburg, Germany.}

\author{Tomohiro Hashizume}
\affiliation{Institute for Quantum Physics, University of Hamburg, Luruper Chaussee 149, 22761 Hamburg, Germany.}
\affiliation{The Hamburg Centre for Ultrafast Imaging, Luruper Chaussee 149, 22761 Hamburg, Germany.}

\author{Pia Siegl}
\affiliation{Institute for Quantum Physics, University of Hamburg, Luruper Chaussee 149, 22761 Hamburg, Germany.}
\affiliation{Institute of Software Methods for Product Virtualization, German Aerospace Center (DLR), Nöthnitzer Straße 46b, 01187 Dresden, Germany.}

\author{Abhishek Setty}
\affiliation{Forschungszentrum Jülich, Institute of Quantum Control (PGI-8), 52425 Jülich, Germany.}
\affiliation{Institute for Theoretical Physics, University of Cologne, D-50937 Cologne, Germany.}

\author{José Diogo da Costa Jesus}
\affiliation{Forschungszentrum Jülich, Institute of Quantum Control (PGI-8), 52425 Jülich, Germany.}
\affiliation{Institute for Theoretical Physics, University of Cologne, D-50937 Cologne, Germany.}

\author{Paul Over}
\affiliation{Institute for Fluid Dynamics and Ship Theory, Hamburg University of Technology, 21073 Hamburg, Germany.}

\author{Sergio Bengoechea}
\affiliation{Institute for Fluid Dynamics and Ship Theory, Hamburg University of Technology, 21073 Hamburg, Germany.}

\author{Muhammad Umer}
\affiliation{Centre for Quantum Technologies, 3 Science Drive 2, Singapore 117543.}

\author{Spyros Tserkis}
\affiliation{School of Electrical and Computer Engineering, Technical University of Crete, Chania 73100, Greece.}
\affiliation{Institute for Quantum Computing and Quantum Technologies, NCSR Demokritos, Greece.}

\author{Eleftherios Mastorakis}
\affiliation{School of Electrical and Computer Engineering, Technical University of Crete, Chania 73100, Greece.}

\author{Tristan Kraft}
\affiliation{Technical University of Munich, TUM School of Natural Sciences, Department of Physics, 85748 Garching, Germany.}
\affiliation{Munich Center for Quantum Science and Technology (MCQST), Schellingstrasse 4, 80799 Munich, Germany.}

\author{Francisco Cárdenas-López}
\affiliation{Forschungszentrum Jülich, Institute of Quantum Control (PGI-8), 52425 Jülich, Germany.}

\author{Leonardo Scandurra}
\affiliation{ENGYS Srl, Via del Follatoio, 12, 34148 Trieste TS, Italy.}

\author{Thomas Rung}
\affiliation{Institute for Fluid Dynamics and Ship Theory, Hamburg University of Technology, 21073 Hamburg, Germany.}

\author{Felix Motzoi}
\affiliation{Forschungszentrum Jülich, Institute of Quantum Control (PGI-8), 52425 Jülich, Germany.}
\affiliation{Institute for Theoretical Physics, University of Cologne, D-50937 Cologne, Germany.}

\author{Belda Yesil}
\affiliation{Institute for Quantum Physics, University of Hamburg, Luruper Chaussee 149, 22761 Hamburg, Germany.}

\author{Barbara Kraus}
\affiliation{Technical University of Munich, TUM School of Natural Sciences, Department of Physics, 85748 Garching, Germany.}
\affiliation{Munich Center for Quantum Science and Technology (MCQST), Schellingstrasse 4, 80799 Munich, Germany.}

\author{Martin Kiffner}
\affiliation{PlanQC GmbH, Münchener Str. 34, 85748 Garching, Germany.}

\author{Dimitris G. Angelakis}
\affiliation{Institute for Quantum Computing and Quantum Technologies, NCSR Demokritos, Greece.}
\affiliation{School of Electronics and Computer Science,
University of Southampton, Southampton SO17 1BJ, UK.}
\affiliation{AngelQ Quantum Computing, 531A Upper Cross Street, \#04-95 Hong Lim Complex, Singapore 051531.}

\author{Eugene de Villiers}
\affiliation{ENGYS Ltd., London SW18 3SX, United Kingdom.}

\author{Dieter Jaksch}
\affiliation{Institute for Quantum Physics, University of Hamburg, Luruper Chaussee 149, 22761 Hamburg, Germany.}
\affiliation{The Hamburg Centre for Ultrafast Imaging, Luruper Chaussee 149, 22761 Hamburg, Germany.}
\affiliation{Clarendon Laboratory, University of Oxford,
Parks Road, Oxford OX1 3PU, UK.}

\begin{abstract}
We present a comprehensive review of quantum approaches for solving partial differential equations (PDEs) arising in computational fluid dynamics (CFD). We examine fully quantum approaches, including quantum linear system algorithms (QLSAs), ranging from the Harrow--Hassidim--Lloyd (HHL) algorithm to quantum singular value transformation (QSVT), Hamiltonian simulation, and quantum lattice Boltzmann methods (QLBMs), while emphasizing hybrid quantum--classical approaches, including quantum physics-informed neural networks (QPINNs) and amplitude-encoded variational PDE solvers. We focus on hardware-agnostic algorithms compatible with present noisy processors and emerging fault-tolerant architectures. For each framework, we analyze the mathematical formulation, algorithmic structure, and principal limitations. We also examine tensor-network (TN) representations, since CFD fields, differential operators, and geometrical information can often be encoded efficiently in low-rank form. The TN formalism bridges CFD discretizations and quantum states, operators, and circuits, enabling compact representations to be translated into tensor-programmable variational quantum algorithms (TP-VQAs). We further review benchmark problems, including Poisson, reaction, diffusion, and nonlinear model equations, and assess how well quantum algorithms capture key features of fluid dynamics. Our analysis highlights that potential quantum advantage is highly problem dependent and governed by condition number, representational complexity, state preparation, and measurement constraints. We outline capabilities, limitations, and challenges toward scalable quantum algorithms for CFD.

\par\vspace{0.5em}

\end{abstract}

% =========================================================
% Keywords
% =========================================================

% \keywords{
% quantum algorithms for fluid dynamics;
% tensor networks;
% variational quantum algorithms;
% quantum linear systems;
% Hamiltonian evolution;
% quantum lattice Boltzmann method
% }

% =========================================================

\maketitle

%%%%%%%%%%%%%%%%%%%%%%%%%%%%%%%%%%%%%%%%%%
% \setcounter{section}{-1} %% Remove this when starting to work on the template.
% \section{How to Use this Template}

% The template details the sections that can be used in a manuscript. Note that the order and names of article sections may differ from the requirements of the journal (e.g., the positioning of the Materials and Methods section). Please check the instructions on the authors' page of the journal to verify the correct order and names. For any questions, please contact the editorial office of the journal or support@mdpi.com. For LaTeX-related questions please contact latex@mdpi.com.%\endnote{This is an endnote.} % To use endnotes, please un-comment \printendnotes below (before References). Only journal Laws uses \footnote.

% % The order of the section titles is different for some journals. Please refer to the "Instructions for Authors” on the journal homepage.

% \newpage 
% \tableofcontents 

% \newpage 
\onecolumngrid
\section{Introduction}
    
Computational fluid dynamics (CFD) is a cornerstone of modern science and engineering, enabling the numerical investigation of complex flow phenomena across aerospace, energy, and environmental applications \cite{ferziger02:CMFD,Anderson1995}. A central challenge in CFD arises from the interaction of multiple physical fields, $\bm{f}=\{\vec{v},\theta,P,\ldots\}$, including velocity $\vec{v}$, temperature $\theta$, and pressure $P$, across a wide range of spatial and temporal scales \cite{Pope2000}. Resolving these interactions accurately is exceptionally demanding and, in some cases, prohibitive even with national-scale supercomputing resources
\cite{alfonsi2011,Slotnick2014CFDV2}. For fully resolved direct
numerical simulation (DNS), the number of degrees of freedom required to capture all dynamically relevant turbulent scales grows approximately as $N\sim Re^{9/4}$ in three dimensions, where $Re=UL/\nu$ is the Reynolds number, $U$ is a characteristic velocity, $L$ is a characteristic length scale, and $\nu$ is the kinematic viscosity \cite{Pope2000,MoinMahesh1998}. This scaling reflects the
need to resolve the Kolmogorov scales that govern energy dissipation and scalar mixing, together with the cascade of energy from large flow structures to small dissipative scales
\cite{Boschung2016,Tennekes1972,Verma2018}. In engineering flows, these small-scale dynamics influence macroscopic quantities such as skin-friction drag, flow separation, and heat transfer \cite{MoinMahesh1998}. Classical CFD therefore relies heavily on approximation hierarchies, including Reynolds-averaged Navier--Stokes (RANS) models and large-eddy simulation (LES) \cite{WILCOX06,Pope2000,LESWBForlandi2000,Sagaut2006,
argyropoulosRecentAdvancesNumerical2015,
durbinRecentDevelopmentsTurbulence2018}, reduced-order models (ROMs) \cite{reducedorder2004,Keiper2018-tf}, and massively parallel computing. Nevertheless, turbulent flows around complex geometries remain computationally demanding \cite{Chapman1979,Choi2012}.

These computational challenges have driven growing interest in quantum computing as a possible route to accelerating scientific simulations. Quantum algorithms can offer polynomial or, under suitable assumptions,
exponential speedups for selected problems in large-scale linear algebra
\cite{Harrow2009,Childs2017}, differential equations \cite{Berry_2017,Childs_2020,Childs_2021}, and nonlinear high-dimensional dynamical systems \cite{Liu_2021}. In CFD, the long-term motivation is to enable larger and more predictive simulations at lower computational cost. This goal is strongly industrially motivated: high-Reynolds turbulence, complex transport processes, and multiscale phenomena often require very fine spatial and temporal resolution, making DNS and scale-resolving LES prohibitively expensive for many realistic configurations. Quantum computing should therefore not be viewed primarily as a near-term replacement for mature RANS-based design pipelines, which are already highly optimized for estimating macroscopic engineering quantities. Rather, the most plausible targets are DNS-like simulations, high-fidelity LES, multidimensional kinetic transport, repeated structured operator evaluations, and new algorithmic approaches for representing, evolving, or querying high-dimensional flow fields. Present quantum hardware does not yet provide the qubit numbers, circuit depths, error correction, or measurement accuracy required for full engineering-scale turbulent flows around complex geometries \cite{Beverland2022,Preskill_2025,Bharti2022,Gidney_2021,
Bluvstein2024,Bluvstein2026}. Current quantum-CFD research therefore focuses on smaller algorithmic building blocks that can be studied using simplified benchmark problems, with the aim of identifying where quantum representations may eventually offer advantages over optimized classical solvers, ROMs, and tensor-network (TN) methods.

After spatial discretization, many CFD problems reduce to large algebraic or dynamical systems. Steady-state formulations and implicit
time-stepping schemes commonly lead to sparse linear systems of the
form $A\mathrm{x}=b$, where $\mathrm{x}$ contains the unknown discrete
flow variables, $b$ includes source terms, boundary-condition
contributions, or data from previous time steps, and $A$ may arise
from finite-difference, finite-volume, finite-element, or spectral
discretizations \cite{ferziger02:CMFD,hirsch2007numerical}.
Time-dependent problems require the application of evolution
operators, while lattice Boltzmann formulations decompose the dynamics
into streaming and collision steps acting on particle distribution
functions that also evolve in time. Nonlinear partial differential equations (PDEs) further require the treatment of nonlinear
residuals, source terms, and constraints. These computational
primitives have motivated several quantum and hybrid
quantum--classical paradigms, including quantum linear system algorithms (QLSAs) for discretized PDEs
\cite{Harrow2009,Childs2017,Gily_n_2019}, Hamiltonian simulation and
quantum differential-equation solvers for time-dependent dynamics
\cite{Berry_2015,Low_2019,Childs_2021}, quantum lattice-gas and
quantum lattice Boltzmann formulations
\cite{Yepez2001,Itani2024,Tiwari2025,Wang2025}, and variational and physics-informed quantum-machine-learning methods for nonlinear problems \cite{Lubasch2020,Kyriienko2021,Setty2025}. These building blocks do not yet constitute complete CFD solvers, but they isolate
the main computational components required in a future quantum-accelerated CFD workflow. An important requirement is the efficient extraction of physically relevant observables from quantum states without reconstructing the complete classical flow field \cite{Paris2004,Aaronson2018}. Recent full-stack proposals have begun to examine how these components could be combined into end-to-end
quantum workflows for Navier--Stokes simulations, including input and
output procedures, circuit synthesis, and fault-tolerant resource estimates \cite{zhuang2025pathwaypracticalquantumadvantage}.

The noisy intermediate-scale quantum (NISQ) era has provided valuable
experimental access to noisy quantum processors \cite{Preskill2018}. However, it is increasingly clear that scalable scientific applications will require some degree of quantum error correction and fault-tolerant control
\cite{Preskill2018,Preskill_2025}. Recent milestones suggest that this
transition is becoming increasingly concrete. Experiments have
demonstrated below-threshold error suppression in superconducting
surface-code memories and programmable logical processors based on reconfigurable neutral-atom arrays \cite{Acharya2025,Bluvstein2024}. In parallel, theoretical developments in quantum low-density parity-check codes and hardware-aware fault-tolerant architectures indicate that the physical-qubit overhead can be substantially reduced \cite{Bravyi2024,Xu2024}. As a concrete illustration of this progress, Cain \textit{et al.}~\cite{Cain2026shorsalgo} estimated that a cryptographically relevant implementation of Shor's algorithm could be performed with approximately $10^4$ reconfigurable neutral-atom qubits by combining high-rate quantum error-correcting codes with hardware-aware fault-tolerant circuit constructions. These advances do not yet constitute large-scale fault-tolerant computation, but they indicate that early error-corrected processors are moving from theoretical proposals toward experimentally accessible devices. 

Progress toward this goal is being pursued across several competing hardware platforms. Trapped-ion processors, originating from early proposals for motion-mediated quantum gates and scalable microtrap arrays
\cite{Cirac1995,Cirac2000}, provide long coherence times and high-fidelity operations, although scaling to larger systems requires multiple processing zones and interconnected modules \cite{Bruzewicz2019,Akhtar2023}. Recent progress includes a 98-qubit
trapped-ion processor with all-to-all connectivity and an average
two-qubit gate infidelity below $10^{-3}$ \cite{Ransford2026}. Neutral-atom and Rydberg-array platforms combine reconfigurable
geometries, long-range interactions, and large atomic registers
\cite{Jaksch2000,Henriet2020quantumcomputing}. Recent experiments have
demonstrated optical-tweezer arrays containing more than $6100$
coherently controlled atomic qubits \cite{Manetsch2025}, together with
continued progress in programmable arrays, high-fidelity entangling
gates, and fast coherent interactions
\cite{Bluvstein2024,Bluvstein2026,Xu2024,
Wintersperger2023,menssen2026strategicplanneutralatom,
Gyger2024,Bojovic2026,Chew2022}. Photonic architectures offer largely room-temperature operation and
natural fibre-based interconnects between distributed modules, but
scalable fault-tolerant computation requires high-quality photon generation, indistinguishable-photon interference, efficient detection,
and strong tolerance to optical loss
\cite{Slussarenko_2019,AghaeeRad2025,Pankovich2024}. Semiconductor spin qubits provide another route based on compact solid-state devices that may leverage established semiconductor manufacturing, while scalability depends on uniform high-fidelity control and the interconnection of large qubit arrays
\cite{GonzalezZalba2021,Burkard2023}. Topological-qubit architectures aim
to suppress errors directly at the hardware level and remain a further
potential route toward scalable fault-tolerant computation
\cite{aasen2025roadmapfaulttolerantquantum}. This diversity motivates quantum-CFD algorithms that are not tied to a single hardware architecture and that can operate on present noisy processors while
remaining compatible with early fault-tolerant devices. The transition from NISQ to fault-tolerant hardware is therefore a central motivation for the present review.

The rapid growth of the field has motivated broad reviews of quantum computing across science and engineering \cite{Dalzell_2025}, as well as more focused surveys of quantum algorithms for CFD and fluid dynamics \cite{Malinverno2026,meng2026geometricencodingturbulenceendtoend} and of quantum machine-learning and quantum-inspired approaches to CFD \cite{Amaral2026}. The present review adopts a complementary perspective by considering where TN and low-rank methods can act not only as quantum-inspired solvers or compression techniques, but also as a representational layer connecting classical discretizations, structured operators, and quantum-circuit construction. This viewpoint is motivated by the multiscale structure of turbulence. Although turbulent flows contain many active degrees of freedom, the Kolmogorov cascade is predominantly local in scale space, so the resulting correlations remain structured rather than fully delocalized \cite{Pope2000}. Such structure is relevant to both TN compression and quantum algorithms because it affects state preparation, ansatz design, operator encoding, and the extraction of physically meaningful observables
\cite{Gourianov2022,Kiffner2023,H_lscher_2025,
meng2026geometricencodingturbulenceendtoend}. A comprehensive review of recent quantum-inspired algorithms is beyond the scope of the present work. We therefore discuss only the classical TN methods that are needed to provide a useful baseline for quantum algorithms and to introduce the concepts required for the developments presented later in the review.

Accordingly, this review develops a unified perspective on quantum CFD, progressing from theoretical algorithms to practical implementation requirements. Sec.~\ref{Sec:FullyQuantumAlgo} examines fully quantum approaches, including QLSAs based on the HHL algorithm and QSVT, Hamiltonian simulation, and quantum lattice Boltzmann methods (QLBMs), and identifies the assumptions behind their potential advantages, such
as efficient state preparation, operator encoding, coherent time
evolution, and economical readout. The resulting limitations motivate
the hybrid quantum--classical approaches discussed in
Sec.~\ref{Sec:Hybrid_Approach}, including quantum physics-informed
neural networks (QPINNs) in Sec.~\ref{Sec:QPINN} and amplitude-encoded variational PDE solvers in Sec.~\ref{Sec:StandardVGA}. These approaches replace explicit matrix inversion or long coherent evolution with variational optimization, but shift the main challenges toward circuit
construction, measurement cost, and trainability. The effectiveness of these algorithms ultimately depends on whether CFD
fields and operators admit structured representations.
Sec.~\ref{Sec:TPQA} therefore introduces TNs, first as classical
compression methods and their relation to quantum algorithms in
Sec.~\ref{sec:Translation_TN_to_QC}, and subsequently as an intermediate programming layer for translating discretized states and operators into quantum circuits. This naturally leads to tensor-programmable variational
quantum algorithms (TP-VQAs) in Sec.~\ref{Sec:TPVQA}, in which TN
representations, matrix product operator (MPO)-to-circuit compilation, and quantum nonlinear processing unit (QNPU)-based treatments of nonlinear terms provide a systematic connection between classical numerical methods and VQAs
\cite{akshay2024tensornetworksbasedquantum,Termanova2024,Siegl_2026}.
Finally, Sec.~\ref{Sec:NISQ_Fault_Tolerant} relates these algorithmic
requirements to NISQ and early fault-tolerant hardware, while
Sec.~\ref{Sec:QuantumAdvantage} synthesizes the preceding results to identify the problem regimes in which an end-to-end quantum advantage appears most and least plausible.

%%%%%%%%%%%%%%%%%%%%%%%%%%%%%%%%%%%%%%%%%%
\section{Fully Quantum, Fault-Tolerant Paradigm}
\label{Sec:FullyQuantumAlgo}

A conventional CFD discretization represents the field value at every
grid point in physical space, resulting in $N=m^d$ degrees of freedom,
where $m$ is the number of grid points per spatial coordinate direction
and $d$ is the dimensionality of the system. This leads to a storage
cost that scales as $\mathcal{O}(m^d)$, causing a rapid increase in
memory usage and computational expense as the grid is refined. Even
when sparse matrix representations are employed, the memory requirement
remains at least linear in the number of degrees of freedom,
$\mathcal{O}(m^d)$ \cite{Saad2003}, since local differential operators
couple each grid point only to a fixed number of neighbors.

In addition, the condition number $\kappa(A)$, defined for a
diagonalizable operator as the ratio between the largest and smallest
nonzero eigenvalue magnitudes, typically worsens with mesh refinement
for discretized elliptic and convection--diffusion operators, slowing
the convergence of iterative solvers unless effective preconditioning
techniques are used~\cite{preconditioning2012,Saad2003}. For
time-dependent problems, explicit schemes are constrained by stability
conditions such as the Courant--Friedrichs--Lewy (CFL) criterion
\cite{CourantFriedrichsLewy1928}, while implicit schemes require
solving large linear systems at each time step.

Fully quantum algorithms seek to address these challenges by encoding discretized fields and operators directly on a quantum processor. A field with $N$ degrees of freedom can, in principle, be represented
using $\lceil\log_2 N\rceil$ qubits through amplitude encoding \cite{Harrow2009}. This compact representation underlies many proposed quantum speedups, but
does not by itself guarantee an end-to-end advantage, which also
depends on efficient state preparation, operator implementation, and
observable extraction.

Several fully quantum paradigms have been proposed for solving PDEs.
QLSAs, including the algorithm
introduced by Harrow, Hassidim, and Lloyd (HHL) \cite{Harrow2009}, prepare quantum states that encode the solutions of large linear systems, subject to assumptions on sparsity, matrix conditioning, and efficient operator access. Nonlinear equations can be incorporated into this framework through lifting techniques such as Carleman linearization \cite{Carleman1932}, although the resulting increase in system dimension introduces additional truncation and computational overhead. Hamiltonian simulation methods provide an alternative route by representing discretized differential operators as generators of quantum time evolution \cite{Brearley2024,Over2024b,Bengoechea2026,Low_2019,Berry_2015,PhysRevLett.114.090502}. More specialized approaches, such as QLBMs, retain the mesoscopic streaming--collision structure of classical lattice Boltzmann schemes and adapt it to quantum computation \cite{Tiwari2025,Wang2025}. The following sections review these approaches, their computational
assumptions, and the main limitations that determine their relevance
for practical CFD applications.

\subsection{Quantum Linear System Algorithms}
\label{Sec:QLSA}

The following construction corresponds to Carleman linearization \cite{Carleman1932}, in which a nonlinear dynamical system is lifted to a linear system on an enlarged space of monomials. Consider a nonlinear dynamical system of the form
\begin{equation}
    \dot{x}(t) = f(x(t)),
\end{equation}
where $x(t)\in\mathbb{R}^N$ is a state vector representing the
discretized physical fields at $N$ grid points. Here,
$f:\mathbb{R}^N\rightarrow\mathbb{R}^N$ is an analytic function encompassing the discretized spatial operators, such as convective and diffusive terms, as well as any nonlinear interactions inherent in the governing equations. If $f$ admits a polynomial expansion, one can define an augmented state vector $y$ containing all monomials of $x$:
\begin{equation}
    y =
    \begin{bmatrix}
        x_1, & \dots, & x_N, &
        x_1x_1, & x_1x_2, & \dots, &
        x_ix_jx_k, & \dots
    \end{bmatrix}^{T}.
\end{equation}

The time derivative of each monomial can be expressed linearly in terms of higher-order monomials, yielding an infinite-dimensional linear system:
\begin{equation}
    \dot{y}(t)=\mathcal{L}y(t),
\end{equation}
where $\mathcal{L}$ is a countably infinite linear operator. Truncating this hierarchy at a maximum polynomial order $K$ produces a finite-dimensional linear system. Despite its utility, the Carleman approach suffers from a combinatorial expansion of the state space, $\mathcal{O}(N^K)$. This curse of dimensionality has traditionally limited the method's feasibility to systems with mild nonlinearities or low-dimensional approximations.

The relevance of Carleman linearization to quantum algorithms was demonstrated by Liu \textit{et al.}~\cite{Liu_2021}, who combined the lifting procedure with a quantum linear differential-equation solver. Their analysis showed that quantum speedups may be obtained for certain weakly nonlinear, dissipative systems, provided that the truncation error, condition number, and state-preparation costs remain controlled.
Krovi~\cite{Krovi_2023} subsequently extended the range of linear and nonlinear differential equations accessible to quantum algorithms by relaxing several assumptions used in earlier analyses.

Applying these methods to CFD remains challenging because the required Carleman truncation order and lifted state-space dimension may increase rapidly with the strength of the nonlinear coupling. Turbulent flows introduce additional difficulties through their broad range of interacting spatial and temporal scales \cite{Bharadwaj2020}. Gonzalez-Conde \textit{et al.}~\cite{GonzalezConde2025} examined this issue specifically for nonlinear fluid dynamics by relating the numerical parameter governing efficient Carleman truncation to physical flow parameters through the Kolmogorov scale. They also extended the formulation to multidimensional vector fields, including the discretization of spatial operators and boundary conditions. Dissipative or reduced-order flow models are generally more tractable, as the damping of higher-order modes can limit the contribution of the higher levels in the Carleman hierarchy \cite{Liu_2021,Malinverno2025}.

Recent work has also explored adaptive and structure-aware extensions of Carleman linearization. Novikau and Joseph~\cite{Novikau2025}
introduced a piecewise formulation in which the linearization region is
updated along the trajectory, extending the range over which the
approximation remains accurate. They also note that a non-adaptive
piecewise variant may be better suited to quantum implementation,
because the fully adaptive scheme requires additional nonlinear
operations. Demirdjian \textit{et al.}~\cite{Demirdjian2026} exploited the sparse structure
of the nonlinear term in the one-dimensional Burgers' equation to
construct a more efficient Carleman-linearized representation of the
resulting enlarged linear system. Related work on parabolic PDEs has
also investigated structure-exploiting discretizations of the higher Carleman levels \cite{Heinzelreiter2025}.

Related quantum approaches have also been explored in several fluid-dynamical settings. Sanavio and Succi~\cite{Sanavio_2024} combined lattice-Boltzmann formulations with polynomial lifting and demonstrated the treatment of moderate-Reynolds-number flows under simplifying assumptions. Todorova and Steijl~\cite{Todorova2020} developed a quantum algorithm for the collisionless Boltzmann equation, illustrating how linear transport dynamics can be represented using quantum circuits. These results indicate that linear and weakly nonlinear flow models can be mapped to QLSA-compatible formulations, whereas extending such methods to strongly nonlinear, high-dimensional CFD remains an open challenge \cite{Malinverno2025}.

Before proceeding further, we establish the use of the Dirac bra-ket notation, which is the standard convention in quantum mechanics. In this notation, the discretized physical fields appearing in CFD problems are mapped to quantum states. Specifically, a field is represented by a \textbf{ket} vector, denoted as $\ket{u}$, while its complex conjugate transpose, or adjoint, is represented by a \textbf{bra} vector, $\bra{u}$. For the purposes of this discussion, we assume these states are unnormalized unless otherwise specified, such that the inner product $\braket{u}{u}$ corresponds to the squared $\ell^2$-norm of the discretized field.

\subsubsection{Harrow Hassidim Lloyd Algorithm}
\label{subsubsec:HHL}

The QLSA introduced by Harrow, Hassidim, and Lloyd provides one of the foundational examples of a quantum algorithm for solving linear systems of equations, $Ax=b$ \cite{Harrow2009}. Known as the HHL algorithm, it implements the spectral transformation $\lambda\mapsto1/\lambda$ of a sparse Hermitian
matrix $A$ by combining Hamiltonian simulation \cite{Berry_2015} with quantum phase estimation (QPE) \cite{Kitaev1995QuantumMA}. Pedagogical introductions and surveys of HHL variants, improvements, and applications can be found in Refs.~\cite{zamanStepbyStepHHLAlgorithm2023,
liuSurveyImprovementApplication2022}. Appendix~\ref{Appendix:HHL} summarizes the main steps of the algorithm.

Under standard sparsity and oracle assumptions, the runtime of the original HHL algorithm for an $s$-sparse matrix
$A\in\mathbb{C}^{D\times D}$ scales as
\begin{equation}
    \mathcal{O}\!\left(
        s^2\kappa^2\frac{\log D}{\epsilon}
    \right),
\end{equation}
where $\kappa(A)$ is the condition number of $A$, $s$ is the sparsity, defined as the maximum number of nonzero entries in any row or column of $A$, $D$ is the dimension of the linear system, and $\epsilon$ is the target precision \cite{Harrow2009}. In a CFD discretization with $N$ total degrees of freedom, one typically has $D=N$; however, $D$ is used here to distinguish the abstract linear-system dimension from the grid notation.

A key feature of HHL is the probabilistic nature of the post-selection step, which relies on an ancillary qubit used to flag whether the desired matrix inversion has been successfully applied. The probability of
measuring this ancilla qubit in the success state $\ket{1}$ scales as $\mathcal{O}(1/\kappa^2)$ \cite{Harrow2009}, leading to a substantial repetition overhead for ill-conditioned systems. Amplitude amplification, introduced by Brassard \textit{et al.}~\cite{Brassard_2002}, reduces this overhead from $\mathcal{O}(\kappa^2)$ to $\mathcal{O}(\kappa)$, although the overall complexity remains strongly dependent on $\kappa$. This quadratic condition-number dependence motivated the development of improved quantum linear solvers. Ambainis~\cite{Ambainis2012}
introduced variable-time amplitude amplification, which reduces the
amplification overhead by allowing different computational branches
to terminate at different times. Subsequently, Childs, Kothari, and
Somma~\cite{Childs2017} proposed a refined QLSA with improved precision scaling and near-linear dependence on $\kappa$. More recent QLSAs build on the broader framework of singular-value transformation, providing a systematic alternative to the phase-estimation-based approach used in HHL.

\subsubsection{Quantum Singular Value Transformation}
\label{subsubsec:QSVT}

Gilyén \textit{et al.}~\cite{Gily_n_2019} developed the framework of Quantum Singular Value Transformation (QSVT), which provides a general approach for applying polynomial transformations to the singular values of a matrix. In particular, matrix inversion can be implemented by approximating the reciprocal function over the relevant singular-value interval. Under the assumption that the required block encoding can be implemented efficiently, QSVT-based linear solvers achieve essentially linear dependence on the condition number $\kappa$ and polylogarithmic dependence on the target precision. A central ingredient of QSVT is block encoding. Since a general matrix $A$ arising from a PDE discretization is not unitary, it cannot in general be implemented directly as a quantum gate. Instead, a normalized representation of $A$ is embedded as a subblock of a larger unitary operator acting on the system and additional ancilla qubits. Formally, a unitary $U_A$ is a
$(\beta,n_{\mathrm{aux}},\epsilon)$-block encoding of $A$ if
\begin{equation}
\label{eq:block-encoding}
   \left\|
   A
   -
   \beta
   \left(\bra{0}^{\otimes n_{\mathrm{aux}}}\otimes I\right)
   U_A
   \left(\ket{0}^{\otimes n_{\mathrm{aux}}}\otimes I\right)
   \right\|
   \leq \epsilon ,
\end{equation}
where $n_{\mathrm{aux}}$ is the number of ancilla qubits, $\beta$ is a
normalization factor, and $\epsilon$ is the block-encoding error
\cite{Gily_n_2019,Low_2019}. For an exact block encoding ($\epsilon=0$), applying $U_A$ to an input state $\ket{\psi}$ gives
\begin{equation}
    U_A\left(
    \ket{0}^{\otimes n_{\mathrm{aux}}}\otimes\ket{\psi}
    \right)
    =
    \ket{0}^{\otimes n_{\mathrm{aux}}}
    \otimes\frac{A\ket{\psi}}{\beta}
    +
    \ket{\Phi_\perp},
\end{equation}
where $\ket{\Phi_\perp}$ contains components for which the ancilla
register is orthogonal to
$\ket{0}^{\otimes n_{\mathrm{aux}}}$. Measuring the ancillas and
postselecting the outcome
$\ket{0}^{\otimes n_{\mathrm{aux}}}$ therefore selects the desired
action of $A$, producing a normalized state proportional to
$A\ket{\psi}$. The corresponding success probability is
\begin{equation}
    \label{eq:success_prob}
    p_{\mathrm{succ}}
    =
    \left\|
    \left(\bra{0}^{\otimes n_{\mathrm{aux}}}\otimes I\right)
    U_A
    \left(
    \ket{0}^{\otimes n_{\mathrm{aux}}}\otimes\ket{\psi}
    \right)
    \right\|_2^2
    =
    \frac{\|A\ket{\psi}\|_2^2}{\beta^2}.
\end{equation}
Alternatively, amplitude-amplification techniques can increase the weight of this successful component without directly postselecting it.
Efficient block encodings are known for several classes of sparse and
structured matrices
\cite{Low_2019,Gily_n_2019,Chakraborty2019,Martyn2021}. For general CFD discretizations, however, constructing such a block encoding at the gate level remains a significant practical challenge, as the implementation cost depends on the matrix sparsity and structure, ancilla requirements, and the cost of accessing and mapping matrix entries within the quantum circuit.

Setty~\cite{1nts-v6y9} addressed this gate-level implementation challenge by introducing a block-encoding framework that bridges abstract oracle-based descriptions with explicit quantum circuits. The framework focuses on index mapping and uses coherent permutation of quantum-state amplitudes to rearrange computational-basis states before controlled operations are applied, enabling shift, delete, and insert operations to be unified and groups of multi-controlled operations to be compressed. The construction applies to general sparse matrices, with efficiency gains for structured cases. Demonstrations on representative examples and resource analysis on IBM superconducting backends showed significant reductions in circuit depth compared with unoptimized implementations.

Setty~\cite{Setty_2026} subsequently made the connection between QSVT-based linear-system solving and differential-equation applications more explicit by combining block encoding with QSVT and applying the resulting framework to CFD-relevant examples, including the heat equation with mixed boundary conditions and a Carleman-linearized Burgers' equation. The study also provides superconducting architecture-level resource
estimates in terms of two-qubit gate depth. For the heat-equation benchmarks on the IBM heavy-hex layout, a nine-grid-point problem required a two-qubit circuit depth of $2.58\times10^{5}$, increasing to $9.03\times10^{5}$ for $17$ grid points, with post-selection probabilities of approximately $4$--$5\%$. For the Carleman-linearized Burgers' equation, a seven-grid-point problem required two-qubit depth of about $3.4\times10^{6}$ on the heavy-hex layout, with a post-selection probability of about $8.7\%$. Using an IBM square-lattice layout reduced these resource requirements by a substantial margin. These results link QLSA-style PDE algorithms to concrete hardware-level resource requirements and illustrate the importance of efficient block encoding, post-selection success probability, circuit depth, and hardware connectivity in guiding future implementations.

\subsubsection{Practical Limitations}

Despite their favourable asymptotic scaling in the system size $N$, QLSAs rely on several assumptions that must be satisfied for a practical quantum advantage in CFD. First, the remaining dependence on the condition number is important because the conditioning of many discretized operators deteriorates under mesh refinement. For representative elliptic subproblems, such as pressure--Poisson and diffusion equations, the spectral condition number of the standard discrete Laplacian on a quasi-uniform mesh scales as
\begin{equation}
    \kappa(A)=\mathcal{O}(h^{-2}),
\end{equation}
where $h$ is the grid spacing \cite{Elman2014,Saad2003}. For a grid
with $m$ points per coordinate direction, $h\sim 1/m$ and therefore $\kappa(A)=\mathcal{O}(m^2)$. In $d$ spatial dimensions, with
$N=m^d$ degrees of freedom, this becomes
\begin{equation}
    \kappa(A)=\mathcal{O}(N^{2/d}).
\end{equation}
Thus, mesh refinement can offset part of the favourable logarithmic-in-$N$ scaling of QLSAs.

Classical iterative solvers are also affected by ill-conditioning. For example, the iteration count of the conjugate-gradient method scales as $\mathcal{O}(\sqrt{\kappa})$ for symmetric positive-definite systems, up to logarithmic factors in the target accuracy \cite{Saad2003}. Consequently, classical CFD relies extensively on preconditioning and multigrid techniques to improve convergence \cite{preconditioning2012}. Quantum preconditioning introduces an additional implementation requirement: the preconditioned operator relies on efficient block encoding. Several block-encoding constructions have been developed for structured operators \cite{Clader2013,Shao2018,Lapworth2025}; however, practically relevant industrial-scale CFD operators are generally unstructured. The block-encoding framework introduced by Setty~\cite{1nts-v6y9} addresses this aspect by targeting general sparse matrices through coherent permutation-based index mapping, while exploiting available structure in the sparsity pattern to reduce circuit complexity.

A second assumption concerns state preparation. Preparing the input
state $\ket{b}$ may require $\mathcal{O}(N)$ operations when the data do not admit a compact representation \cite{Childs2017,Montanaro_2016}. This is particularly relevant in CFD, where the right-hand side may contain discretized field data, source terms, or boundary-condition contributions. Chen et
al.~\cite{Chen2021} showed that repeated data encoding can constitute a substantial fraction of the total cost in quantum finite-volume workflows. Efficient state preparation is therefore justified only
when the input can be generated from a compact analytical description, a structured circuit, or another efficient data-access procedure. Clader \textit{et al.}~\cite{Clader2013}, for example, demonstrated reduced
state-preparation costs for problems with efficiently computable right-hand sides.

A related limitation arises at the output stage. QLSAs return a quantum state proportional to the solution rather than an explicit classical solution vector. Recovering the complete field requires an amount of measurement and classical-output data that scales with the system size
and can remove the advantage of the compact quantum representation. QLSA-based CFD methods are therefore most promising when the desired outputs are selected observables, flow statistics, or other reduced quantities of interest rather than the full instantaneous solution field \cite{Griffin2019,Malinverno2025}.

Quantum random access memory (QRAM) has been proposed as a possible means of accelerating data loading \cite{Giovannetti2008}. However,
because a scalable physical realization of QRAM remains an open challenge, complexity claims that rely on QRAM should be distinguished from approaches in which the input state is generated directly from problem structure.

Experimental implementations of HHL have so far remained restricted to small, highly structured systems. Cai \textit{et al.}~\cite{Cai2013}
implemented a compiled photonic version for a $2\times2$ system, while Pan \textit{et al.}~\cite{Pan2014} demonstrated the algorithm on a four-qubit nuclear magnetic resonance processor. In both cases, shallow circuits were obtained by exploiting prior knowledge of the matrix spectrum and simplifying the phase-estimation procedure.

These demonstrations establish the feasibility of the basic HHL protocol, but their gate counts, typically of the order of tens of gates, should not be interpreted as estimates for scalable implementations. General QLSAs require fault-tolerant implementations of block encoding, Hamiltonian simulation, phase estimation, and
controlled rotations, with resource requirements determined by the problem size, condition number, target precision, and cost of implementing the underlying operator.
%%%%%%%%%%%%%%%%%%%%%%%%%%%%%%%%%%%%%%%%%%%%%%%%%%%%%%%%%%%%
\subsection{Hamiltonian Simulation and Quantum Time-Stepping}
\label{Sec:QTS}

Hamiltonian simulation methods directly encode the time evolution of a system into a unitary operator of the form $e^{-\mathrm{i}Ht}$, where $H$ is a Hermitian operator. This approach is naturally suited to linear, conservative systems whose dynamics preserve the $\ell^2$-norm. Rather than reformulating the dynamics as a linear system, the quantum device simulates the continuous-time evolution of the system state. These methods therefore avoid matrix inversion and postselection, and instead rely on efficient implementations of unitary time evolution using techniques such as Trotterization, qubitization, or quantum
signal processing
\cite{Berry_2015,Low_2019,Low2017,Brearley2024,Over2024b,Bengoechea2026}.

This distinction is particularly relevant for CFD, where at least linear scaling with evolution time is required for practical feasibility, since CFD simulations typically involve many successive time steps, particularly when treating nonlinear and diffusive dynamics through quantum time stepping. QLSA-based methods can address a broader range of linear or linearized problems, including elliptic equations, whereas Hamiltonian simulation is most directly applicable when the discretized dynamics admit an efficient unitary representation. In such cases, it provides a natural framework for simulating time-dependent evolution. Several recent works illustrate how this framework can be extended to transport and dissipative dynamics. This is demonstrated in \cite{Bengoechea2026}, where the non-unitary character of diffusion is addressed, while the treatment of advection is demonstrated in \cite{Brearley2024}. Ref.~\cite{Over2024b} combines diffusion and advection. All three
approaches retain linear scaling with evolution time.

\subsubsection{Quantum Algorithm for the Advection Equation}

Brearley and Laizet~\cite{Brearley2024} proposed a Hamiltonian-based quantum algorithm for the linear advection equation,
\begin{equation}
    \frac{\partial u}{\partial t}
    + c\frac{\partial u}{\partial x}=0,
\end{equation}
where $c$ is the constant advection velocity. In general, after spatial
discretization and periodic boundary conditions, the evolution can be written as
\begin{equation}
    \frac{d\mathbf{u}}{dt}=L\mathbf{u},
\end{equation}
where the discrete derivative operator $L$ is sparse. In case of an anti-Hermitian operator, the Hamiltonian $H=\mathrm{i}L$
therefore yields
\begin{equation}
    \mathbf{u}(t)=e^{-\mathrm{i}Ht}\mathbf{u}(0),
\end{equation}
allowing the dynamics to be implemented directly using Hamiltonian
simulation. The discretized field is represented through amplitude encoding,
\begin{equation}
    \ket{\psi(t)}
    =
    \frac{1}{\|\mathbf{u}(t)\|}
    \sum_{j=0}^{N-1}u_j(t)\ket{j},
\end{equation}
requiring $n=\lceil\log_2N\rceil$ qubits for a grid with $N$ points.
The resulting evolution,
\begin{equation}
    \ket{\psi(t)}
    =
    e^{-\mathrm{i}Ht}\ket{\psi(0)},
\end{equation}
is unitary and preserves the norm of the encoded field. Brearley and Laizet~\cite{Brearley2024} exploit the sparsity of the finite-difference operator and formulate its time evolution using established sparse-Hamiltonian simulation methods \cite{Berry_2015,Low_2019}. Assuming that the nonzero entries of the sparse discretized operator can be queried efficiently by the quantum algorithm, the complexity scales nearly linearly with the evolution time and polylogarithmically with the system size.

The method is therefore well suited to linear, near norm-preserving transport problems for which the discretized generator admits an efficient Hamiltonian representation. Its extension to nonlinear or dissipative flows requires additional reformulation. Subsequent work extended this quantum time-marching framework to multidimensional linear transport problems with arbitrary boundary conditions \cite{Bengoechea2026} for which the success probabilities are intrinsic to the problem. The method adapts the linear-combination-of-unitaries (LCU) framework to block-encode the diffusive dynamics, while the boundary conditions are enforced through the method of images. The resulting algorithm achieves optimal success probabilities with linear time complexity. Related Hamiltonian-based methods for linear differential equations are discussed in Refs.~\cite{Berry_2017,Childs_2021}.

\subsubsection{Advection–Diffusion and Optimal Success Probability}

The advection–diffusion equation augments conservative transport with a dissipative mechanism,
\begin{equation}
    \frac{\partial u}{\partial t}
    + c \frac{\partial u}{\partial x}
    = \nu \frac{\partial^2 u}{\partial x^2},
\end{equation}
where $\nu > 0$ is the diffusion coefficient. After spatial discretization on a grid with $N$ points, the equation can again be written as a linear system of ordinary differential equations,
$
\frac{d\mathbf{u}}{dt} = L \mathbf{u},
$
where $L = L_{\text{adv}} + L_{\text{diff}}$ consists of an anti-Hermitian advection operator and a diffusion operator arising from the discrete Laplacian. While $L_{\text{adv}}$ can be treated as in \cite{Brearley2024} to generate unitary evolution, the diffusion term is symmetric negative semidefinite, as is standard for discrete Laplacian operators arising from finite-difference discretizations \cite{LeVeque2007}. Consequently, the evolution operator
$
\mathbf{u}(t) = e^{Lt}\mathbf{u}(0)
$
is non-unitary and contracts the $\ell^2$-norm of the solution, reflecting the dissipative nature of the underlying physics.

This loss of unitarity prevents the direct application of Hamiltonian simulation techniques, which require evolution under a Hermitian generator. To address this, Over \textit{et al.}~\cite{Over2024b} recombine the advective and diffusive contribution of $L$ using the LCU method, for which the advection part is implemented with the method proposed in \cite{Brearley2024}. The weighting factors are selected such that the subnormalization factors are canceled out and thus the algorithm achieves optimal, problem intrinsic, success probabilities. 

A central result of \cite{Over2024b} is the characterization and optimization of the success probability. 
Because diffusion leads to norm contraction, the probability of successfully recovering the dissipatively evolved state is necessarily less than one and depends on the evolution time and diffusion strength. Over \textit{et al.}~\cite{Over2024b} showed that their method achieves asymptotically optimal scaling of the success probability, avoiding the exponentially small success rates that arise in more naive embeddings. This reduces the overhead associated with amplitude amplification as the time complexity is maintained linear and makes long-time simulations more tractable.

More generally, this approach fits within a broader class of quantum algorithms for non-unitary dynamics, including LCU methods and block-encoding techniques for open-system evolution \cite{Berry_2015, Low_2019, Childs_2021}. These methods similarly embed dissipative processes into unitary circuits, but often incur significant overhead in success probability or ancilla resources. The construction of Over \textit{et al.}~\cite{Over2024b} demonstrates that, for
advection--diffusion systems, it is possible to achieve near-optimal scaling in this respect. Their approach therefore extends Hamiltonian-based simulation to dissipative PDEs while retaining
controlled and asymptotically optimal success-probability scaling. 

An alternative perspective arises from engineered dissipation and quantum-cooling algorithms, where energy is removed from the system
through interactions with auxiliary degrees of freedom or engineered reservoirs. Such approaches provide a different route to simulating
non-unitary dynamics and to preparing low-energy states
\cite{Harrington_2022,Mi2024,Matthies_2024,Lloyd2025,Molpeceres2025}. More recently, Molpeceres \textit{et al.}~\cite{Molpeceres2026} compared cooling, adiabatic, and optimization-based ground-state preparation algorithms in the presence of noise. Their results show that the relative performance depends on the physical regime, with multi-frequency cooling becoming competitive or superior when
gap-closing limits adiabatic preparation, while also exhibiting enhanced robustness to parameter imperfections. Although these results
concern ground-state preparation rather than PDE time evolution, they
illustrate how engineered open-system dynamics can provide useful alternatives to purely unitary quantum algorithms.

\subsubsection{Advantages and Limitations of Hamiltonian-Based Time-Stepping}
\label{subsubsec:Adv_Lim_Hamil}
Modern Hamiltonian-simulation techniques can implement
$e^{-\mathrm{i}Ht}$ with complexity that is nearly linear in the evolution time $t$ and polylogarithmic in the system dimension $N$, provided that the generator admits an efficient quantum representation
\cite{Berry_2015,Low_2019}. This avoids explicit classical time-stepping and replaces the associated temporal discretization error with the approximation error of the quantum simulation procedure.

The practical applicability of this approach depends primarily on the structure of the discretized generator. Efficient simulation requires that the generator admit a suitable sparse representation or block encoding. While this condition can be satisfied for certain local and structured discretizations, it is not guaranteed for general unstructured CFD operators.

A further limitation arises for dissipative systems, whose generators are generally non-Hermitian. Such dynamics cannot be implemented directly as unitary quantum evolution and must instead be embedded into
a larger Hermitian system \cite{Over2024b}. This increases the qubit and circuit-depth requirements and may introduce a probabilistic step in recovering the physical solution. Although the dependence on the system dimension can be polylogarithmic, the simulation cost scales at least linearly with the evolution time. Long-time simulations may therefore remain computationally demanding even on a quantum computer \cite{Berry_2017,Childs_2021}. The general requirements associated with efficient state preparation and restricted quantum-state readout also apply to Hamiltonian-based methods, as discussed previously for QLSAs.

On near-term quantum devices, the intended evolution $e^{-iHt}$ is typically realized either as the effective evolution of a controlled many-body system in analog devices or approximated using quantum
circuits in digital devices. In both cases, the implemented evolution generally deviates from the ideal operator due to coherent control errors, dissipation, higher-order Trotter errors, and other hardware imperfections.

Hamiltonian and Lindbladian learning techniques
\cite{Bairey2019,Olsacher2025,Kraft2026} provide a toolbox for efficiently reconstructing the implemented generator; see also the references therein. This includes, for instance, the effective Trotterized Hamiltonian and associated Trotter errors \cite{Pastori2022}. This makes it possible to verify whether the intended operator has been realized and, when deviations occur, to identify and diagnose errors in the computation. Analogously, Lindbladian learning can characterize the effective dissipative dynamics. The resulting uncertainties in $H$ and $\mathcal{L}$ can then be propagated to simulated observables \cite{Kraft2026}, yielding quantitative error bounds for Hamiltonian-simulation and quantum time-stepping approaches. This provides a route toward certified analog and digital quantum simulations whose predictions are accompanied by experimentally derived error bounds. These approaches were experimentally demonstrated by Kraft \textit{et al.}~\cite{Kraft2026} on a trapped-ion quantum simulator implementing long-range Ising interactions with up to 51 ions.

Overall, Hamiltonian simulation is particularly promising for structured linear PDEs whose discretized generators admit efficient unitary representations, while dissipative dynamics, long evolution times, complex operator constructions, and reliable verification of the implemented dynamics remain important practical challenges.

%%%%%%%%%%%%%%%%%%%%%%%%%%%%%%%%%%%%%%%%%%%%%%%%%%%%%%%%%%%%
\subsection{Quantum Lattice Boltzmann Methods}
\label{Sec:QLBM}

The lattice Boltzmann method (LBM) is a mesoscopic approach to CFD in which macroscopic fluid variables
are recovered from the evolution of particle distribution functions defined on a discrete lattice \cite{Succi2001,Kruger2017}. Rather than directly discretizing the Navier–Stokes equations,
LBM evolves a set of distribution functions
$f_i(x,t)$ associated with discrete velocities
$\mathrm{c}_i$ according to a streaming–collision scheme:
\begin{equation}
    f_i(x+\mathrm{c}_i \Delta t, t+\Delta t)
    =
    f_i(x,t)
    - \frac{\Delta t}{\tau}
    \left(
        f_i(x,t)
        - f_i^{\mathrm{eq}}(x,t)
    \right),
\end{equation}
where $\tau$ is a relaxation parameter and
$f_i^{\mathrm{eq}}$ denotes the local equilibrium distribution. Macroscopic quantities are obtained by taking moments of the distribution functions,
\begin{equation}
    \rho = \sum_i f_i,
    \qquad
    \rho u = \sum_i \mathrm{c}_i f_i.
\end{equation}
In the hydrodynamic limit, and under appropriate scaling assumptions, the method recovers the incompressible Navier–Stokes equations \cite{Y.H.Qian_1992,Chen1998}. The appeal of LBM in classical CFD stems from the locality of its updates, the absence of global Poisson solves, and its natural parallelizability, since streaming and collision operations are performed locally on the lattice.

\subsubsection{Quantum Formulation}

QLBMs encode the lattice populations in a quantum state and implement
the streaming--collision dynamics using quantum circuits. For a
one-dimensional lattice, a representative amplitude encoding is
\begin{equation}
    \ket{\Psi(t)}
    =
    \sum_{j,i}
    \psi_i(x_j,t)\ket{x_j}\ket{i},
\end{equation}
where $\ket{x_j}$ labels the lattice site, $\ket{i}$ labels the
discrete velocity, and the amplitudes encode suitably normalized
distribution data. The position register requires $\lceil\log_2 N\rceil$ qubits for a lattice with $N$ sites, together with additional qubits for the velocity and ancillary registers \cite{Montanaro_2016}. The streaming step is naturally compatible with quantum computation, since
\begin{equation}
    \ket{x}\ket{i}
    \mapsto
    \ket{x+\mathrm{c}_i}\ket{i}
\end{equation}
is a permutation of computational basis states and is therefore unitary. Controlled-shift and quantum-walk constructions of this type were already used in early quantum lattice-gas and lattice-Boltzmann-inspired methods \cite{Aharonov2002,Ambainis2014}.

The collision step presents a greater challenge and could involve nonlinearities depending on the equilibrium formulation. Classical Bhatnagar--Gross--Krook collision is dissipative and cannot generally be implemented directly as a unitary operation. Existing QLBM formulations therefore employ approximations, enlarged unitary embeddings, LCU constructions, or linearizations in which collision becomes compatible with quantum evolution
\cite{Stinespring1955,Watrous_2018,Gily_n_2019}.

Budinski developed QLBM algorithms for advection--diffusion and later for Navier--Stokes-type problems using quantum-walk and
LCU constructions \cite{Budinski2021,Budinski2022}. These studies demonstrated that streaming and collision updates can be represented by quantum circuits, but commonly required measurement and state reinitialization between time steps, limiting the efficiency of long-time simulations.

More recent work has addressed specific components of this workflow.
Georgescu and M{\"o}ller \cite{georgescu2026efficientexpressiveboundaryconditions} developed
more efficient constructions for imposing boundary conditions in
QLBMs, showing how boundary updates can be incorporated through
localized circuit operations rather than a global reconstruction of
the flow field. Duong \textit{et al.}~\cite{duong2026quantumlatticeboltzmanndenoising} introduced denoising collision operators that avoid tomography-based collision updates by reformulating the relaxation process as a projection onto a linearized equilibrium representation. 

L{\u a}c{\u a}tu{\c s} and M{\"o}ller~\cite{Lacatus2025} recently proposed a learned surrogate quantum circuit for the nonlinear BGK collision operator on the D2Q9 lattice, namely the
two-dimensional lattice Boltzmann model with nine discrete velocities
\cite{Y.H.Qian_1992}. The circuit is designed to preserve mass conservation, scale equivariance, and $D_8$ symmetry, while momentum conservation is promoted through the training loss. In contrast to LCU-based collision constructions, the approach does not require ancilla qubits, postselection, or multiple copies of the state to reproduce dissipative and nonlinear effects. The final 15-block architecture was compiled to the IBM Heron native gate set $\{RZ,SX,CZ\}$, assuming all-to-all connectivity, and required $724$ native gates, including $95$ two-qubit $CZ$ gates. Since the collision circuit acts locally on the velocity register, this gate count is independent of the number of lattice sites. The method was validated on Taylor--Green vortex (TGV) decay and lid-driven cavity benchmarks at $\mathrm{Re}=10$ and $\mathrm{Re}=50$, but currently still requires measurement and reinitialization at each time step.

Itani, Sreenivasan, and Succi~\cite{Itani2024} addressed nonlinear BGK
collision through a Carleman-type embedding. The lifted dynamics were
represented in a bosonic Fock space, allowing streaming and collision to be treated within a unified unitary framework, at the cost of an enlarged state space and truncation overhead.

Wawrzyniak \textit{et al.}~\cite{Wawrzyniak2025} developed explicit circuit
constructions for initialization, collision, streaming, and
macroscopic measurement for the linear advection--diffusion equation
in one, two, and three dimensions. Their analysis also showed how the
choice of encoding affects gate complexity and sampling efficiency.
In subsequent work, they introduced a dynamic-circuit formulation that
uses mid-circuit measurements and adaptive circuit operations to reduce
the qubit and circuit overhead associated with earlier collision
implementations \cite{Wawrzyniak2025b}.

Wang \textit{et al.}~\cite{Wang2025} proposed a QLBM formulation for nonlinear
fluid dynamics in which the collision process remains linear and
quantum-compatible while preserving a mesoscopic interpretation.
Agreement with classical benchmark solutions demonstrated that QLBMs
can extend beyond purely linear transport, although the construction
still relies on a problem-specific reformulation of the collision
dynamics.

Tiwari \textit{et al.}~\cite{Tiwari2025} identified practical bottlenecks in
earlier QLBM proposals, including tomography, data loading, observable
estimation, and circuit depth, and introduced modifications intended
to improve hardware realizability. Their implementation solved a
two-dimensional advection--diffusion problem with a Gaussian initial
condition on a $16\times16$ grid using IonQ Forte trapped-ion
hardware. The circuit used $19$ qubits and approximately $260$
two-qubit gates. By using a matrix product state (MPS) based loading procedure, the state preparation for the ten-step hardware demonstration required only $28$ Controlled-NOT (CNOT) gates, while the full circuit for each time step required approximately $250$ CNOT gates. The measured density reconstruction achieved more than $92\%$ fidelity at all ten time steps, and more than $97\%$ fidelity for the first seven time steps. Their one-hot encoding of the velocity register also reduced the streaming-operator
two-qubit gate count; for example, for the D2Q5 model on a $16\times16$ lattice, the count was reduced from $480$ to $244$ CX gates, and for D2Q9 on the same lattice from $1152$ to $488$ CX gates. As in other amplitude-encoded quantum PDE algorithms, the end-to-end advantage therefore depends on efficient initialization and on extracting physically relevant observables without reconstructing the complete distribution field.

\subsection{Computational Trade-offs of Fully Quantum PDE Approaches}

The fully quantum approaches reviewed above differ mainly in how they
represent and evolve the discretized PDE. QLSAs encode the problem as
a global linear system and are therefore most naturally suited to
linear or linearized equations. Their practical performance depends
on efficient state preparation, operator access, matrix conditioning,
and the cost of extracting useful observables
\cite{Harrow2009,Childs2017,Gily_n_2019}. Hamiltonian simulation
methods instead implement continuous-time evolution generated by a
Hermitian operator, and are best suited to linear dynamics that are
unitary or admit an efficient Hermitian embedding
\cite{Berry_2015,Low_2019,Childs_2021}. QLBMs retain the mesoscopic
streaming--collision structure of the classical lattice Boltzmann
method and replace global matrix inversion with repeated local
updates. Their resource requirements are therefore governed primarily
by the implementation of the collision operator, the number of time
steps, and the cost of extracting the required observables
\cite{Budinski2021,Itani2024,Wawrzyniak2025,Tiwari2025}.

These algorithms rely on several assumptions that are important in
CFD applications. The first is efficient state preparation: the
right-hand side, initial condition, or flow field must be loaded into
a quantum state without an $\mathcal{O}(N)$ overhead that would remove
the benefit of amplitude encoding \cite{Aaronson2015,Dalzell_2025}.
A second assumption concerns operator encoding. QLSA and Hamiltonian
methods require sparse-access or block-encoded representations of the
discretized operators; for general CFD matrices, constructing these
representations can be as important as the quantum solver itself
\cite{Gily_n_2019,Low_2019}. Generic QRAM-based loading would provide
one possible access model, but it is a strong hardware assumption and
is not required in approaches based on structured state preparation or
TN encodings \cite{Dalzell_2025,Termanova2024}. A third
assumption is that the relevant CFD data possess exploitable
structure, such as sparsity, locality, or approximate low-rank
compressibility. TN studies of turbulent and multiscale
flows provide evidence that such structure can occur in selected
regimes, although it is not a universal property of all turbulent CFD
problems
\cite{Gourianov2022,Kiffner2023,H_lscher_2025}.

The main bottleneck is therefore shifted to different parts of the
solution procedure. QLSAs emphasize matrix conditioning, state
preparation, and block encoding; Hamiltonian methods require suitable
generators and sufficiently long coherent evolution; and QLBMs
emphasize collision implementation and repeated time stepping. These
trade-offs motivate hybrid variational and TN-based
methods. VQAs do not implement a global matrix inverse through HHL or
QSVT, and therefore do not inherit the explicit condition-number
dependence of QLSA complexity bounds. Instead, the PDE residual,
energy, or time-stepping error is incorporated into a variational cost
function. This does not remove conditioning entirely, since
ill-conditioned problems may still affect optimization convergence,
but it changes the dominant resource requirements from global matrix
inversion to ansatz expressivity, measurement cost, and trainability
\cite{McClean2016,Cerezo2021,Lubasch2020}.

An additional consideration is the gate-level implementation of the
resulting quantum operators. Arithmetic and controlled operations
appearing in quantum PDE circuits can involve multi-controlled Toffoli (MCT) gates, whose decomposition into elementary gates may introduce substantial circuit depth. Tserkis \textit{et al.}~\cite{Tserkis_2026_MCT} recently proposed a teleportation-based decomposition that implements an arbitrary MCT gate with unit Toffoli depth, independent of the number of controls, at the cost of a linear overhead in ancillary qubits. This construction requires that certain qubits share an initial entangled state. While this is achievable in any computational paradigm, it is especially natural in distributed quantum computing
architectures, where entanglement links between separate quantum
processing units are a native resource.

TN and tensor-programmable formulations provide a concrete way to exploit the structure required for this strategy.
MPS-based state preparation can avoid generic QRAM when the field has moderate bond dimension, while MPOs provide compact representations of structured differential operators and finite-difference stencils
that can be compiled into quantum circuits
\cite{Termanova2024,Siegl_2026}. Nonlinear terms can then be treated through variational cost functions or shallow operator circuits, rather than through a fully linearized global system. These features make VQA and TP-VQA methods a complementary route for quantum CFD, especially in regimes where the flow is sufficiently structured to be encoded compactly but too complex for classical tensor contractions
to remain efficient.

%%%%%%%%%%%%%%%%%%%%%%%%%%%%%%%%%%%%%%%%%%%%%%%%%%%%%%%%%%%%
\section{Hybrid Quantum-Classical Algorithms}
\label{Sec:Hybrid_Approach}
Variational quantum circuits provide a common framework for several hybrid quantum approaches to differential equations. A parameterized quantum circuit is a sequence of quantum gates whose action depends on a set of tunable classical parameters $\boldsymbol{\theta}$. Acting on an initial state, it prepares
\begin{equation}
    \ket{\psi(\boldsymbol{\theta})}
    =
    U(\boldsymbol{\theta})\ket{0},
\end{equation}
where $U(\boldsymbol{\theta})$ denotes the corresponding parameterized unitary circuit. The parameters are subsequently adjusted by a
classical optimizer according to a problem-dependent objective function. The quantum processor estimates the quantities entering a problem-dependent objective function, such as expectation values, overlaps, or residual norms. These estimates are supplied to a classical optimizer, which updates the parameters $\boldsymbol{\theta}$ and closes the variational loop
\cite{McClean2016,Cerezo2021,Jaksch2023}.

Different variational PDE approaches mainly differ in how the physical
solution is encoded and how the governing equation enters the
objective function. In physics-informed quantum machine learning,
spatial or temporal coordinates are encoded through a feature map and
the circuit acts as a function approximator. In amplitude-encoded
variational solvers, by contrast, the discretized field itself is
encoded in the amplitudes of the quantum state.

\subsection{Quantum Physics-Informed Neural Networks}
\label{Sec:QPINN}

\begin{figure}[h]
    \centering
    \includegraphics[width=0.60\linewidth]
    {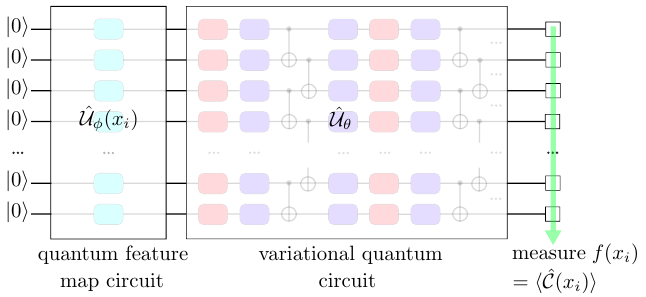}
    \caption{Circuit for physics-informed quantum machine learning. The circuit has a feature-map block $\phi(x_i)$, which encodes the input variable $x_i$ through a predefined nonlinear feature map. This is followed by a variational block $\hat{\mathcal{U}}_{\theta}$, where $\theta$ denotes the trainable parameters optimized with respect to a cost function. Finally, the target function $f(x_i)$ in the differential equation is obtained as the expectation value of the circuit with respect to a chosen operator $\hat{\mathcal{C}}$.
    Adapted from \cite{Setty2025}.}
    \label{fig:QPINNs}
\end{figure}

Quantum Physics-Informed Neural Networks (QPINNs) use parameterized quantum circuits as coordinate-dependent function approximators, providing a natural quantum analogue of physics-informed neural
networks \cite{Raissi2019, Kyriienko2021,Setty2025,Siegl2025,Tam2026-arxiv}. Kyriienko \textit{et al.}~\cite{Kyriienko2021} introduced a differentiable quantum-circuit
framework that approximates a target function using Chebyshev polynomials, as illustrated in Figure~\ref{fig:QPINNs}. The circuit consists of three main components: (i) a quantum feature map that encodes the input coordinates, (ii) a variational ansatz optimized with respect to a physics-informed cost function, and (iii) measurements used to extract the corresponding classical function values.

For a single independent variable $x_i$, the quantum state prepared by the circuit is given by $|f_{\phi, \theta} (x_i)\rangle = \hat{\mathcal{U}}_{\theta}\hat{\mathcal{U}}_{\phi}(x_i) |0 \rangle$. The corresponding real-valued classical function $f(x_i)$ is obtained through expectation value of an operator $\hat{\mathcal{C}}$, 
\begin{equation}
f(x_i) = \langle f_{\phi , \theta}(x_i) | \hat{\mathcal{C}} | f_{\phi, \theta}(x_i)\rangle.
\end{equation}
To impose the governing differential equation through the physics-informed loss, derivatives of the predicted function with respect to the independent variable,
$\frac{df(x)}{dx}$, are required. For this circuit, such derivatives depend on the feature map $(\phi(x))$, with $\frac{d \hat{\mathcal{U}}_{\phi}(x)}{d x} = \sum_j \hat{\mathcal{U}}_{d \phi ,j}(x).$ The Chebyshev feature map, which encodes Chebyshev polynomials into the circuit is given by 
\begin{equation}\label{eq:Chebyshev}
	\hat{\mathcal{U}}_{\phi}(x) = \bigotimes_{j=1}^n R_{Y,j}(2j \arccos{x}),    
\end{equation}
where $j$ is the qubit index of $n$-qubit circuit. The solution of the differential equation is then obtained by evaluating the residual and boundary losses at the selected collocation points and minimizing the total loss function with a classical optimizer.

Setty \textit{et al.}~\cite{Setty2025} extended this formulation by introducing trainable self-adaptive weights to balance the competing residual ($\mathcal{L}_f$), and boundary-condition ($\mathcal{L}_b$), objectives:
\begin{equation}
\mathcal{L}(\boldsymbol{\theta},\boldsymbol{\lambda}_f,\boldsymbol{\lambda}_b)
=
\mathcal{L}_f(\boldsymbol{\theta},\boldsymbol{\lambda}_f)
+
\mathcal{L}_b(\boldsymbol{\theta},\boldsymbol{\lambda}_b).
\end{equation}
Here, the weights $\boldsymbol{\lambda}_f$ and $\boldsymbol{\lambda}_b$ are optimized jointly with the circuit parameters $\boldsymbol{\theta}$. Assigning weights to individual collocation and boundary points enables the algorithm to dynamically emphasize constraints that are more difficult to satisfy. This self-adaptive physics-informed quantum machine-learning formulation is particularly relevant to CFD, where PDE
residuals and boundary conditions can exhibit substantially different scales and convergence rates.

Setty \textit{et al.}~\cite{Setty2025} demonstrated this framework across a diverse suite of problems, nonlinear Riccati and Duffing equations, coupled systems of differential equations, a second-order linear equation, and the two-dimensional Poisson equation. Furthermore, they showed that increasing the expressibility of the variational quantum circuit through additional entangling layers and
incorporating quantum-correlated measurements to capture correlations in the loss can improve convergence and solution accuracy, offering a promising path toward evaluating differential equations on near-term quantum hardware.

Tam \textit{et al.}~\cite{Tam2026-arxiv} further incorporated the geometric symmetry of the problem into the circuit architecture itself, ensuring that the solution respects the underlying invariances by construction. Across a range of benchmark PDEs—including the Poisson, diffusion, and Burgers' equations—these symmetry-aware circuits achieve higher final accuracy than standard QPINNs at equivalent number of training parameters.

Beyond physics-informed variational approaches, a related data-driven direction is the quantum Koopman method of Zhang \textit{et al.}~\cite{zhang2025datadrivenquantumkoopmanmethod}, which combines Koopman operator theory with deep autoencoder embeddings to represent
nonlinear dynamics as linear unitary evolution in higher-dimensional observable spaces. Although this approach is not a residual-minimizing VQA, it is relevant to quantum CFD because it addresses the same nonlinearity bottleneck and was benchmarked on reaction--diffusion systems, shear flows, and two-dimensional turbulence.

\subsection{Amplitude-Encoded Variational PDE Solvers}
\label{Sec:StandardVGA}

A second class of variational PDE algorithms represents the discretized solution field directly in the amplitudes of a parameterized quantum state. In contrast to QPINNs, where the circuit acts as a
coordinate-dependent function approximator, the complete discretized field is encoded in the quantum state. Differential and nonlinear
operators are then evaluated through dedicated quantum circuits and incorporated into a problem-dependent objective function.

\begin{figure}[t]
    \centering
    \includegraphics[width=0.75\linewidth]
    {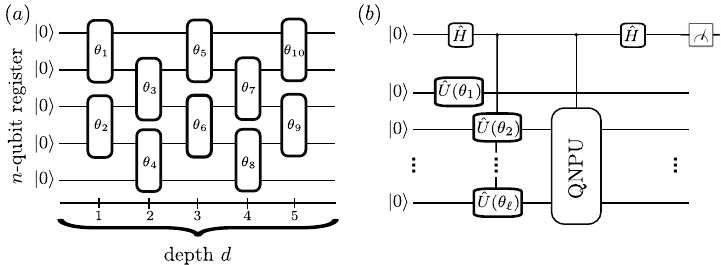}
    \caption{(a) A five-layer ($d=5$) unitary network in which the
    rotational parameters $\theta_i$ determine the action of the
    underlying two-qubit gates. (b) Schematic representation of a
    hybrid VQA for nonlinear problems, where parameterized unitary
    operations $\hat{U}(\theta_i)$ are applied to the quantum register
    and evaluated using a QNPU.
    Adapted from \cite{Lubasch2020}.}
    \label{fig:VQC}
\end{figure}

Unlike direct gate-based implementations of time evolution, amplitude-encoded VQAs do not attempt to realize the complete sequence
of propagators or inverse operators explicitly. This is attractive for PDE applications, where long-time evolution and repeated application of non-unitary operators may otherwise require deep circuits, leading to substantial noise accumulation and, in probabilistic
implementations, decreasing overall success probabilities \cite{Preskill2018,Low_2019}.
A representative parameterized quantum circuit and the corresponding
hybrid QNPU workflow for nonlinear PDEs are illustrated in Figure~\ref{fig:VQC}.

In the amplitude-encoded setting, the variational state $\ket{\psi(\boldsymbol{\theta})}$ represents the discretized solution field. The quantum processor evaluates quantities entering the objective function, such as expectation values, overlaps, or residual norms, which are then supplied to the classical optimizer
\cite{McClean2016,Cerezo2021,Jaksch2023}. The structure of these measurements depends on the differential and nonlinear operators appearing in the governing equation.

In benchmark problems, where a reference solution is available, the quality of the variational solution can be quantified using the fidelity. For two discretized fields $u_1$ and $u_2$, treated as vectors, we define
\begin{equation}
    F
    =
    \frac{
    \left|u_1^{\dagger}u_2\right|^2
    }{
    \|u_1\|_2^2 \, \|u_2\|_2^2
    },
\end{equation}
where $(\cdot)^\dagger$ denotes the conjugate transpose. This
quantity measures the normalized overlap between the two fields:
$F=1$ indicates perfect agreement up to an overall normalization or
phase, while smaller values indicate a less accurate approximation.
The corresponding infidelity is defined as
$\epsilon_F=1-F$. In practical CFD applications, where an exact
reference solution is usually unavailable, residual norms, conserved
quantities, relative $\ell^2$ errors, or physically relevant observables
provide complementary accuracy measures.

For PDEs, the objective function may be constructed through residual
minimization, energy-based variational principles, or space--time embeddings
\cite{Lubasch2020,Trahan2023,Sato2021Poisson,Barison2022,Pool2024}.
For a discretized equation of the form $\mathcal{L}[u]=g$, a representative residual-based objective is
\begin{equation}
    C(\boldsymbol{\theta})
    =
    \left\|
    \mathcal{L}
    [u_{\boldsymbol{\theta}}]
    -
    g
    \right\|_2^2,
\end{equation}
where $u_{\boldsymbol{\theta}}$ denotes the discretized trial solution encoded in the parameterized quantum state. Boundary conditions and conservation constraints may be incorporated through additional contributions to the objective function \cite{Over2024a}. In practice, the objective is estimated using quantum measurements rather than by reconstructing all amplitudes of the solution state. The resulting computational cost is instead governed by repeated measurements, non-convex optimization, and the expressivity of the chosen ansatz.

The optimization proceeds through a hybrid loop in which the circuit parameters are updated using measurement-based estimates of the objective function. Depending on the availability of gradients, the noise level, and the cost of circuit evaluations, derivative-free,
stochastic, or gradient-based methods may be employed \cite{NocedalWright2006,Spall1992,Spall1998ANOO}. For time-dependent problems, the parameters obtained at one time step can be used to initialize the optimization at the next. Such warm-start strategies can accelerate convergence when the solution evolves smoothly \cite{Lubasch2020,Jaksch2023}.

A representative framework for nonlinear PDEs was introduced by
Lubasch \textit{et al.}~\cite{Lubasch2020}. The discretized solution is encoded
in the amplitudes of a parameterized quantum state, while a QNPU, shown schematically in Figure~\ref{fig:VQC}b, evaluates nonlinear contributions using multiple independently prepared copies of the variational state. This provides a mechanism for incorporating nonlinear operators into the objective
function without implementing the nonlinear dynamics as a deep
time-evolution circuit. Related work by Bengoechea \textit{et al.}~\cite{Bengoechea_2025} extended this perspective to generalized linear and nonlinear transport phenomena, including spatially variable material constants and upwind-biased discretization schemes.

More recently, da Costa Jesus \textit{et al.}~\cite{Jesus2026} experimentally implemented this variational time-propagation framework for the viscous and inviscid Burgers equations on superconducting quantum hardware. The nonlinear convective term is evaluated directly through higher-order overlaps between amplitude-encoded states, avoiding an explicit linearization of the nonlinear dynamics. The authors demonstrated iterative propagation over multiple time steps, including a convection-dominated viscous regime with an effective Reynolds number of approximately $100$, where nonlinear steepening competes strongly with diffusion.

Building on this framework, Umer \textit{et al.}~\cite{Umer_2025b} developed a
problem-tailored method for the efficient evaluation and optimization
of VQA cost functions. Their approach represents a parameterized
quantum circuit as a weighted combination of distinct unitary
operators, enabling the evaluation of nonlocal properties of the cost
function and its higher-order derivatives. The resulting sequential
optimization protocol uses this additional information to navigate the optimization landscape more efficiently.

The method was demonstrated for two nonlinear problems with different
objective functions. For the one-dimensional viscous Burgers' equation, the objective was formulated as the squared residual between the variational state and a time-evolved target state. For the nonlinear Schr\"odinger equation (NLSE), an energy expectation value was minimized to approximate the ground state energy. In both cases,
the proposed optimization procedure improved convergence speed and
accuracy relative to conventional optimization methods
\cite{Umer_2025b}.

These two problems moreover provide complementary benchmarks for quantifying the quantum resources required by different nonlinear operators. The NLSE contains a local density-dependent interaction, whereas the Burgers' equation contains a convection-type nonlinearity that couples the field to its spatial derivative. Their corresponding QNPU constructions therefore require different numbers of independently prepared state registers.

\begin{figure}[t]
    \centering
    \includegraphics[width=0.90\linewidth]
    {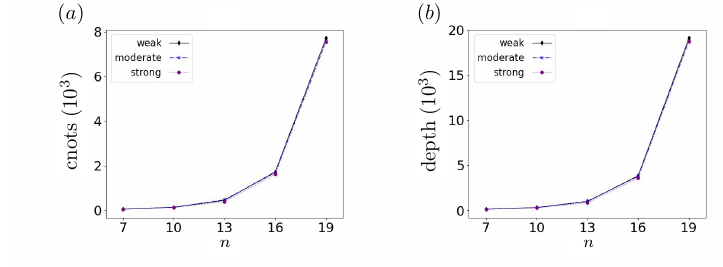}
    \caption{Estimated circuit-resource requirements for the NLSE. The CNOT-gate count and circuit depth are shown as functions of the total number of qubits for weak, intermediate, and strong nonlinear interaction regimes. The estimates are based on variational circuits achieving solution fidelities above $99\%$. Adapted from
    \cite{Umer_2025b}.}
    \label{fig:NLSE_resources}
\end{figure}

For the nonlinear interaction in the NLSE, three independently prepared field registers and one ancillary qubit are required, giving a total circuit width of
\begin{equation}
    n_{\mathrm{NLSE}}
    =
    3n+1.
\end{equation}
For $n=2$, $3$, and $4$ system qubits, corresponding to $N=2^n=4$, $8$, and $16$ grid points, this gives total circuit widths of
$n_{\mathrm{NLSE}}=7$, $10$, and $13$ qubits, respectively (see Figure~\ref{fig:NLSE_resources}). 

Gate-level simulations showed that the required CNOT count and circuit depth increase rapidly with problem size. The resource estimates for the weak, moderate, and strong nonlinear regimes remain very similar
over the investigated range, indicating only a weak dependence on the nonlinearity strength. The resulting resource trends were obtained for variational states achieving fidelities above $99\%$ with respect to the corresponding reference solutions.

The Burgers' equation provides a more directly fluid-dynamical
benchmark. Its nonlinear convection term, $u\partial_xu$, captures
important features associated with nonlinear transport, steep
gradients, and shock formation. Evaluation of this term requires two
independently prepared field registers and one ancillary qubit,
resulting in

\begin{equation}
    n_{\mathrm{Burgers}}
    =
    2n+1.
\end{equation}

\begin{figure}[t]
    \centering
    \includegraphics[width=0.30\linewidth]
    {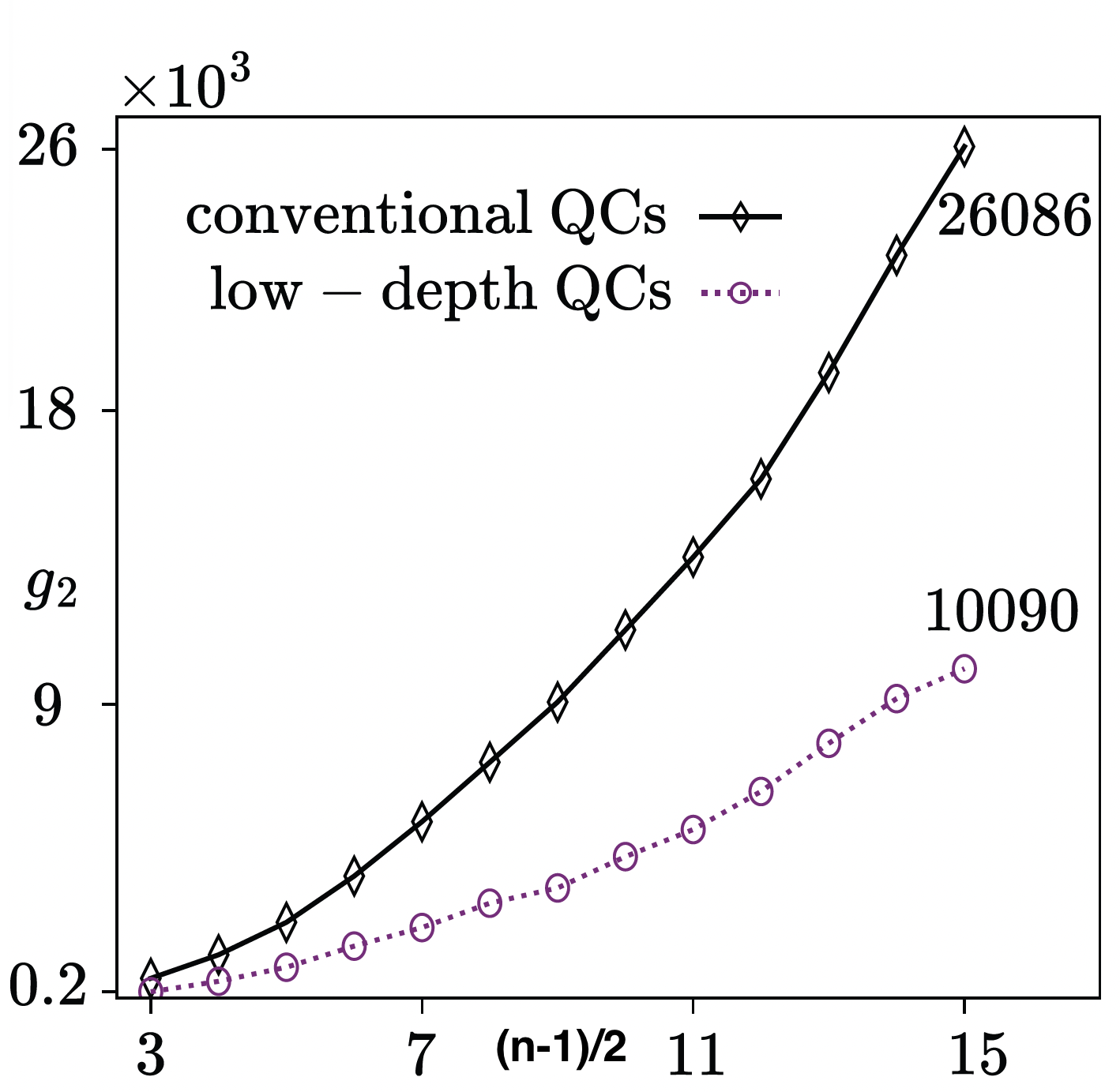}
    \caption{Scaling of the two-qubit gate count $g_2$ with problem size, expressed in terms of the total number of qubits. The black and purple curves correspond to the conventional and optimized low-depth circuit constructions, respectively. The optimized implementation reduces the two-qubit gate requirement by
    approximately a factor of three over the investigated system
    sizes. Adapted from
    \cite{Mastorakis_2026}.}
    \label{Fig:Burgers_cnot}
\end{figure}

Mastorakis \textit{et al.}~\cite{Mastorakis_2026}
developed an optimized low-depth Hadamard-test implementation for the Burgers' equation. For the investigated benchmark configurations, the optimized construction reduced the number of two-qubit gates by approximately a factor of three relative to the conventional circuit
while retaining high-fidelity representations of the nonlinear
dynamics. As shown in Figure~\ref{Fig:Burgers_cnot}, this reduction persists as the problem size increases.

These case studies indicate that the quantum resources required for nonlinear PDEs depend not only on the number of spatial degrees of freedom, but also on the form of the nonlinear operator and the circuit construction used to evaluate it. For the specific QNPU implementations
considered here, the NLSE construction uses one additional amplitude-encoded field register compared with the Burgers' construction. This difference should not be interpreted as a universal
resource relation between the two equations, since the required circuit width depends on the chosen formulation and implementation. Moreover, circuit width alone does not determine the total computational cost, which also depends on circuit depth, entangling-gate count, measurement requirements, and optimization convergence.

Measurement cost can also be reduced through the design of the variational ansatz itself. Umer and Angelakis~\cite{Umer2026Adaptive} recently developed an adaptive variational framework in which hardware-connectivity and Hadamard-test circuit constraints are incorporated directly into the candidate gate pool. By reducing the
number of candidate gates that must be evaluated at each iteration, the approach reduces the estimated measurement resources by at least
$25\%$, with reductions exceeding $50$--$55\%$ during the early iterations. Using the NLSE ground-state problem as a benchmark, the
authors showed that these constraints retain the ability of the adaptive procedure to construct expressive, low-depth ans\"atze. These results illustrate that resource requirements can be reduced by considering problem structure, hardware connectivity, and circuit design jointly rather than treating ansatz construction and hardware implementation as separate steps.

Fully quantum algorithms have also been proposed for the Burgers' equation using Carleman linearization, Cole--Hopf transformations, and direct circuit-based formulations \cite{Liu_2021,Wu_2025,Uchida_2026,Esmaeilifar2024}. However, most of these studies report asymptotic complexity estimates or proof-of-principle implementations rather than gate counts and circuit depths evaluated under the same spatial resolution, time horizon, and accuracy requirements considered here. Consequently, the available results do not yet support a direct resource comparison with the variational implementation, and establishing such a common benchmark remains an interesting direction for future work.

\begin{figure}[t]
    \centering
    \includegraphics[width=0.99\linewidth]
    {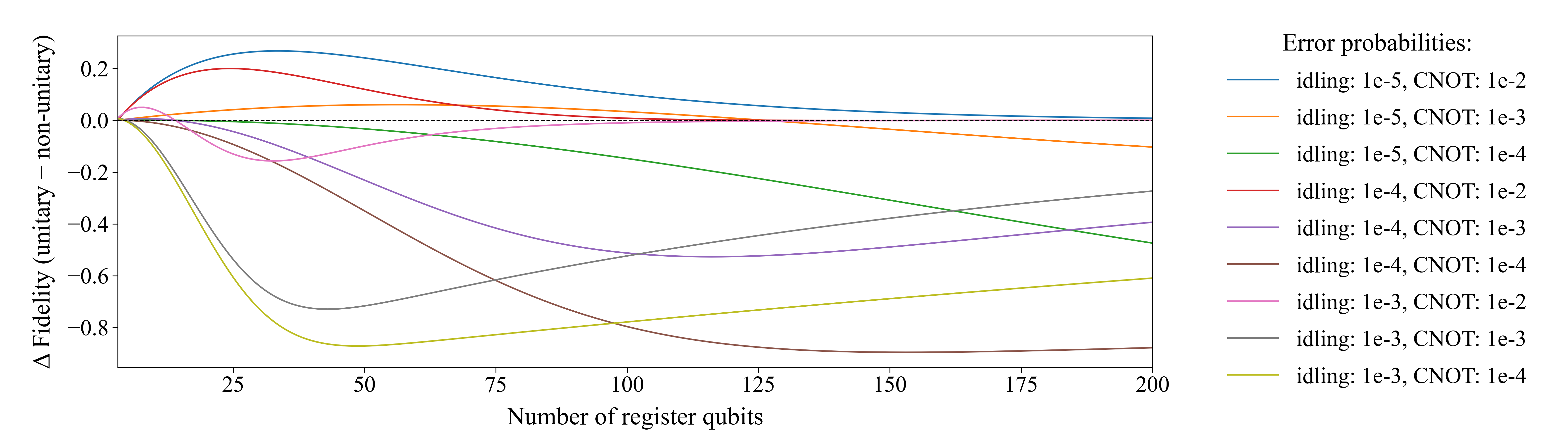}
    \caption{Hardware-resource trade-off between the conventional unitary CNOT ladder and non-unitary circuit constructions. Here, $\Delta$ Fidelity denotes the difference between the process fidelities of the conventional unitary and non-unitary implementations, respectively. The relative advantage depends on the balance between two-qubit-gate and idling errors. Adapted from \cite{Tserkis_2026}.}
    \label{fig:Tserkis_hardware_tradeoff}
\end{figure}

Further work has examined how circuit design interacts with the limitations of current quantum hardware. Tserkis \textit{et al.}~\cite{Tserkis_2026} replaced conventional CNOT ladders, whose
two-qubit depth grows linearly with the register size, with measurement-assisted constructions of constant CNOT depth. For an $n$-qubit register, the proposed construction reduces the CNOT depth from $n-1$ to $2$, at the cost of increasing the circuit width to $2n-3$ qubits and introducing $n-3$ mid-circuit measurements. Although this analysis was not performed for a Burgers' equation solver, the results are relevant to VQAs because the preferred circuit construction depends strongly on the relative magnitude of two-qubit-gate and idling errors. As illustrated in Figure~\ref{fig:Tserkis_hardware_tradeoff}, the non-unitary circuit becomes advantageous when the reduction in circuit depth compensates for the additional qubits, measurements, and feed-forward operations.

\begin{figure}[t]
    \centering
    \includegraphics[width=0.62\linewidth]
    {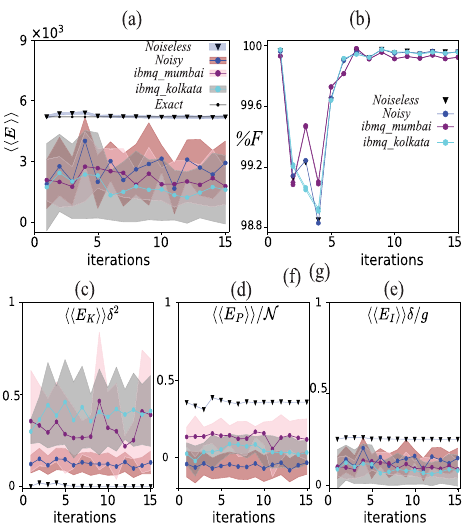}
    \caption{Performance of the NLSE VQA on
    superconducting quantum hardware. (a) Total energy during the optimization and (b) fidelity of the prepared state with
    respect to the exact ground state. (c)--(e) Kinetic,
    potential, and nonlinear interaction contributions to the objective
    function, respectively. Noiseless and noisy simulations are compared
    with results from the \textit{ibmq\_mumbai} and
    \textit{ibmq\_kolkata} devices; the shaded regions indicate one
    standard deviation. Although the prepared state retains high fidelity,
    the energy estimates exhibit substantial deviations and fluctuations,
    highlighting the greater sensitivity of the cost-function evaluation
    circuits to hardware noise. Adapted from \cite{Umer_2025a}.}
    \label{fig:Umer_hardware_results}
\end{figure}

The hardware results shown in Figure~\ref{fig:Umer_hardware_results}
make this distinction explicit. Here,
$\langle\!\langle E\rangle\!\rangle$ denotes the total energy used as
the variational cost function, while $\%F$ measures the agreement
between the measured trial-state probabilities and the exact
ground-state distribution. The lower panels separate the normalized
kinetic, potential, and nonlinear interaction contributions,
$\langle\!\langle E_K\rangle\!\rangle\delta^2$,
$\langle\!\langle E_P\rangle\!\rangle/\mathcal{N}$, and
$\langle\!\langle E_I\rangle\!\rangle\delta/g$, respectively, where
$\delta$ is the grid spacing, $\mathcal{N}$ is the norm used to encode
the potential, and $g$ is the nonlinear interaction strength.

Although the probability fidelity remains above approximately
$98.8\%$ and returns close to $100\%$ after optimization, the total
energy measured on the noisy simulator and quantum processors is
strongly shifted from the exact value and exhibits large fluctuations.
The individual energy contributions show that these errors arise during
the deeper QNPU calculations: the kinetic term requires an adder
circuit, while the potential and nonlinear terms require additional
registers for potential encoding or copies of the variational state.
Thus, the main hardware bottleneck is not the preparation of a
high-fidelity variational state, but the reliable evaluation of the
cost function supplied to the classical optimizer.

Together, these studies illustrate several complementary routes toward improving hardware performance. Measurement-assisted circuit constructions can exchange additional qubits and mid-circuit
measurements for reduced entangling-gate depth, whereas hardware experiments show that the circuits used to evaluate nonlinear
cost-function terms may remain substantially more demanding than the underlying state-preparation circuit.

\begin{figure}[t]
    \centering
    \includegraphics[width=0.95\linewidth]
    {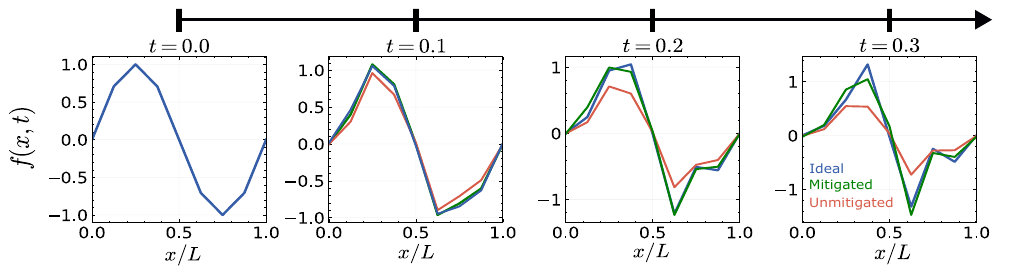}
    \caption{Effect of error mitigation on iterative variational time
    propagation of the viscous Burgers equation on superconducting quantum
    hardware. The comparison illustrates how hardware errors accumulate
    over successive time steps and how the mitigation procedure stabilizes
    the nonlinear evolution. Adapted from \cite{Jesus2026}.}
    \label{fig:Jose_hardware_results}
\end{figure}

The iterative character of variational time propagation introduces an additional challenge, since hardware errors can propagate from one time step to the next. Quantum error-mitigation techniques provide a route for reducing such errors without full quantum error correction \cite{Cai2023}. Da Costa Jesus \textit{et al.}~\cite{Jesus2026} showed that errors in the reconstructed state and its normalization can feed back into subsequent time steps and accumulate throughout the nonlinear evolution. To stabilize the propagation, they introduced a
Hadamard-test-specific depolarization-mitigation procedure that estimates the effective noise without increasing the circuit depth through gate folding. For the investigated Burgers simulations, this substantially reduced the accumulation of errors over successive time steps, as shown in Figure~\ref{fig:Jose_hardware_results}.

A further challenge for amplitude-encoded variational approaches is the trainability of increasingly expressive quantum circuits. In particular, barren plateaus can arise when gradients become exponentially small,
making the classical optimization increasingly difficult
\cite{McClean2018,Cerezo2021,Larocca2023}. Hashizume et
al.~\cite{hashizume2026} recently proposed a regularization strategy to
mitigate this problem based on the concept of quantum sparsity. Their approach relates poor trainability to excessive scrambling of
local information across the quantum circuit and introduces a regularizer based on topological entanglement entropy (TEE) to suppress unnecessary nonlocal information spreading. In this picture, trainable
circuits should operate near the ``edge of chaos'': sufficiently expressive to represent complex states, while avoiding excessive
scrambling that leads to vanishing gradients. The regularized objective
takes the form
\begin{equation}
    C_{\mathrm{reg}}(\boldsymbol{\theta})
    =
    C(\boldsymbol{\theta})
    +
    \lambda_{\mathrm{reg}} R_{\mathrm{TEE}}(\boldsymbol{\theta}),
\end{equation}
where $C(\boldsymbol{\theta})$ denotes the original variational cost function and $R_{\mathrm{TEE}}$ penalizes excessive nonlocal information
spreading. Hashizume \textit{et al.}~\cite{hashizume2026} showed that this regularization can guide the optimization toward more trainable parameter regimes. Numerically, they reported improved convergence and
accuracy for the amplitude encoding of complex data, including turbulent flow fields, as well as for ground-state preparation tasks. These results indicate that controlling the information structure of the variational circuit may provide an important route toward more
trainable amplitude-encoding schemes for complex CFD fields.

Beyond standard variational PDE solvers, recent work has explored more systematic approaches for translating discretized operators into quantum circuits. Using a variational compilation framework, Over \textit{et al.}~\cite{Over2026} transform arbitrary discrete operators into compact quantum circuits using only a single ancilla qubit for block encoding. Hardware connectivity and long-range interaction constraints can also be incorporated directly into the learning process.

A complementary approach is provided by tensor-programmable quantum circuits. Termanova \textit{et al.}~\cite{Termanova2024} proposed representing differential operators through TN decompositions, such as MPOs, and
compiling them into structured quantum circuits. Building on this approach, Siegl \textit{et al.}~\cite{Siegl_2026} developed a
tensor-programmable variational framework in which TNs act as an intermediate programming layer between classical discretizations and
quantum hardware. This enables the systematic construction of complex
differential operators while retaining the shallow-circuit structure targeted by near-term and early fault-tolerant quantum devices.

%%%%%%%%%%%%%%%%%%%%%%%%%%%%%%%%%%%%%%%%%%%%%%%%%%%%%%%%%%%%
\section{Tensor-programmable Quantum Circuits}
\label{Sec:TPQA}

Tensor-programmable variational quantum algorithms (TP-VQAs) provide a systematic framework for solving PDEs by combining variational optimization with TN operator representations. Unlike conventional VQAs, where problem-specific quantum circuits are typically
designed manually, TP-VQAs exploit TNs as an intermediate representation
to systematically translate discretized differential operators into executable quantum circuits. This abstraction can accommodate a broad class of numerical operators, including high-order finite-difference
stencils, boundary conditions, and nonlinear terms, while maintaining the shallow circuit depths.

The key concept underlying this framework is the close correspondence between TN representations and quantum states, operators, and circuits
\cite{Vidal2003,Oseledets2011,Schollwock2011,Orus2013}. TNs therefore provide a natural intermediate layer between classical PDE discretizations and their quantum implementation. In the following, we first introduce the relevant TN structures before discussing their compilation into
tensor-programmable quantum circuits.

\subsection{Tensor Network Methods}
\label{Sec:TN}

Originally developed in the context of quantum many-body physics to efficiently describe low-entanglement states
\cite{White1992,mccullochDensitymatrixRenormalizationGroup2007, orusTensorNetworksComplex2019a,Schollwock2011,Orus2013,Bridgeman_2017}, TNs have since evolved into powerful computational tools for the solution of PDEs on classical computers
\cite{Kiffner2023,Gourianov2022,Gourianov2022b,van_H_lst_2026,
vanhülst2026quantuminspiredsimulation2dturbulent,Ye2024,
connor2025tensornetworkmethodsgrosspitaevskii,Arenstein2026,
arenstein2026gridsolutionmultiassetoptions,H_lscher_2025,Gross_2026, Horner2026}. Tensor networks can adopt different geometries. Among the simplest are chain-like representations, referred to as MPS in quantum many-body physics or more generally, as tensor trains (TT) in numerical linear algebra and scientific computing \cite{Oseledets2011,Schollwock2011}. Other architectures include Tree Tensor Networks (TTNs), Projected Entangled-Pair States (PEPS), and the Multiscale Entanglement Renormalization Ansatz (MERA) \cite{Orus2013,Bridgeman_2017}. These architectures offer different trade-offs between expressive power, the correlations that can be represented efficiently, and the contraction complexity. 

The MPS decomposition provides a compact representation of discretized
multidimensional fields when the required bond dimensions remain moderate \cite{Oseledets2011}. MPS-based methods have also been successfully applied to high-dimensional parametric and stochastic elliptic PDEs \cite{Khoromskij_2011}. In this review, we focus primarily on the MPS structure because its one-dimensional chain geometry admits well-established algorithms for tensor contraction, truncation, and operator representation.
%%%%%%%%%%%%%%%%%%%%%%%%%%%%%%%%%%%%%%%%%%%%%%%%%%%%%%%%%%%%

\subsubsection{Matrix Product State Representation}

An MPS expresses a high-dimensional tensor $u_{i_1 i_2 \dots i_n}$ as a sequence of interconnected low-order tensors. Specifically, an $n$-dimensional tensor can be written as
\begin{equation}
u_{i_1 i_2 \dots i_n}
\approx
\sum_{\alpha_1=1}^{\chi_1}
\cdots
\sum_{\alpha_{n-1}=1}^{\chi_{n-1}}
G^{(1)}_{1,i_1,\alpha_1}
G^{(2)}_{\alpha_1,i_2,\alpha_2}
\cdots
G^{(n)}_{\alpha_{n-1},i_n,1},
\label{eq:TT}
\end{equation}
where $i_k=1,\ldots,I_k$ denotes the physical index of the $k$th tensor, with $I_k$ its physical dimension. In the quantum-state interpretation, $I_k$ corresponds to the dimension of the local Hilbert space; for qubit systems, $I_k=2$. The index
$\alpha_k=1,\ldots,\chi_k$ denotes the bond index connecting neighboring
tensors, with $k=1,\ldots,n-1$ for the nontrivial internal bonds. Each MPS tensor core therefore has dimensions
\begin{equation}
    G^{(k)} \in
    \mathbb{R}^{\chi_{k-1}\times I_k\times\chi_k},
\end{equation}
with $\chi_0=\chi_n=1$. The overall bond dimension of the MPS is defined as $\chi = \max_{k=1,\ldots,n-1} \chi_k$.
The role of the bond dimension in controlling correlations and entanglement has also been studied in quantum many-body settings. In particular, across any bipartition of an MPS, the Schmidt rank is bounded by the bond dimension $\chi$, implying an upper bound on the entanglement entropy of $S\leq\log\chi$ \cite{Vidal2003}. For ensembles of MPS generated by random quantum circuits, correlation properties can further be characterized through the spectrum of the associated transfer matrix \cite{Loio2025}. These results illustrate more generally how the bond dimension controls the amount of structure that can be represented efficiently by an MPS.

Restricting the size of the bond dimension allows to replace the exponential storage cost of a full tensor with a compressed representation. This allows significant reductions in memory and computational cost, while maintaining high accuracies for many applications \cite{Oseledets2011,Grasedyck2013}.

For qubit systems, each physical index has dimension $I_k=2$. An MPS with $n$ sites and $I_k=2$ therefore directly represents an $n$-qubit quantum state. This correspondence provides a natural connection between
classical MPS representations and quantum algorithms, while also making MPS methods a useful classical baseline when assessing potential
quantum advantage.

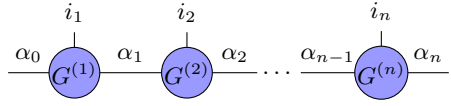
\begin{figure}[t]
    \centering
    % \begin{tikzpicture}[tensornetwork, scale=1.0, xscale=0.9, baseline=(A1.base)]
    \begin{tikzpicture}[tensornetwork,scale=1.0,xscale=0.9,baseline=(A1.base),atensor/.append style={fill=blue!40} % <-- add this line
    ]
        % Nodes
        \node[atensor] (A1) at (1, 0) {$G^{(1)}$};
        \node[atensor] (A2) at (3.2, 0) {$G^{(2)}$};
        \node (dots) at (5.0, 0) {$\dots$};
        \node[atensor] (An) at (7.0, 0) {$G^{(n)}$};
        % Physical Legs (Standard: label at the very top)
        \draw (A1.north) -- +(0, 0.4) node[above] {$i_1$};
        \draw (A2.north) -- +(0, 0.4) node[above] {$i_2$};
        \draw (An.north) -- +(0, 0.4) node[above] {$i_n$};
        % Virtual Bonds (Standard: label midway)
        \draw (A1) -- node[above] {$\alpha_1$} (A2);
        \draw (A2) -- node[above] {$\alpha_2$} (dots);
        \draw (dots) -- node[above] {$\alpha_{n-1}$} (An);
        % Boundary virtual bonds
    \draw (A1.west) -- +(-0.8,0) node[midway, above] {$\alpha_0$};
    \draw (An.east) -- +(0.8,0) node[midway, above] {$\alpha_n$};
    \end{tikzpicture} 
    \caption{Graphical representation of a MPS: the circles represent the tensors $G^{(1)},...,G^{(n)}$, horizontal lines correspond to the internal ranks ($\alpha_n$), and the upward legs denote the physical dimensions (indices of size $i_n$).}
    \label{fig:TT-vec}
\end{figure}

An important tool for the development of TN algorithms is its graphical notation  which provides a clear, visual framework for conceptualizing TNs. The graphical representation of the MPS in Eq.~\ref{eq:TT} is depicted in Figure~\ref{fig:TT-vec}.

For discretized fields, a qubit-compatible MPS representation can be obtained by decomposing each grid index into binary digits, resulting in tensor modes with physical dimension two. A coordinate direction containing $N=2^n$ grid points can then be represented using $n$ binary
indices, while other grid sizes can be handled through padding. For multidimensional fields, the binary indices associated with different spatial coordinates may be grouped or interleaved. The resulting bond dimensions can depend strongly on the chosen ordering, so the most efficient arrangement is generally problem dependent
\cite{Kazeev2014}. In practice, the initial condition is chosen to have a low-rank MPS representation, and the evolution is then performed directly in MPS form. As the solution evolves, the bond dimension is allowed to grow to capture the increasing complexity of the field, while truncation is used to keep it below a prescribed maximum value. In this way, the full multidimensional field need not be explicitly
formed and subsequently decomposed into MPS format.

%%%%%%%%%%%%%%%%%%%%%%%%%%%%%%%%%%%%%%%%%%%%%%%%%%%%%%%%%%%%
\subsubsection{Matrix Product Operator}

\begin{figure}[t]
    \centering
    % \begin{tikzpicture}[tensornetwork, scale=1.0, xscale=0.9, baseline=(A1.base)]
    \begin{tikzpicture}[tensornetwork,scale=1.0,xscale=0.9,baseline=(A1.base),atensor/.append style={fill=red!40} % <-- add this line
    ]
        % Nodes
        \node[atensor] (A1) at (1, 0) {$A^{(1)}$};
        \node[atensor] (A2) at (3.2, 0) {$A^{(2)}$};
        \node (dots) at (5.0, 0) {$\dots$};
        \node[atensor] (An) at (7.0, 0) {$A^{(n)}$};
        % Physical Legs (Standard: label at the very top)
        \draw (A1.north) -- +(0, 0.4) node[above] {$i'_1$};
        \draw (A2.north) -- +(0, 0.4) node[above] {$i'_2$};
        \draw (An.north) -- +(0, 0.4) node[above] {$i'_n$};
        % Physical Legs (Standard: label at the very top)
        \draw (A1.south) -- +(0, -0.4) node[below] {$i_1$};
        \draw (A2.south) -- +(0, -0.4) node[below] {$i_2$};
        \draw (An.south) -- +(0, -0.4) node[below] {$i_n$};
        % Virtual Bonds (Standard: label midway)
        \draw (A1) -- node[above] {$\alpha^{(o)}_1$} (A2);
        \draw (A2) -- node[above] {$\alpha^{(o)}_2$} (dots);
        \draw (dots) -- node[above,] {$\alpha^{(o)}_{n-1}$} (An);
        % Boundary virtual bonds
    \draw (A1.west) -- +(-0.8,0) node[midway, above] {$\alpha^{(o)}_0$};
    \draw (An.east) -- +(0.8,0) node[midway, above] {$\alpha^{(o)}_n$};
    \end{tikzpicture} 
    \caption{Graphical representation of a MPO, namely circles represent the cores $A^{(1)},...,A^{(n)}$, horizontal lines are the bond dimensions ($\alpha^{(o)}_n$), and the upward legs and downward legs denote the physical dimensions of the input and output indices respectively.}
    \label{fig:TT-mat}
\end{figure}
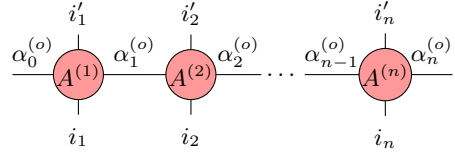

The operator analogue of an MPS is the MPO, in which a linear operator
is represented as a chain of order-$4$ tensor cores. Specifically, an
operator $A$ can be written as
\begin{equation}
    A_{i'_1\dots i'_n,\,i_1\dots i_n}
    \approx
    \sum_{\alpha^{(o)}_1=1}^{\chi^{(o)}_1}
    \cdots
    \sum_{\alpha^{(o)}_{n-1}=1}^{\chi^{(o)}_{n-1}}
    W^{(1)}_{1,i_1,i'_1,\alpha^{(o)}_1}
    W^{(2)}_{\alpha^{(o)}_1,i_2,i'_2,\alpha^{(o)}_2}
    \cdots
    W^{(n)}_{\alpha^{(o)}_{n-1},i_n,i'_n,1}.
    \label{eq:MPO}
\end{equation}
Here, $i_k=1,\ldots,I_k$ and $i'_k=1,\ldots,I'_k$ denote the input
and output physical indices of the $k$th core, respectively. Thus, $I_k$ is the dimension of the local input space and $I'_k$ the dimension of the corresponding output space. For an operator acting within the same local space, $I_k=I'_k$; for qubit systems,
$I_k=I'_k=2$. The index $\alpha^{(o)}_k$ denotes the MPO bond index with dimension $\chi^{(o)}_k$, with $\chi^{(o)}_0=\chi^{(o)}_n=1$. Each MPO tensor core therefore has dimensions
\begin{equation}
    W^{(k)} \in
    \mathbb{R}^{\chi^{(o)}_{k-1}\times I_k\times I'_k\times\chi^{(o)}_k},
\end{equation}
with $\chi^{(o)}_0=\chi^{(o)}_n=1$. The graphical representation resembles that of an MPS but contains two physical legs per core, corresponding to the input and output indices; see Figure~\ref{fig:TT-mat}.

In the MPS representation, key linear algebra operations, including matrix-vector products and inner products, can be performed through local contractions of individual cores or variationally through DMRG-like sweeps. The corresponding computational scalings are summarized in Table~\ref{table:operation_cost}. The cost of these operations is governed primarily by the bond dimension $\chi$, rather than directly by the total number of grid points $N$. If the underlying fields and operators admit low-rank representations, this can lead to substantial savings in both memory usage and computational time compared with operations performed on the fully contracted representation.

\begin{table*}[b]
    \centering
    \small
    \setlength{\tabcolsep}{6pt}
    \setlength{\aboverulesep}{0.6ex}
    \setlength{\belowrulesep}{0.6ex}
    \renewcommand{\arraystretch}{1.25}
    
    \begin{tabular}{@{}
      p{0.20\textwidth}
      p{0.42\textwidth}
      p{0.28\textwidth}
    @{}}
    \toprule
    \textbf{Operation} &
    \textbf{Method} &
    \textbf{Leading cost (after rounding to target rank)} \\
    \midrule
    
    \multirow{2}{*}{Addition $\,c=a+b$}
      & Direct & $O\!\big(\chi^3\big)$ \\
      & Variational & $O\!\big(\chi^3\big)$ \\
     % & TP-VQA \cite{Siegl_2026} & $O\!\big(\chi^2\big)$ \\
    \midrule
    
    \multirow{4}{*}{Hadamard $\,c=a\odot b$}
      & Direct & $O\!\big(\chi^6\big)$ \\
      & Variational & $O\!\big(\chi^4\big)$ \\
      & Tensor-Train Multiplication (TTM) \cite{Michailidis2024}
      & $O\!\big(\chi^3\big)$ \\
      & Recursive Sketched Interpolation (RSI) \cite{meng2026recursivesketchedinterpolationefficient}
      & $O\!\big(\chi^3\big)$ \\
      %& TP-VQAs & $O\!\big(\chi^2\big)$ \\
    \midrule
    
    \multirow{2}{*}{Matvec $\,b=Ma$}
      & Direct & $O\!\big(\chi_{_O}^{3}\,\chi^{3}\big)$ \\
      & Variational & $O\!\big(\chi_{_O}\,\chi^{3}\big)$ \\
      %& TP-VQA & $O\!\big(\chi_{_O}^{2}\big)$ \\
    \midrule
    
    \multirow{2}{*}{Linear solve $\,Ma=b$}
      & Variational with local exact LU
      & $O\!\big(\chi^{6}\big)$ \\
      & Variational with local iterative (fast matvec)
      & $O\!\big(\chi_{_O}\,\chi^{3}
        +\chi_{_O}^{2}\,\chi^{2}\big)$ \\
      %& TP-VQA (per quantum matvec evaluation)
      %& $O\!\big(\chi_{_O}^{2}\big)$ \\
    \bottomrule
    \end{tabular}
    
    \caption{Scaling table for MPS algorithmic operations. Here, $\chi_{_O}$ denotes the bond dimension of the MPO $M$, and $\chi$ the bond dimension of MPS $a$, $b$, and $c$ (assumed equivalent for simplicity) \cite{van_H_lst_2026}.}
    \label{table:operation_cost}
\end{table*}

For structured discretizations, many standard differential operators can be constructed directly in MPO form from known tensor cores, Kronecker-product structures, or shift operators, without first forming and decomposing the full matrix
\cite{Kazeev2012}. As a representative example, consider the discrete Laplacian, which admits an explicit low-rank representation in multiple spatial dimensions; in one dimension, even its inverse admits an explicit low-rank representation \cite{Kazeev2012}. The discrete $d$-dimensional Laplacian obtained from a tensor-product discretization takes the form
\begin{equation}
    \Delta=\sum_{k=1}^d I^{\otimes(k-1)} \otimes
    \mathfrak{D} \otimes I^{\otimes(d-k)},
\end{equation}
where $\mathfrak{D}$ denotes the one-dimensional discrete Laplacian and
$I$ the corresponding identity operator. Explicit low-rank representations of the discrete Laplace operator have also been derived in arbitrary spatial dimensions, while in one dimension an explicit low-rank representation of its inverse is available
\cite{Kazeev2012}.

When combined with the binary tensorization introduced above, this
Kronecker-sum structure admits a compact MPO representation whose bond
dimensions remain bounded independently of the spatial dimension and grid resolution \cite{Kazeev2012}. For $n$ grid points per spatial dimension, the corresponding storage scales as
$\mathcal{O}(d\log n)$, while tensor-network operations retain a logarithmic dependence on the grid resolution, with their full cost also depending on the bond dimensions of the operands.

These favourable rank properties rely strongly on the structured Cartesian discretizations considered here. Extending TN representations to arbitrary unstructured polyhedral meshes, as commonly used in industrial CFD, is substantially less straightforward, since the resulting connectivity and operators generally do not retain the same simple Kronecker-product structure. Developing efficient low-rank representations for such discretizations therefore remains an important open research direction.

%%%%%%%%%%%%%%%%%%%%%%%%%%%%%%%%%%%%%%%%%%%%%%%%%%%%%%%%%%%

\subsubsection{Tensor Networks for Computational Fluid Dynamics}
The practical efficiency of TN methods in the context of CFD depends on how the required bond dimension grows with increasing flow complexity and simulation time. While scale-local interactions suggest
compressibility, increasingly complex flows introduce a broader hierarchy of active scales, which may increase the rank needed to
accurately capture inter-scale correlations. Recent numerical studies have investigated how efficiently turbulent flow fields can be represented in TN form, showing that substantial compressibility can persist despite their multiscale structure
\cite{Gourianov2022,H_lscher_2025}.

Beyond the representation of individual flow fields, an important question is how the required bond dimension scales with physically relevant flow parameters. For two-dimensional Rayleigh--Bénard convection, van Hülst et
al.~\cite{vanhülst2026quantuminspiredsimulation2dturbulent} found that the bond dimension required to represent the full flow fields increases with the Rayleigh number without showing saturation. However, in dynamical TN simulations, the bond dimension required to recover physical observables such as the mean Nusselt number grows substantially more slowly with increasing Rayleigh number.

Complex geometries provide an important test of whether TN compressibility survives beyond idealized benchmark domains. Gross et
al.~\cite{Gross_2026} recently addressed this issue by applying a TN formulation to three-dimensional structured-media and vascular geometries. In their approach, both the flow variables and the geometry information are represented in MPS format, with boundaries incorporated through geometry masks. The method faithfully replicates classical reference simulations, delivering compression ratios greater than two orders of magnitude for multiple instances. The study includes nontrivial three-dimensional examples such as flow through an aneurysm geometry and a pin-fin heat-sink configuration,
demonstrating that TN compression can accommodate complex geometries within structured Cartesian discretizations. Related CFD studies using TN have also treated complex objects, including construction of NACA-airfoil into MPS format using TT-Cross algorithm \cite{Peddinti2024,OSELEDETS2010TTCROSS,Ritter24} and body-fitted curvilinear grids for flows around immersed objects \cite{van_H_lst_2026}. Despite this progress, the efficient treatment of complex geometries and genuinely unstructured-grid discretizations remains an active research direction in TN-based CFD. Recent approaches have considered
geometry-dependent and isogeometric discretizations
\cite{Ion2022,Tran2026}, while tensor-train operator-inference methods have also been demonstrated using CFD data generated on unstructured
meshes \cite{Danis2025b}. A general low-rank formulation of arbitrary unstructured CFD operators, however, remains largely open.

Taken together, these results show that TN methods can provide compact representations for a range of CFD problems, although their classical cost increases as larger bond dimensions are required. This may become a limiting factor for increasingly complex flows and geometries. At the same time, the close correspondence between TNs and quantum states and operators provides a natural route for transferring tensor network representations to quantum circuits, which are expected to provide at least polynomial advantages. This motivates the investigation of tensor programmable quantum circuits and their application to CFD. 

\subsection{Translating Tensor Networks into Quantum Circuits}
\label{sec:Translation_TN_to_QC}

% ------------------------------------------------------------
\begin{figure}[t]
    \centering
    \includegraphics[width=0.85\linewidth]{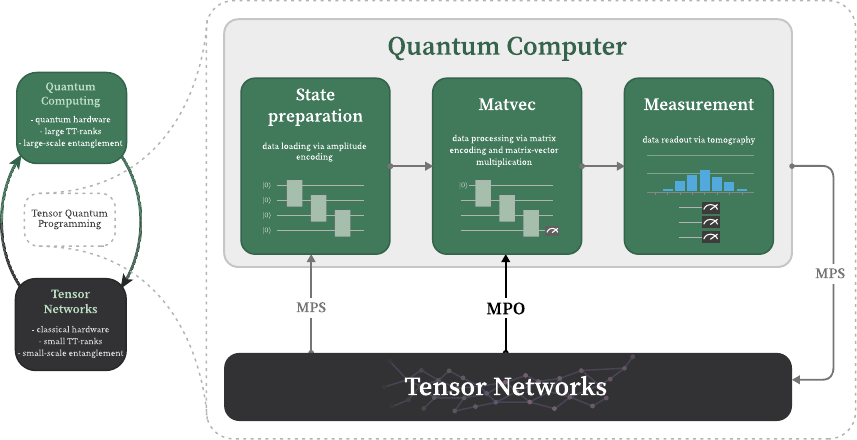}
    \caption{Quantum tensor programming workflow introduced by Termanova \textit{et al.}~\cite{Termanova2024}. The numerical algorithm is expressed in TN form. MPS encoding prepares the input state on the quantum processor, MPO encoding implements the matrix--vector multiplication, and MPS tomography returns the output state to a TN representation.}
    \label{fig:TQP_pipeline}
\end{figure}
% ------------------------------------------------------------

Quantum tensor programming provides a structured route from classical fields and operators to executable quantum circuits
\cite{Termanova2024,Siegl_2026,vanhülst2025quantumsolverspredictiveaeroacoustic}. By first expressing the discretized state and operators in TN form, the corresponding circuits can be generated systematically and avoid the manual circuit design which was implemented in earlier variational
time-stepping approaches \cite{Lubasch2020,Over2024a}. This becomes particularly useful when incorporating more complex ingredients such as higher-order derivatives, different boundary conditions, or non-Cartesian grids.

As introduced in Sec.~\ref{subsubsec:QSVT}, block encoding provides a general framework for embedding non-unitary operators into unitary quantum circuits. Quantum tensor programming exploits the low-rank structure of MPOs to construct such embeddings directly from their local tensor cores, rather than from the full dense operator. Nibbi and Mendl~\cite{Nibbi2024} developed a block-encoding strategy based on the individual tensors of a MPO, providing a direct mapping to a quantum circuit. However, the associated success probability decreases exponentially with system size, with some mitigation strategies proposed for finite systems.

Termanova \textit{et al.}~\cite{Termanova2024} introduced the broader Tensor
Quantum Programming framework illustrated in
Figure~\ref{fig:TQP_pipeline}. They proposed an MPS-based quantum workflow in which TN states and operators are translated into quantum circuits. For state preparation, the local MPS cores can be mapped sequentially to quantum gates, providing an efficient loading strategy when the MPS
bond dimension remains small
\cite{Schon2005,Ran2020,Termanova2024}. The corresponding circuit depth scales as $\mathcal{O}(\chi^2)$ \cite{Lubasch2020}. Related work has also explored how measurements can reduce the circuit depth required to manipulate MPS states. In this context, Gunn \textit{et al.}~\cite{Gunn2025} showed that measurements and classical feedforward can enable symmetric transformations between symmetry-constrained MPSs using short-depth quantum circuits, even when the allowed operations are themselves constrained to respect the symmetry. This illustrates how measurement-assisted protocols can
reduce the circuit depth required for manipulating structured TN states while preserving symmetry constraints. Within the quantum tensor programming workflow, the operator, represented as an MPO, is translated into a quantum circuit and applied to the encoded state. After the quantum operation, the output state can be reconstructed as an MPS using MPS tomography \cite{Cramer_2010,Lanyon_2017}. 

% ------------------------------------------------------------
\begin{figure}[t]
    \centering
    \includegraphics[width=0.75\linewidth]{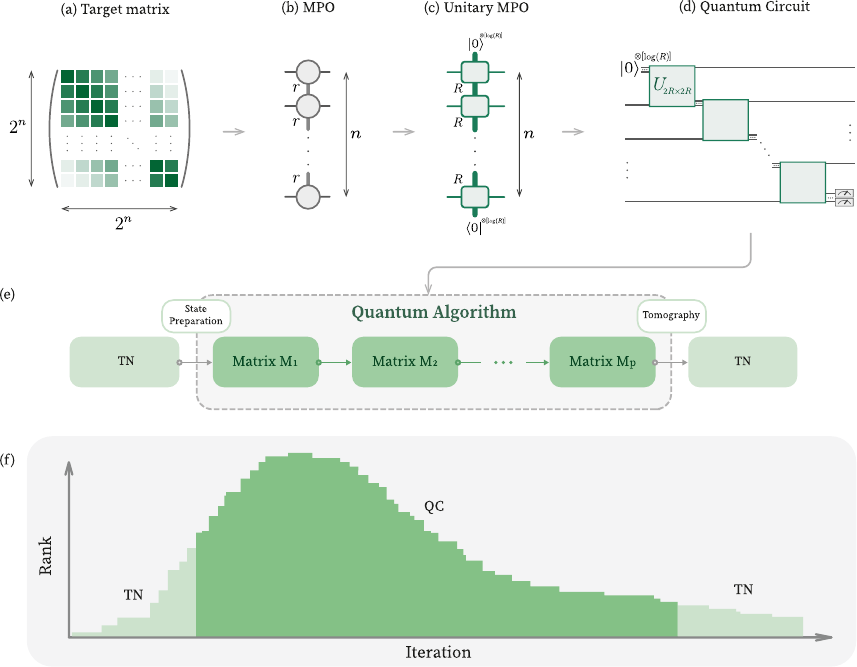}
    \caption{Schematic representation of the MPO-to-circuit encoding strategy of Termanova \textit{et al.}~\cite{Termanova2024}. A target matrix is first written as an MPO, approximated by a unitary MPO with enlarged bond dimension, and finally converted into a quantum circuit using auxiliary qubits and postselection.}
    \label{fig:Termanova_MPO_encoding}
\end{figure}
% ------------------------------------------------------------

In contrast to the construction of Nibbi and Mendl~\cite{Nibbi2024},
Termanova \textit{et al.}~\cite{Termanova2024} construct a unitary representation of an MPO by Riemannian optimization over the manifold of isometric tensor cores \cite{Absil2007,Hauru2021,Luchnikov2021}. The resulting unitary cores can be interpreted directly as multi-qubit gates acting sequentially on the system and auxiliary registers, as illustrated in Figure~\ref{fig:Termanova_MPO_encoding}. For MPO bond dimension $\chi_O$, the
auxiliary register requires
\begin{equation}
    n_{\rm aux}=\left\lceil \log_2 \chi_O \right\rceil
\end{equation}
qubits. The classical optimization remains polynomial in $\chi_O$, with the dominant tensor contractions scaling as $\mathcal{O}(\chi_O^3)$ per core.

Because the target operator is generally non-unitary, its action is recovered probabilistically by post-selecting the auxiliary register in
$\ket{0}^{\otimes n_{\rm aux}}$. The corresponding success probability is
given by Eq.~\ref{eq:success_prob}. The block-encoding normalization factor $\beta$ can be optimized, thereby improving the post-selection probability and, for the differential operators considered in
Refs.~\cite{Termanova2024,Siegl_2026}, avoiding the exponential suppression
that would otherwise arise.

This provides a compiler-like route from a structured MPO representation to a quantum circuit, without requiring a separate hand-crafted circuit construction for each operator. Termanova \textit{et al.}~\cite{Termanova2024}
demonstrated the encoding of operators including diagonal and Laplace matrices on systems of up to 50 qubits, with circuit depth scaling
linearly with the number of qubits for fixed MPO bond dimension. This becomes relevant for large CFD problems, where, for example, resolving the Kolmogorov scale in three-dimensional turbulence requires
$N_{\mathrm{grid}}\sim Re^{9/4}$ degrees of freedom \cite{Pope2000,MoinMahesh1998}. 

\subsection{Tensor-programmable VQA}
\label{Sec:TPVQA}

\begin{figure}[t]
    \centering
    \includegraphics[width=0.75\linewidth]{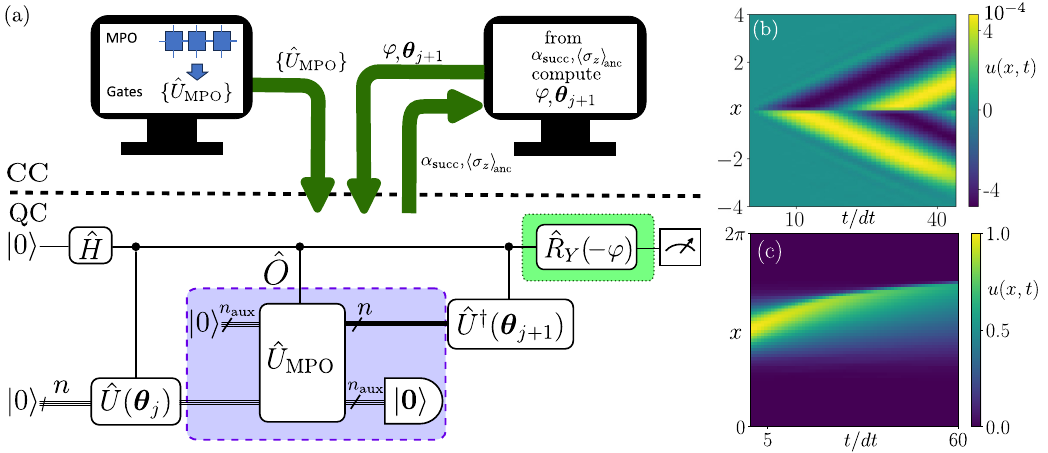}
    \caption{Tensor-programmable variational quantum algorithm of Siegl \textit{et al.}~\cite{Siegl_2026}. (a) TP-VQA workflow. The MPO
    representation of the discretized PDE operator $\hat{O}$ is compiled
    into the quantum circuit $\hat{U}_{\mathrm{MPO}}$ and applied to the variational state prepared by $\hat{U}(\boldsymbol{\theta}_j)$, where
    $\boldsymbol{\theta}_j$ denotes the variational parameters at step $j$. The quantum register contains $n$ system qubits and $n_{\mathrm{aux}}$
    auxiliary qubits initialized in
    $\ket{0}^{\otimes n_{\mathrm{aux}}}$. Measurements of the auxiliary register provide
    $\alpha_{\mathrm{succ}}$, which denotes the block-encoding
    postselection success probability corresponding to $p_{\mathrm{succ}}$ defined in Eq.~\ref{eq:success_prob}, while
    measurement of the global ancilla provides the expectation value
    $\langle\sigma_z\rangle_{\mathrm{anc}}$. These quantities are used in the classical feedback loop to determine the rotation angle $\varphi$ and update the parameters to $\boldsymbol{\theta}_{j+1}$. Here,
    $R_Y(-\varphi)$ denotes a rotation about the $y$-axis, while CC and QC indicate the classical and quantum parts of the hybrid loop. (b,c)
    Representative solution fields $u(x,t)$, plotted as functions of the spatial coordinate $x$ and the rescaled time $t/\Delta t$, where $\Delta t$ denotes the time-step size used in the variational time evolution, for the linearized Euler equations with absorbing boundaries and the nonlinear Burgers' equation, respectively.}
    \label{fig:Siegl_QNPU_workflow}
\end{figure}

Siegl \textit{et al.}~\cite{Siegl_2026} extended quantum tensor programming from matrix-vector multiplication to full PDE time evolution. In this framework, the discretized PDE operators are first represented as MPOs and then compiled into unitary quantum circuits that are incorporated directly into a variational time-stepping scheme, as illustrated in Figure~\ref{fig:Siegl_QNPU_workflow}. A parameterized quantum circuit prepares the trial solution at a given time step. As in the variational approaches discussed in Sec.~\ref{Sec:StandardVGA}, the cost function quantifies how well the trial state satisfies the discretized governing equation. In this work, it is evaluated through an adapted Hadamard-test circuit, and a classical
optimizer updates the ansatz parameters to obtain the next time step. The formulation also includes a state-dependent norm correction. This is essential for CFD, because diffusion, absorbing boundaries, and many time-stepping operators are not unitary. The framework also allows several terms of a PDE to be grouped into a single operator circuit, reducing the number of separate cost-function contributions that must be measured compared with earlier variational PDE solvers.

In earlier variational PDE solvers, boundary conditions have typically been incorporated by modifying the discretized operator and source terms, or through penalty terms in the variational cost function. In tensor-programmable approaches, local boundary
conditions can be represented through modifications of the corresponding MPOs. Such modifications often increase the MPO bond dimension only slightly, or not at all, so the associated circuit depth can remain essentially unchanged. More complex geometries, nonlocal boundary conditions, or state-dependent nonlinear boundary models may, however, require larger MPO bond dimensions and therefore deeper quantum
circuits. In particular, models such as wall functions depend on the evolving local flow state and may require the boundary operator or source terms to be updated during the simulation. Their efficient implementation therefore remains an important topic for future work \cite{Termanova2024,Siegl_2026}. 

% ------------------------------------------------------------
% ------------------------------------------------------------

The TP-VQA framework of Siegl \textit{et al.}~\cite{Siegl_2026} was demonstrated on both linear and nonlinear PDEs. For the linearized Euler equations,
the authors included absorbing boundary conditions through a sponge operator and showed that the velocity and pressure dynamics can be
reproduced using the variational quantum solver. For the nonlinear Burgers' equation, an initial Gaussian profile evolves into a shock-like structure, providing a simple test case for nonlinear transport and
diffusion. These examples are important not because they represent full-scale CFD simulations, but because they demonstrate that TP-VQA can incorporate linear operators, non-unitary terms, boundary
treatments, and nonlinear contributions within the same
tensor-programmable framework.

The probabilistic operator implementation in TP-VQAs makes the scaling of the algorithm directly depend on the success probability of the operator implementation. Importantly, given the quantum--classical feedback loop, there is no exponential decrease with the number of variational time steps, as can occur in fully quantum time-marching schemes with repeated probabilistic operator applications
\cite{Fang_2023}. The postselection overhead  per time step was evaluated quantitatively by Siegl \textit{et al.}~\cite{Siegl_2026}. For the investigated linear finite-difference operators, including Laplacian and first-derivative
stencils, the mean success probability remained approximately size-independent, with values between $0.35$ and $0.50$. For the nonlinear pointwise multiplication, the situation is more subtle: For fields with a high degree of  localization, the success probability remains nearly constant until the spatial features are resolved and decreases only beyond this resolution threshold \cite{Lubasch2020, Siegl_2026}. However, this favorable scaling of the nonlinear treatment does not
generally persist for arbitrary flow fields when state-based implementations are used \cite{Siegl2026-arxiv}.

\subsubsection{Efficient Treatment of Non-Linearity with TP-VQA}
Although TP-VQA provides a systematic framework for compiling discretized PDE operators into quantum circuits, the efficient treatment of nonlinear terms remains a central challenge. Quantum operations act linearly on state amplitudes, whereas nonlinear CFD terms such as
$(\mathbf{u}\cdot\nabla)\mathbf{u}$ require products of the evolving field with itself. Previous variational approaches evaluate such terms using multiple copies of the amplitude-encoded state and probabilistic
postselection. While this avoids explicitly linearizing the nonlinear dynamics, the associated success probability can decrease exponentially with system size, leading to a rapidly increasing measurement overhead.
Siegl \textit{et al.}~\cite{Siegl2026-arxiv} address this scalability bottleneck by replacing the state-based implementation with a diagonal block encoding of the evolving field.

\begin{figure}[t]
    \centering
    \includegraphics[width=0.60\linewidth]{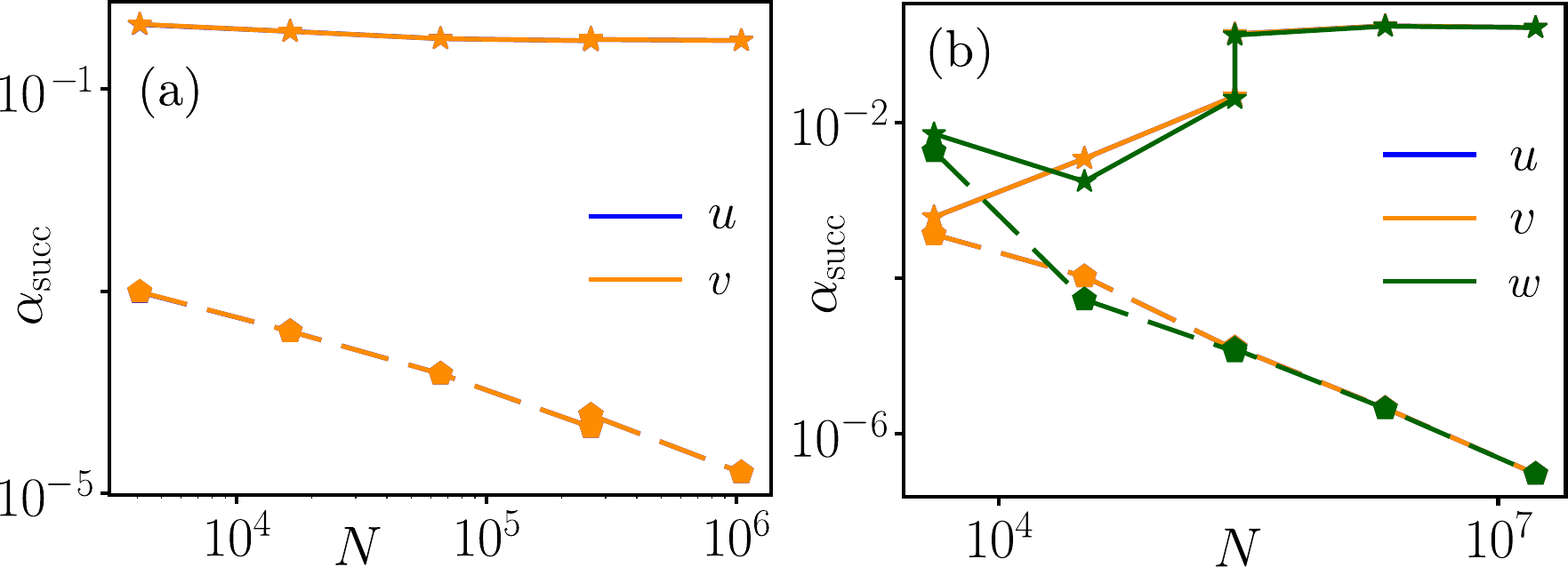}
    \caption{Success probability for one Euler time step of the Burgers' equation using the state-based (dashed lines) and diagonal
    block-encoding (solid lines) treatments of the nonlinearity, adapted from Siegl \textit{et al.}~\cite{Siegl2026-arxiv}. (a) Two-dimensional temporally developing jet (TDJ) and (b) three-dimensional Taylor--Green vortex (TGV) for increasing system size $N$. Here, $u$, $v$, and $w$ denote the Cartesian velocity components, and
    $\alpha_{\mathrm{succ}}$ denotes the success probability of the
    nonlinear operator implementation.}
    \label{fig:Siegl_success_probs}
\end{figure}

The diagonal-block encoding can be built from an approximate classical MPS reconstruction of the quantum state, transforming it into a diagonal MPO, followed by the compilation into unitary gates using quantum tensor programming. This alternative implementation of the non-linearity has a strong impact on the success probability of the non-linear computation as shown in Figure~\ref{fig:Siegl_success_probs}. 

The difference in success probability between the two implementations of the nonlinearity originates from the normalization of the corresponding block encodings, defined as in
Eq.~\ref{eq:block-encoding}. The success probability is given by
Eq.~\ref{eq:success_prob} and depends on the operator, the state on which it acts, and the normalization constant $\beta$. For the diagonal block encoding of a field $\phi$, the optimal normalization can be chosen as
$\beta_{\rm opt}=\|\hat{D}_{\phi}\|=\max_i |\phi_i|,$ where $\hat{D}_{\phi}=\mathrm{diag}(\phi_1,\ldots,\phi_N)$ denotes the diagonal operator constructed from the field values. This corresponds
to the spectral norm of the diagonal operator. In contrast, the state-based implementation of the nonlinearity can be understood as a block encoding of $\hat{D}_{\phi}$ that uses the normalization $\beta_{\rm vec}=\|\phi\|_2$. While $\beta_{\rm opt}$ is typically independent of grid resolution, $\beta_{\rm vec}$ generally increases with the number of degrees of freedom, leading to a rapidly decreasing success probability for the state-based implementation. For the investigated turbulent flow fields, the diagonal block encoding therefore maintains an approximately constant success probability with
increasing grid resolution and Reynolds number, whereas the state-based Hadamard product exhibits a strong decrease (see Figure~\ref{fig:Siegl_success_probs}).

The construction of the diagonal-block encoding requires an additional classical tensor network step, which needs to be considered when studying the overall scaling of the approach. Several strategies can be used to reconstruct an MPS from a quantum state. Siegl \textit{et al.}~\cite{Siegl2026-arxiv} compare three such approaches in terms of their bond-dimension requirements, computational scaling, and
measurement overhead: tensor cross interpolation (TCI)
\cite{NumezFernandez2025}, efficient state tomography (EST) \cite{Cramer_2010,Lanyon_2017}, and MPS-based classical simulation of the parameterized quantum circuit. Both TCI and EST are expected to require the evaluation of $\mathcal{O}(\chi^2)$ expectation values to reconstruct the target MPS.
However, Ref.~\cite{Siegl2026-arxiv} shows that a large number of measurement shots may still be required to reach a fixed accuracy,
particularly at higher Reynolds numbers, although this overhead appears to saturate. The MPS based simulation of the parametrized circuit therefore provides a viable alternative. While scaling as $\chi^3$, this is equivalent to the tensor programming compilation step, and therefore, does not increase the overall asymptotic cost. 
As discussed in the following, this additional classical processing still provides scaling benefits over fully classical tensor-network routines and does not negate the potential quantum advantage.

\subsection{Scaling Analysis and Comparison to Classical Tensor Networks}

An analysis of the computational scaling of the tensor-programmable approach was carried out by Siegl et
al.~\cite{Siegl_2026,Siegl2026-arxiv}, including a comparison with tensor-network methods. Classically, the memory required to store an MPS scales as $\mathcal{O}(n\chi^2)$, while typical contraction costs scale at least as $\mathcal{O}(n\chi^3)$. The deterministic evaluation of nonlinear Hadamard products using conventional MPS contractions scales as $\mathcal{O}(\chi^4)$
\cite{Michailidis2025,meng2026recursivesketchedinterpolationefficient}. More recent algorithms, including the tensor-train multiplication method of Michailidis \textit{et al.}~\cite{Michailidis2025} and the
Recursive Sketched Interpolation method of Meng \textit{et al.}~\cite{meng2026recursivesketchedinterpolationefficient}, reduce the asymptotic scaling of the Hadamard product to $\mathcal{O}(\chi^3)$. In particular, RSI is designed to mitigate the growth of intermediate bond dimensions by keeping the representation compressed throughout the computation, although large intermediate ranks can still limit the practical computational gain as shown for turbulent flow fields in \cite{Siegl2026-arxiv}. A summary of the leading bond-dimension scalings and practical
limitations of the classical and TP-VQA implementations of the
nonlinear Hadamard product is provided in
Table~\ref{tab:Hadamard_comparison}.

\begin{table*}[h]
    \centering
    \small
    \setlength{\tabcolsep}{6pt}
    \renewcommand{\arraystretch}{1.25}

    \begin{tabular}{@{}
        p{0.23\textwidth}
        p{0.18\textwidth}
        p{0.20\textwidth}
        p{0.31\textwidth}
    @{}}
        \toprule
        \textbf{Method}
        & \textbf{Bond-dimension scaling}
        & \textbf{Success probability}
        & \textbf{Main issue} \\
        \midrule

        Classical variational MPS
        & $\mathcal{O}(\chi^4)$
        & Deterministic
        & High $\chi$-dependence \\

        TTM
        & $\mathcal{O}(\chi^3)$
        & Deterministic
        & Intermediate-rank considerations \\

        RSI
        & $\mathcal{O}(\chi^3)$
        & Randomized
        & Approximation accuracy depends on sketching \\

        TP-VQA, state-based
        & $\mathcal{O}(\chi^2)$
        & Decreases with system size
        & Large measurement overhead \\

        TP-VQA, diagonal encoding
        & $\mathcal{O}(\chi^3)$ classical overhead
        & $\mathcal{O}(1)$ in investigated regimes
        & Requires MPS reconstruction \\

        \bottomrule
    \end{tabular}

    \caption{Comparison of classical tensor-network and TP-VQA
    implementations of the nonlinear Hadamard product. The quoted
    scalings indicate the leading dependence on the bond dimension.
    For TP-VQA, the overall computational cost also depends on the
    postselection success probability and associated measurement
    overhead.}
    \label{tab:Hadamard_comparison}
\end{table*}

\begin{figure}[t]
    \centering
    \includegraphics[width=0.90\linewidth]{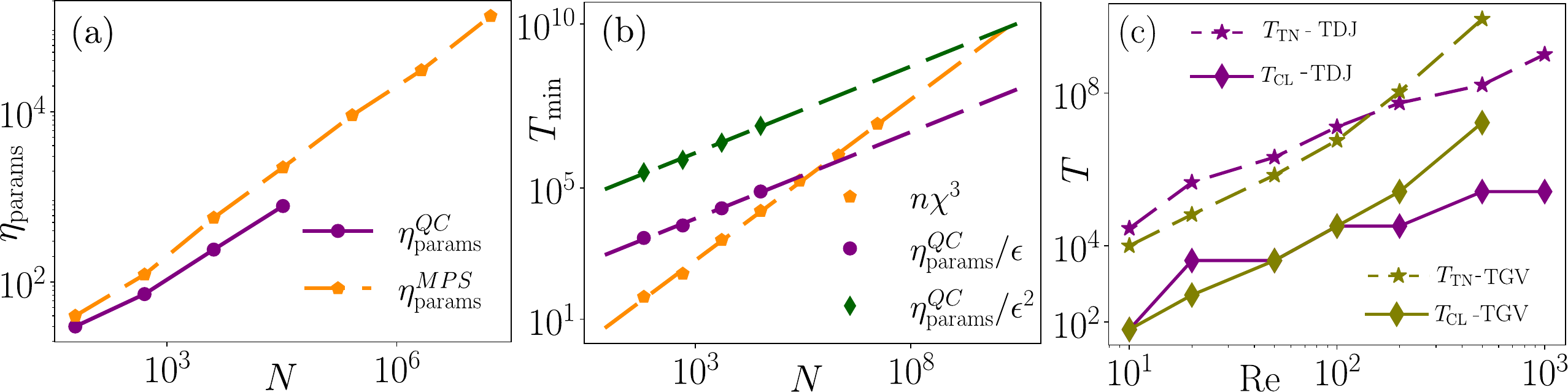}
    \caption{Scaling results for TP-VQA, adapted from Siegl \textit{et al.}~\cite{Siegl_2026,Siegl2026-arxiv}. (a) Parameter counts of the
    variational quantum circuit and classical MPS representation for
    increasingly large subdomains of a three-dimensional turbulent field. (b) Estimated computational cost as a function of system size $N$. (c) Classical tensor-network cost $T_{\mathrm{TN}}$ and classical reconstruction cost $T_{\mathrm{CL}}$ of the hybrid scheme versus
    Reynolds number for the temporally developing jet (TDJ) and TGV.}
    \label{fig:Siegl_scaling}
\end{figure}

For TP-VQA, several contributions must be considered when estimating the total computational cost. First, we consider the cost of the quantum circuit $T_{\rm QC}$ and its measurement. The dominant contribution is controlled by the number of parameters in the ansatz circuit $\eta^{\rm QC}_{\rm params}$, together with the sampling cost required to estimate expectation values, which depends on the target accuracy $\epsilon$ and the success probability $\alpha_{\rm succ}$ as
\begin{equation}
    T_{\rm QC}
    \sim
    \frac{\eta^{\rm QC}_{\rm params}}
    {\epsilon^2 \alpha_{\rm succ}} .
\end{equation}
Here, $\alpha_{\rm succ}$ denotes the block-encoding success probability,
corresponding to $p_{\rm succ}$ defined in Eq.~\ref{eq:success_prob}, following the notation of Ref.~\cite{Siegl_2026}. Using a three-dimensional turbulent flow field from the Johns Hopkins turbulence database \cite{Li_2008}, Siegl \textit{et al.}~\cite{Siegl_2026} compared the number of parameters required by a classical MPS representation with those required by a variational quantum circuit. At a fixed infidelity threshold of $\epsilon_F=0.01$, the fitted parameter counts scaled as 
\begin{equation}
    \eta^{\mathrm{MPS}}_{\text{params}}\propto N^{0.65},
    \qquad
    \eta^{\mathrm{QC}}_{\text{params}}\propto N^{0.52},
\end{equation}
where the increase in $N$ was obtained by considering progressively
larger subdomains of the same turbulent flow field (see Figure~\ref{fig:Siegl_scaling}a). The variational quantum representation therefore required fewer parameters than the corresponding MPS representation, with the difference increasing as the system size grew. More recent approaches suggest that further reductions may be possible by exploiting the
structure of turbulent fields directly in the circuit construction. For example, the approach in Ref.~\cite{meng2026geometricencodingturbulenceendtoend} exploits the multiscale self-similarity of turbulence to construct linear-depth quantum circuits, while Hashizume \textit{et al.}~\cite{hashizume2026} introduced a quantum-sparsity regularization strategy that improves the trainability
and accuracy of variational encodings of turbulent flow fields, as discussed previously in Sec.~\ref{Sec:StandardVGA}.

Using the resource model adopted in Ref.~\cite{Siegl_2026}, the classical MPS contraction cost scaled approximately as
$T_{\mathrm{MPS}}\propto N$, whereas the estimated measurement cost of the VQA scaled approximately as $T_{\mathrm{VQA}}\propto\sqrt{N}$ over
the investigated range (see Figure~\ref{fig:Siegl_scaling}b). For the VQA measurement cost, the available finite-size data could also be fitted by a polylogarithmic curve, namely
$\mathcal{O}((\log N)^3)$.  This resource model, however, did not explicitly account for the additional costs associated with nonlinear terms, including decreasing success probabilities in TP-VQA and the $\mathcal{O}(\chi^4)$ scaling, or potentially large intermediate bond dimensions, of the corresponding classical TN routines. These results should therefore be interpreted as preliminary scaling evidence rather than an asymptotic complexity proof. At the circuit level, using the MPO-to-circuit construction of Termanova \textit{et al.}~\cite{Termanova2024}, the operator complexity is controlled by the bond dimension of the unitary MPO approximation
rather than directly by the full spatial resolution. Since each MPO core is converted into a multi-qubit gate acting on
$\lceil \log_2 \chi_O\rceil+1$ qubits, the corresponding two-qubit gate count scales approximately as $\mathcal{O}(n\chi_O^2)$, where $n$ is the number of system qubits and $\chi_O$ is the bond dimension of the
unitary MPO approximation.

For the treatment of nonlinear terms, the Hadamard-product evaluation typically remains the dominant computational contribution in both the classical and hybrid tensor-network approaches. In the hybrid scheme, the MPS reconstruction and unitary-compilation steps scale as $\mathcal{O}(\chi^3)$, providing a polynomial reduction relative to the fully classical time-evolution routine. Since the classical reconstruction step scales more strongly with bond dimension than the
quantum-circuit depth, it is expected to become the dominant component of the hybrid cost for increasingly turbulent flows, where larger bond dimensions are required.

Figure~\ref{fig:Siegl_scaling}(c) compares the cost of the classical tensor-network routine with the classical reconstruction cost in the hybrid scheme for evaluating the nonlinear term in the Burgers' equation at an accuracy of $0.01$. The comparison considers two canonical flow problems, the two-dimensional TDJ and the three-dimensional TGV, over increasing Reynolds numbers. In both cases, the classical reconstruction cost in the hybrid quantum--classical approach remains two to three orders of magnitude below the corresponding fully classical computation.

Taken together, these works establish TP-VQA as a systematic interface between classical TN discretizations and quantum circuits
\cite{Termanova2024,Siegl_2026,Siegl2026-arxiv}. PDE states and operators can be formulated in familiar TN representations and systematically translated into quantum circuits, rather than requiring a separate gate-level construction for each problem.

\subsection{Computational Characteristics of Hybrid Variational Methods}
\label{sec:computational_considerations}

A direct quantitative comparison between the different approaches outlined in this review article is difficult, since they employ different representations, address different classes of problems, and are typically evaluated using different benchmarks. Their computational characteristics can nevertheless be compared
qualitatively by identifying the main sources of cost in each formulation. As with fully quantum algorithms, hybrid variational methods require quantum-circuit execution and measurement, but they introduce the additional cost of a classical optimization loop, whose number of iterations and convergence behavior are generally difficult to predict a priori.

The dominant contributions depend on the variational formulation. In QPINNs, the circuit must be evaluated, together with the required
derivatives, at a set of collocation and boundary points, so the computational cost depends strongly on the number of sampling points and derivative evaluations
\cite{Kyriienko2021,Setty2025}. In amplitude-encoded approaches such as TP-VQA, the cost additionally depends on state preparation, operator
implementation, postselection, and measurement. For TP-VQA in
particular, the required resources are closely linked to the bond dimensions of the MPS representations of the solution fields and the MPO representations of the operators, which control both the tensor-network overhead and the complexity of the compiled quantum circuits.

A practical advantage of the tensor-programmable framework is that much of the operator construction can be automated. The discretized
differential operators are represented as MPOs and systematically translated into quantum circuits
\cite{Termanova2024,Siegl_2026}. For fixed linear operators, these circuits can be constructed once and reused throughout the variational optimization. For state-dependent nonlinear operators, however, the
operator representation or circuit parameters may need to be updated, and in some formulations recompiled, as the solution evolves.

Compared with fully quantum approaches, hybrid variational methods can
operate with comparatively shallow circuits and avoid explicit global matrix inversion or long coherent time evolution. QPINNs also avoid encoding the complete discretized solution field into a single quantum state, since the variational circuit instead acts as a coordinate-dependent function approximator. TP-VQA, by contrast, retains a global amplitude-encoded
representation of the field and therefore enables structured operator
manipulations directly on the encoded state. It does not inherit the explicit
dependence on the condition number $\kappa$ appearing in QLSA complexity, although poor conditioning may still affect the optimization
landscape and convergence rate \cite{Harrow2009,Siegl_2026}.

Siegl \textit{et al.}~\cite{Siegl_2026} also introduced a fidelity-based measure for monitoring variational convergence without requiring a classically computed reference solution and observed improved trainability for the
operator-based cost-function construction considered. An additional advantage is the direct treatment of nonlinear terms. Among the approaches reviewed here, TP-VQA is one of the few that implements nonlinear field products directly within the quantum workflow, while recent diagonal block-encoding strategies show favorable success-probability scaling for the investigated turbulent flow fields
\cite{Siegl_2026,Siegl2026-arxiv}.

These practical benefits are accompanied by additional computational costs that must be considered when assessing hybrid variational methods. For QPINNs, the number of collocation points and the repeated evaluation of spatial or temporal derivatives can lead to substantial sampling and
optimization costs. For TP-VQA and related amplitude-encoded methods, the total cost depends on the number of optimization iterations, the measurement shots required
to estimate the variational objective, and, where necessary, postselection and classical reconstruction. In all cases, the ansatz must remain sufficiently expressive while avoiding excessive circuit depth and associated trainability problems. General trainability
issues, including barren plateaus and quantum-sparsity regularization strategies, are discussed in Sec.~\ref{Sec:StandardVGA}. In the TP-VQA time-stepping demonstrations considered so far, such trainability problems have not been observed, which may be aided by warm starts in which the optimized parameters from one time step initialize the next.

Overall, the practical performance of hybrid variational CFD algorithms
depends on the interplay between ansatz expressivity and trainability, measurement and sampling costs, and classical optimization. QPINNs and amplitude-encoded approaches distribute these costs differently. More
generally, the computational viability of quantum CFD methods depends strongly on the structure of the underlying problem, including the complexity of state preparation and operator implementation.

\section{From NISQ to Early Fault-Tolerant Implementations}
\label{Sec:NISQ_Fault_Tolerant}

The hardware results discussed in the preceding sections show that noise remains one of the main obstacles for present-day VQAs, especially when nonlinear cost-function terms have to be evaluated reliably. This limitation is central for quantum CFD, where the solution quality depends not only on preparing a variational state, but also on accurately estimating residuals, nonlinear operators, and time-stepping contributions. An additional challenge for scientific quantum computing is to quantify how hardware imperfections affect the predicted observables. As discussed in Sec.~\ref{subsubsec:Adv_Lim_Hamil}, Hamiltonian and Lindbladian learning can be used to characterize the effective device dynamics and propagate the resulting uncertainties to physical observables \cite{Kraft2026}. Recent CFD demonstrations have nevertheless shown that hybrid quantum algorithms can already be executed on present-day hardware. Chen \textit{et al.}~\cite{Chen2024} demonstrated a hybrid
quantum-classical linear-solver approach on superconducting quantum hardware for steady Poiseuille flow and unsteady acoustic-wave
propagation. Meng \textit{et al.}~\cite{Meng2024} experimentally simulated two-dimensional unsteady flows on a superconducting quantum processor using a Hamiltonian-simulation approach based on the hydrodynamic
formulation of the Schr\"odinger equation, including a compressible diverging flow and a decaying vortex. Early fault-tolerant quantum computers are expected to provide an intermediate regime between noisy intermediate-scale quantum devices and large fully fault-tolerant machines. In this regime, logical qubits become available, with each logical qubit encoded in several physical qubits, but the number of logical qubits and the size of executable circuits remain limited by the overhead of quantum error correction \cite{Katabarwa2024}. 

A further limitation concerns the implementation of the continuous
rotation angles that appear in variational ans\"atze. In a
fault-tolerant setting, arbitrary single-qubit rotations are usually
approximated by sequences from a discrete universal gate set, such as
Clifford+$T$. The Solovay--Kitaev theorem guarantees that such
approximations are possible with only polylogarithmic overhead in the
target precision, while more specialized Clifford+$T$ synthesis
methods can reduce the required $T$ count for single-qubit rotations
\cite{Dawson2005blj,Ross2014okw}. However, fault-tolerant quantum circuits cannot usually implement arbitrary rotation angles directly. Instead, rotations must be approximated using a finite set of fault-tolerant gates, often the Clifford+$T$ gate set. The Clifford gates are comparatively inexpensive in many error-corrected architectures, whereas the non-Clifford $T$ gate is a costly resource because it is commonly implemented using magic-state injection and distillation \cite{Bravyi2005,Campbell2017}. As a result, the number of available $T$ gates limits how accurately a continuous rotation angle can be approximated. This is relevant for VQAs, because the classical optimizer proposes continuously varying parameters, while the quantum processor can only realize finite-precision approximations of those angles. Recent neutral-atom experiments have illustrated this effect by showing that the set of accessible rotation angles becomes denser as the allowed number of $T$ gates is increased \cite{Bluvstein2026}. VQAs are nevertheless well matched to early fault-tolerant settings because they use relatively shallow quantum circuits that are repeatedly evaluated within a classical optimization loop.

Circuit width must also be controlled. As discussed in the context of
the quantum Nyquist--Shannon sampling theorem, the number of qubits
needed for amplitude encoding is set by the range of spatial scales in
the represented field, and in particular by the smallest physically
relevant wavelength that must be resolved \cite{hashizume2026}.
Adding qubits beyond this resolution requirement does not add useful
physical information, but it does increase the hardware and
error-correction overhead. Early fault-tolerant VQAs for CFD should
therefore aim for circuits that are expressive enough to represent the
relevant flow scales, while remaining narrow and shallow enough to be
compatible with the available logical hardware.

Several recent studies suggest that variational algorithms can be adapted to fault-tolerant architectures. Sayginel \textit{et al.}~\cite{Sayginel2024} studied a fault-tolerant implementation of the
variational quantum eigensolver in which continuously parameterized
quantum gates are approximated using the Clifford+$T$ gate set. For spin-model examples with up to $16$ qubits, adapting the approximation accuracy
during the optimization led to convergence behaviour similar to that of the original continuously parameterized circuits. This shows that variational optimization can remain effective after fault-tolerant
gate compilation, although the study does not provide a CFD-specific resource estimate or a quantum-advantage threshold.

Dangwal \textit{et al.}~\cite{Dangwal2025} considered VQAs on an early
fault-tolerant architecture with approximately $10^4$ physical qubits
and projected two-qubit error rates of about $10^{-3}$. Their approach
applies error correction only to selected parts of the circuit, thereby
reducing some of the overhead associated with fully fault-tolerant
implementations. Across the benchmarks considered, the method achieved
an average $9.27\times$ improvement over NISQ execution according to
the performance metric used in the study, and reduced circuit latency
by approximately a factor of two. These results indicate that partial
error correction may already be useful before large fully
fault-tolerant quantum computers are available.

CFD-specific quantum--classical crossover estimates are also beginning
to appear, although they currently refer to different algorithmic
settings. Zhuang et
al.~\cite{zhuang2025pathwaypracticalquantumadvantage} developed an
end-to-end, fully fault-tolerant Navier--Stokes solver based on a
quantum linear-system algorithm. For a two-dimensional
$2^{40}\times2^{40}$ grid, corresponding to
$2^{80}\approx1.21\times10^{24}$ cells, their
fault-tolerant resource model predicts that the proposed algorithm
would require approximately $8.71\times10^{6}$ physical qubits and
$42.6$ days of runtime, assuming a physical error rate of
$5\times10^{-4}$. The corresponding classical runtime was estimated
to be approximately $130$ years. This result provides an important
reference point for quantum CFD, but it relies on several favourable
assumptions, including sparse spectral representations, efficient
quantum input and output procedures without generic QRAM, specialized
operator encodings, and full quantum error correction. It should
therefore be viewed as a fully fault-tolerant QLSA-based benchmark,
rather than as an estimate for early fault-tolerant VQA-based CFD.

For variational quantum CFD (VQCFD), Syamlal et
al.~\cite{Syamlal2024VQCFD} introduced the performance metric
\begin{equation}
    Q_{5E7}
    =
    \frac{T_{\mathrm{VQCFD}}}
         {T_{\mathrm{classical}}},
\end{equation}
which compares the runtimes of VQCFD and classical
CFD for a representative industrial problem with $5\times10^{7}$ grid
points. A value of $Q_{5E7}\leq1$ would indicate that the quantum
method is at least as fast as the classical solver. For their current
unoptimized two-dimensional prototype, they obtained
\begin{equation}
    Q_{5E7}
    =
    1.3\times10^{12}.
\end{equation}
Their performance model was calibrated using circuits executed on IBM
quantum processors. The largest circuits used up to $120$ qubits, and
the deepest reported circuit contained $112\,910$ sequential gate
layers at $96$ qubits. These results provide a direct performance
baseline for VQA-based CFD and quantify the scale of improvement still
required. However, they do not yet determine the number of logical
qubits or the fault-tolerant circuit depth needed for VQA-based CFD to
outperform classical CFD.

Early fault-tolerant devices could nevertheless substantially improve
the prospects of VQA-based CFD. More reliable logical operations would
allow deeper and more expressive ans\"atze, more stable evaluation of
QNPU-based nonlinear terms, and longer variational time integrations
than are practical on present noisy hardware. The NLSE and Burgers' equation discussed in Sec.~\ref{Sec:StandardVGA}, therefore, provide
useful circuit-level benchmarks for width, entangling-gate count, and
depth. At present, however, these requirements have not yet been
translated into logical gate counts, logical circuit depths, or
physical-qubit requirements for a specific error-correction
architecture.

The current literature therefore provides two complementary pieces of
evidence: a projected crossover for a specialized fully
fault-tolerant CFD formulation, and a direct performance baseline for
VQA-based CFD on present hardware. What is still missing is a
corresponding crossover estimate for VQA-based CFD on early
fault-tolerant devices. Establishing such an estimate will require an
end-to-end analysis that includes quantum error correction, state
preparation, measurement repetitions, optimization cost, and comparison
with optimized classical CFD solvers.

\section{Towards Quantum Advantage in CFD}
\label{Sec:QuantumAdvantage}

Quantum CFD approaches investigated above differ substantially in their algorithms, input assumptions, output requirements, and hardware demands, therefore no universal crossover estimate can be applied across the field. Nevertheless, the available results make it possible to identify regimes in which quantum advantage is more or less plausible. A useful distinction is between a representation advantage, whereby a field with $N$ degrees of freedom can, in principle, be encoded using $\mathcal{O}(\log N)$ qubits, and an end-to-end advantage in the total computational cost required to achieve a prescribed accuracy. Although compact encoding is common to quantum CFD formulations, it does not by itself imply a practical speedup. Such an advantage is obtained only if the complete workflow, including state preparation, operator implementation, evolution or optimization, measurement, and classical post-processing, outperforms the corresponding classical CFD method.

Quantum methods appear most promising for very large, structured problems whose states and operators can be implemented efficiently and whose outputs consist of a limited number of observables or coarse-grained quantities. Reconstructing an arbitrary fine-grid field with $N$ components would require at least $\mathcal{O}(N)$ classical output and would therefore remove the exponential benefit of amplitude encoding. Quantum PDE algorithms consequently often formulate the output in terms of selected quantities of interest. For example, Montanaro and Pallister estimate linear functionals of a finite-element solution, including spatial averages over prescribed regions \cite{Montanaro_2016}. Other approaches aim to extract only the large-scale features of a quantum-encoded field rather than reconstructing every fine-grid value. Terraneo \textit{et al.}~\cite{Terraneo2005} showed that coarse information can be obtained directly or after applying a quantum wavelet transform, which separates broad spatial variations from small-scale details. More recent wavelet algorithms make this separation more efficient \cite{Bagherimehrab_2024}. Related multilevel methods for elliptic and parabolic PDEs estimate quantities of interest by combining results from coarse and fine grids, thereby reducing the need to access the highest-resolution solution directly \cite{li2026efficientendtoendquantumelliptic,li2026exponentialreductionmeshdependence}. These approaches do not make the recovery of an arbitrary low-resolution field cost-free, since producing $M$ classical values still requires an output cost that grows with $M$, but they show that physically relevant reduced information can be extracted without reconstructing the full quantum state. The fault-tolerant crossover projected by Zhuang \textit{et al.}~\cite{zhuang2025pathwaypracticalquantumadvantage} illustrates the resulting opportunity: an end-to-end advantage may emerge at sufficiently large problem sizes, provided that state preparation, operator implementation, observable extraction, and error-correction costs remain favourable.

For VQAs and TP-VQAs, the most favourable regime is one in which CFD fields and operators retain sufficient structure for efficient quantum-state preparation and compact operator circuits, while classical TN contractions have already become expensive. Turbulence is therefore not necessarily unfavourable. Its multiscale correlations can support compact TN representations and provide structure that can be exploited in quantum-state construction \cite{Gourianov2022,H_lscher_2025,Siegl_2026}. As discussed in Sec.~\ref{Sec:TPVQA}, Siegl \textit{et al.}~\cite{Siegl_2026} found that the number of variational parameters grew more slowly with problem size than for the corresponding MPS representation at fixed accuracy. For a fixed MPO bond dimension, the auxiliary-register size is independent of the number of system qubits, while the operator gate count grows only polynomially in this logarithmic register size. Hashizume \textit{et al.}~\cite{hashizume2026} similarly showed that the quantum resources required (namely qubit requirement $q_c$) to amplitude-encode a sufficiently regular field are governed primarily by its smallest physically relevant wavelength $\lambda_{\min}$. Defining
\begin{equation}
    q_c=\left\lceil\log_2\left(\pi/\lambda_{\min}\right)\right\rceil,
\end{equation}
the required circuit width and depth scale as $\mathcal{O}(q_c)$ and $\mathcal{O}(q_c^2)$, respectively. These results concern turbulent snapshots rather than complete turbulent-flow evolution and therefore do not imply constant circuit width or depth as the Reynolds number increases. Nevertheless, they suggest that quantum resources may grow with the range of physically relevant scales rather than directly with the total number of grid points. This is encouraging for fault-tolerant implementations, where moderate increases in logical width and depth may be manageable, provided that operator complexity, measurement cost, and optimization remain under control \cite{Katabarwa2024,Bravyi2005,Campbell2017}.

Current quantum CFD demonstrations remain too small to exhibit a practical advantage over optimized classical solvers. The relevant question is therefore how state-preparation, circuit, measurement, and optimization costs scale with problem size, and whether these trends remain favourable when extrapolated to classically demanding regimes. In this context, the value $Q_{5E7}=1.3\times10^{12}$ reported by Syamlal \textit{et al.}~\cite{Syamlal2024VQCFD} provides a useful baseline for the current VQCFD prototype. It shows that substantial reductions in circuit-execution, measurement, and optimization costs are still required, while also identifying the prefactors that future algorithmic and hardware improvements must address before an end-to-end crossover can be achieved.

\section{Conclusion}

This review has examined fully quantum, hybrid quantum--classical, and tensor-programmable approaches to CFD. Fully quantum methods include QLSAs, ranging from the HHL algorithm to more recent QSVT-based formulations, as well as Hamiltonian simulation and QLBMs, which provide distinct routes for encoding and evolving discretized flow equations.
Hybrid approaches instead use parameterized quantum circuits together with classical optimization, including QPINNs and amplitude-encoded variational PDE solvers
\cite{Harrow2009,Childs_2021,Itani2024,Cerezo2021,Kyriienko2021,Setty2025}.
Among these approaches, variational methods are particularly relevant to the transition from present noisy devices to early fault-tolerant
quantum hardware, while tensor-programmable formulations provide a
structured route for translating CFD states and operators into such
circuits.

A central conclusion is that quantum advantage in CFD cannot be
assessed from asymptotic complexity alone. The complete computational
workflow must include state preparation, operator construction,
circuit depth and width, measurement cost, classical optimization,
and post-processing. Amplitude encoding can represent a field with
$N$ degrees of freedom using only
$n=\lceil\log_2N\rceil$ qubits, but this compression is useful only
when the state and operators can be prepared efficiently and when the
required output can be obtained without reconstructing the complete
classical field  \cite{Aaronson2015,Dalzell_2025}. Moreover, a meaningful assessment of quantum advantage requires comparison with optimized classical methods exploiting the same problem structure, including state-of-the-art CFD algorithms, ROMs, and TN
methods \cite{Malinverno2026,Amaral2026}.

TNs provide an important framework for making this comparison. Their
computational efficiency is, however, conditional on the transient CFD fields and operators retaining exploitable approximate low-rank structure, such that the required bond dimensions remain moderate or grow sufficiently slowly with Reynolds number, simulation time, grid refinement, and target accuracy. When this condition is satisfied, TN
methods offer efficient classical compression and provide a demanding baseline for quantum advantage
\cite{Schollwock2011,Oseledets2011}. The same structure can also be used to construct quantum states and circuits. In particular, MPS representations provide compact descriptions of weakly entangled states, while MPOs can encode structured differential operators
\cite{Termanova2024,Siegl_2026}. The required bond dimension therefore acts as both a measure of classical compressibility and an indicator of the resources needed for quantum implementation.

The transition from NISQ to early fault-tolerant hardware is likely to
be especially important for VQA-based CFD. Present hardware
demonstrations show that shallow variational states can be prepared,
but the deeper circuits required to evaluate nonlinear cost functions
remain sensitive to noise. Early fault-tolerant devices could provide
more reliable logical operations and support deeper or more expressive
circuits, while still retaining the hybrid structure of VQAs
\cite{Katabarwa2024,Sayginel2024,Dangwal2025}. Circuit width must also
remain controlled so that the encoded state resolves the physically
relevant flow scales without introducing unnecessary logical-qubit and
error-correction overhead.

Existing studies provide useful quantitative reference points, but not yet a definitive early-fault-tolerant crossover for VQA-based CFD. A fully fault-tolerant QLSA-based Navier--Stokes study has projected an advantage under specific assumptions and with millions of physical qubits, whereas current VQCFD performance models show that substantial reductions in circuit-execution, measurement, and optimization costs are still required \cite{zhuang2025pathwaypracticalquantumadvantage,Syamlal2024VQCFD}. Determining the logical-qubit count, circuit depth, and runtime at which VQA-based CFD can outperform optimized classical solvers therefore remains an open problem.

Progress toward this goal requires several concrete research priorities. First, quantum CFD must move beyond one-dimensional model equations toward a common hierarchy of increasingly realistic benchmarks, including two- and three-dimensional turbulent flows, complex geometries, realistic boundary conditions, and long-time evolution. These studies should report scaling with grid resolution, Reynolds number, simulation time, and target accuracy. Second, efficient representations must be developed for complex geometries and operators, including geometry masks, body-fitted or unstructured grids, nonlinear terms, and preconditioners. The trainability of increasingly expressive variational circuits must also be assessed systematically as the Reynolds number, system size, and simulation time increase. Third, methods for extracting physical observables and coarse-grained flow information must be developed without reconstructing the complete high-resolution field. Finally, resource estimates should include the complete workflow, state preparation, operator compilation, postselection, measurement repetitions, optimization, and quantum error correction and should be compared with optimized CPU, GPU, ROMs, and TN solvers under equivalent accuracy requirements.

Among the approaches reviewed here, TP-VQAs provide a particularly structured framework in which to pursue several of these priorities.
By using TNs as an intermediate representation between classical CFD
discretizations and quantum circuits, they combine structured state preparation, operator compilation, and variational optimization
\cite{Termanova2024,Siegl_2026}. As more reliable quantum processors become available, increasingly ambitious studies should become possible, progressing from small proof-of-principle demonstrations to multidimensional flows, longer
simulations, nonlinear dynamics, and more realistic geometries. The next decisive step is therefore a sequence of reproducible scaling studies that exploit improving hardware while maintaining fair comparisons with classical methods. Demonstrating robust treatment of complex geometries, turbulent multiscale structure, nonlinear operators, and physically relevant outputs, together with credible early-fault-tolerant resource estimates, would mark a decisive step toward end-to-end CFD simulation algorithms with quantum advantage.

% =========================================================
% Author Contributions
% =========================================================

\section*{Author Contributions}

Conceptualization, M.G.C. and D.J.; formal analysis, M.G.C.; investigation, M.G.C.; writing and original draft preparation, M.G.C.; writing, review and editing, M.G.C., N.-L.v.H., T.H., P.S., A.S., J.D.d.C.J., P.O., S.B., M.U., S.T., E.M., T.K., F.C.-L., L.S., T.R., F.M., B.K., M.K., D.G.A., E.d.V. and D.J.; supervision, D.J.; project administration, B.Y.; funding acquisition, D.J. All authors contributed to the interpretation of the results, reviewed the manuscript, and have read and agreed to the published version of the manuscript.

% =========================================================
% Funding
% =========================================================

\section*{Funding}

This research was funded by the European Union’s Horizon Europe Research and Innovation Programme under grant agreement No. 101080085 (QCFD). Additional funding was provided by the Deutsche Forschungsgemeinschaft (DFG) through the project ``Quantencomputing mit neutralen Atomen'' (JA 1793/1-1, Japan-JST-DFG-ASPIRE 2024), the Cluster of Excellence ``Advanced Imaging of Matter'' (EXC 2056, project ID 390715994), and the Hamburg Quantum Computing Initiative (HQIC) EFRE project.

% =========================================================
% Data Availability
% =========================================================

\section*{Data Availability}

No new data were created in this study.
Data sharing is not applicable to this article.

% =========================================================
% Acknowledgments
% =========================================================

\begin{acknowledgments}

The authors also gratefully acknowledge the members of the QCFD consortium for valuable scientific exchanges and collaborative discussions that contributed to this work. During the preparation of this manuscript, the authors used AI to assist with language editing, text refinement, and improving clarity of presentation. The authors reviewed and edited all generated content and take full responsibility for the final manuscript.

\end{acknowledgments}

% =========================================================
% Conflict of Interest
% =========================================================

\section*{Conflict of Interest}

The authors declare no conflicts of interest.

% =========================================================
% Abbreviations
% =========================================================
% \clearpage

\section*{Abbreviations}

The following abbreviations are used in this manuscript:

\begin{table*}[h]
    \centering
    \small
    \setlength{\tabcolsep}{8pt}
    \renewcommand{\arraystretch}{1.15}

    \begin{tabular}{@{}ll@{}}
        \toprule
        \textbf{Abbreviation} & \textbf{Definition} \\
        \midrule
        RANS & Reynolds-Averaged Navier–Stokes \\
        LES & Large-Eddy Simulation \\
        LCU & Linear-Combination-of-Unitaries \\
        TN & Tensor Network \\
        CNOT & Controlled-NOT \\
        NISQ & Noisy Intermediate-Scale Quantum \\
        MPS & Matrix Product State\\
        MPO & Matrix Product Operator\\
        TT & Tensor Train \\
        TCI & Tensor Cross Interpolation \\ 
        EST & Efficient State Tomography \\
        VQA & Variational Quantum Algorithm \\
        TP-VQA & Tensor-Programmable Variational Quantum Algorithm \\
        MCT & Multi-Controlled Toffoli \\
        HHL & Harrow, Hassidim, and Lloyd \\
        QLSA & Quantum Linear System Algorithm \\
        QSVT & Quantum Singular Value Transformation \\
        CFD & Computational Fluid Dynamics \\
        PDE & Partial Differential Equations \\
        ROM & Reduced-Order Model \\
        DNS & Direct Numerical Simulation \\
        QLBM & Quantum Lattice Boltzmann Method \\
        QNPU & Quantum Nonlinear Processing Unit \\
        VQCFD & Variational Quantum Computational Fluid Dynamics \\
        TGV & Taylor--Green vortex \\
        TDJ & Temporally Developing Jet \\
        CPU & Central Processing Unit \\
        GPU & Graphics Processing Unit \\
        \bottomrule
    \end{tabular}
\end{table*}

% =========================================================
% Appendix
% =========================================================

\appendix

\section{The HHL Algorithm}
\label{Appendix:HHL}

Let the spectral decomposition of $A$ be
\begin{equation}
    A = \sum_j \lambda_j \ket{u_j}\bra{u_j},
\end{equation}
and suppose the right-hand side has been prepared as
\begin{equation}
    \ket{b} = \sum_j \beta_j \ket{u_j}.
\end{equation}

The objective is to transform this state into one proportional to
\begin{equation}
    A^{-1}\ket{b}
    =
    \sum_j \frac{\beta_j}{\lambda_j}\ket{u_j}.
\end{equation}
The algorithm achieves this through the following stages:

\begin{enumerate}

\item \textbf{State Preparation:}

The classical vector $b$ is encoded as a normalized quantum state
$\ket{b}$ in the computational basis.

\item \textbf{Quantum Phase Estimation (QPE):}

Because $A$ is Hermitian, the unitary operator $e^{iAt}$ has
eigenvectors $\ket{u_j}$ with eigenvalues $e^{i\lambda_j t}$.
Efficient Hamiltonian simulation \cite{Berry_2015} enables
implementation of controlled applications of $e^{iAt}$.
QPE \cite{Kitaev1995QuantumMA} then coherently extracts
approximations $\tilde{\lambda}_j$ of the eigenvalues, yielding
\begin{equation}
    \sum_j \beta_j \ket{u_j}\ket{0}
    \longrightarrow
    \sum_j \beta_j
    \ket{u_j}\ket{\tilde{\lambda}_j}.
\end{equation}
Thus, eigenvalue information becomes available in a quantum register.

\item \textbf{Controlled Rotation:}

An ancilla qubit is rotated by an angle conditioned on the eigenvalue
register, implementing
\begin{equation}
    \ket{\tilde{\lambda}_j}\ket{0}
    \mapsto
    \ket{\tilde{\lambda}_j}
    \left(
        \sqrt{
        1-\frac{C^2}{\tilde{\lambda}_j^2}
        }\ket{0}
        +
        \frac{C}{\tilde{\lambda}_j}\ket{1}
    \right),
\end{equation}
where $C$ is a normalization constant chosen such that
$|C/\lambda_j|\leq 1$ for all eigenvalues $\lambda_j$ of $A$.
This step encodes the reciprocal $1/\lambda_j$ in the amplitude
of the ancilla state $\ket{1}$.

In practice, the choice of $C$ requires prior knowledge of a lower
bound on the spectrum of $A$. If the eigenvalues satisfy
$\lambda_j\in[\lambda_{\min},\lambda_{\max}]$, one typically sets
$C=\mathcal{O}(\lambda_{\min})$ to ensure that the rotation remains
well defined for all $j$. For matrices rescaled such that
$\lambda_j\in(0,1]$, this corresponds to
$C=\mathcal{O}(1/\kappa)$, where
$\kappa=\lambda_{\max}/\lambda_{\min}$ is the condition number.

Such spectral bounds are often available in practice. For example,
in discretized PDEs, analytical estimates of the operator spectrum,
such as those for the Laplacian, or bounds derived from the problem
structure can be used. Alternatively, preconditioning or normalization
of the matrix $A$ may be employed to ensure that its spectrum lies
within a known interval.

In experimental demonstrations of the HHL algorithm, the eigenvalues
are typically known in advance and the controlled rotations are
compiled directly, effectively fixing $C$ based on the specific
problem instance. While this choice guarantees that the transformation
is physically valid, it also directly affects the algorithm's success
probability. Since the amplitude of the $\ket{1}$ state scales as
$C/\lambda_j$, the overall probability of successful postselection
scales as $\mathcal{O}(C^2)$, which leads to the characteristic
$\mathcal{O}(1/\kappa^2)$ dependence in the original HHL algorithm.

\item \textbf{Postselection:}

Following the controlled rotation, the system register containing the
eigenvectors $\ket{u_j}$, the eigenvalue register, and the ancilla
qubit are in an entangled state. The inverse QPE procedure is therefore
applied to erase the eigenvalue information, returning the auxiliary
register to the $\ket{0}$ state and disentangling it from the system.

Finally, the ancilla qubit is measured. Conditional on observing the
$\ket{1}$ state, the system register collapses into the desired
solution state proportional to
\begin{equation}
    \ket{x}
    =
    A^{-1}\ket{b}
    =
    \sum_j
    \frac{\beta_j}{\lambda_j}
    \ket{u_j}.
\end{equation}

\end{enumerate}

% =========================================================
% References
% =========================================================

\bibliographystyle{apsrev4-2}
\bibliography{main}

@book{ferziger02:CMFD,
  author    = {Ferziger, Joel H. and Peri\'{c}, Milovan and Street, Robert L.},
  title     = {Computational Methods for Fluid Dynamics},
  publisher = {Springer},
  year      = {2002},
  edition   = {3rd},
  address   = {Berlin},
  isbn      = {978-3-540-42074-3},
  doi       = {10.1007/978-3-642-56026-2}
}

@article{Nibbi2024,
  title = {Block encoding of matrix product operators},
  author = {Nibbi, Martina and Mendl, Christian B.},
  journal = {Phys. Rev. A},
  volume = {110},
  issue = {4},
  pages = {042427},
  numpages = {14},
  year = {2024},
  month = {Oct},
  publisher = {American Physical Society},
  doi = {10.1103/PhysRevA.110.042427} 
}

@misc{Tam2026-arxiv,
      title={Geometric Quantum Physics Informed Neural Network}, 
      author={Wai-Hong Tam and Reza Safari and Hiromichi Matsuyama},
      year={2026},
      eprint={2605.02352},
      archivePrefix={arXiv},
      primaryClass={quant-ph},
      url={https://arxiv.org/abs/2605.02352}, 
}

@Article{Siegl2025,
  author   = {Siegl, Pia and Wassing, Simon and Mieth, Dirk Markus and Langer, Stefan and Bekemeyer, Philipp},
  journal  = {CEAS Aeronautical Journal},
  title    = {Solving transport equations on quantum computers--potential and limitations of physics-informed quantum circuits},
  year     = {2025},
  issn     = {1869-5590},
  number   = {1},
  pages    = {63--80},
  volume   = {16},
  doi      = {10.1007/s13272-024-00774-2},
  refid    = {Siegl2025},
  url      = {https://doi.org/10.1007/s13272-024-00774-2},
}

@article{Ransford2026,
    author={Ransford, Anthony
    and Allman, M. S.
    and Arkinstall, Jake
    and Campora, J. P.
    and Cooper, Samuel F.
    and Delaney, Robert D.
    and Dreiling, Joan M.
    and Estey, Brian
    and Figgatt, Caroline
    and Hall, Alex
    and Husain, Ali A.
    and Isanaka, Akhil
    and Kennedy, Colin J.
    and Kotibhaskar, Nikhil
    and Madjarov, Ivaylo S.
    and Mayer, Karl
    and Milne, Alistair R.
    and Park, Annie J.
    and Reed, Adam P.
    and Ancona, Riley
    and Andersen, Molly P.
    and Andres-Martinez, Pablo
    and Angenent, Will
    and Argueta, Liz
    and Arkin, Benjamin
    and Ascarrunz, Leonardo
    and Baker, William
    and Barnes, Corey
    and Bartolotta, John
    and Berg, Jordan
    and Besand, Ryan
    and Bjork, Bryce
    and Blain, Matt
    and Blanchard, Paul
    and Blume-Kohout, Robin
    and Bohn, Matt
    and Borgna, Agust{\'i}n
    and Botamanenko, Daniel Y.
    and Boutelle, Robert
    and Brown, Natalie
    and Buckingham, Grant T.
    and Burdick, Nathaniel Q.
    and Burton, William Cody
    and Carey, Varis
    and Carron, Christopher J.
    and Chambers, Joe
    and Chan, Jia Wen
    and Children, John
    and Colussi, Victor E.
    and Crepinsek, Steven
    and Cureton, Andrew
    and Davies, Joe
    and Davis, Daniel
    and DeCross, Matthew
    and Deen, David
    and Delaney, Conor
    and DelVento, Davide
    and DeSalvo, B. J.
    and Dominy, Jason
    and Drotar, Sydney
    and Duncan, Ross
    and Eccles, Vanya
    and Edgington, Alec
    and Erickson, Neal
    and Erickson, Stephen
    and Ertsgaard, Christopher T.
    and Esposito, Jay
    and Evans, Bruce
    and Evans, Tyler
    and Fabrikant, Maya I.
    and Fischer, Andrew
    and Foltz, Cameron
    and Foss-Feig, Michael
    and Francois, David
    and Freyberg, Brad
    and Gao, Charles
    and Garay, Robert
    and Garvin, Jane
    and Gaudiosi, David M.
    and Gilbreth, Christopher N.
    and Giles, Josh
    and Glynn, Erin
    and Graves, Jeff
    and Hansen, Azure
    and Hayes, David
    and Heidemann, Lukas
    and Higashi, Bob
    and Hilbun, Tyler
    and Hines, Jordan
    and Hlavaty, Ariana
    and Hoffman, Kyle
    and Hoffman, Ian M.
    and Holliman, Craig
    and Hooper, Isobel
    and Horning, Bob
    and Hostetter, James
    and Hothem, Daniel
    and Houlton, Jack
    and Hout, Jared
    and Hutson, Ross
    and Jacobs, Ryan T.
    and Jacobs, Trent
    and Johannsen, Melf
    and Johansen, Jacob
    and Jones, Loren
    and Julian, Sydney
    and Jung, Ryan
    and Keay, Aidan
    and Klein, Todd
    and Koch, Mark
    and Kondo, Ryo
    and Kong, Chang
    and Kosto, Asa
    and Lawrence, Alan
    and Liefer, David
    and Lollie, Michelle
    and Lucchetti, Dominic
    and Lysne, Nathan K.
    and Lytle, Christian
    and MacPherson, Callum
    and Malm, Andrew
    and Mather, Spencer
    and Mathewson, Brian
    and Maxwell, Daniel
    and McCaffrey, Lauren
    and McDougall, Hannah
    and Mendoza, Robin
    and Miller, David B.
    and Mills, Michael
    and Morrison, Richard
    and Narmour, Louis
    and Nguyen, Nhung
    and Nugent, Lora
    and Olson, Scott
    and Ouellette, Daniel
    and Parks, Jeremy
    and Peters, Zach
    and Peterson, Timothy A.
    and Petricka, Jessie
    and Pino, Juan M.
    and Polito, Frank
    and Potter, Andrew C.
    and Preidl, Matthias
    and Price, Gabriel
    and Proctor, Timothy
    and Pugh, McKinley
    and Ratcliff, Noah
    and Raymondson, Daisy
    and Rhodes, Peter
    and Roman, Conrad
    and Roy, Craig
    and Ryan-Anderson, Ciaran
    and Sanchez, Fernando Betanzo
    and Sangiolo, George
    and Sawadski, Tatiana
    and Schaffer, Andrew
    and Schow, Peter
    and Sedlacek, Jon
    and Semenenko, Henry
    and Shevchuk, Peter
    and Shore, Susan
    and Siegfried, Peter
    and Singhal, Kartik
    and Sivarajah, Seyon
    and Skripka, Thomas
    and Sletten, Lucas
    and Spaun, Ben
    and Sprenkle, R. Tucker
    and Stoufer, Paul
    and Tader, Mariel
    and Taylor, Stephen F.
    and Thompson, Travis H.
    and Tobey, Raanan
    and Tran, Anh
    and Tran, Tam
    and Vittorini, Grahame
    and Volin, Curtis
    and Walker, Jim
    and White, Sam
    and Williams, Garrett R.
    and Wilson, Douglas
    and Wolf, Quinn
    and Wringe, Chester
    and Young, Kevin
    and Zheng, Jian
    and Zuraski, Kristen
    and Baldwin, Charles H.
    and Chernoguzov, Alex
    and Gaebler, John P.
    and Sanders, Steven J.
    and Neyenhuis, Brian
    and Stutz, Russell
    and Bohnet, Justin G.},
title={A 98-qubit trapped-ion quantum computer with all-to-all connectivity},
journal={Nature},
year={2026},
month={Jul},
day={01},
volume={655},
number={8121},
pages={81-86},
issn={1476-4687},
doi={10.1038/s41586-026-10676-4},
url={https://doi.org/10.1038/s41586-026-10676-4}
}

@misc{Heinzelreiter2025,
      title={Carleman Linearization of Parabolic PDEs: Well-posedness, convergence, and efficient numerical methods}, 
      author={Bernhard Heinzelreiter and John W. Pearson},
      year={2025},
      eprint={2510.00722},
      archivePrefix={arXiv},
      primaryClass={math.NA},
      url={https://arxiv.org/abs/2510.00722} 
}

@article{Demirdjian2026,
  title = {Efficient decomposition of the Carleman linearized Burgers' equation},
  author = {Demirdjian, Reuben and Hogancamp, Thomas and Gunlycke, Daniel},
  journal = {Phys. Rev. A},
  volume = {113},
  issue = {3},
  pages = {032408},
  numpages = {16},
  year = {2026},
  month = {Mar},
  publisher = {American Physical Society},
  doi = {10.1103/g27q-r2gk},
  url = {https://link.aps.org/doi/10.1103/g27q-r2gk}
}

@misc{Novikau2025,
      title={Globalizing the Carleman linear embedding method for nonlinear dynamics}, 
      author={Ivan Novikau and Ilon Joseph},
      year={2025},
      eprint={2510.15715},
      archivePrefix={arXiv},
      primaryClass={quant-ph},
      url={https://arxiv.org/abs/2510.15715}, 
}

@article{Manetsch2025,
    author={Manetsch, Hannah J.
    and Nomura, Gyohei
    and Bataille, Elie
    and Lv, Xudong
    and Leung, Kon H.
    and Endres, Manuel},
    title={A tweezer array with 6,100 highly coherent atomic qubits},
    journal={Nature},
    year={2025},
    month={Nov},
    day={01},
    volume={647},
    number={8088},
    pages={60-67},
    issn={1476-4687},
    doi={10.1038/s41586-025-09641-4},
    url={https://doi.org/10.1038/s41586-025-09641-4}
}

@misc{Siegl2026-arxiv,
      title={Efficient Treatment of Non-Linearity in Quantum Computational Fluid Dynamics Using Hybrid Tensor Networks}, 
      author={Pia Siegl and Nis-Luca van H{\"u}lst and Maximilian Mandelt Buxadé and Tomohiro Hashizume and Dieter Jaksch},
      year={2026},
      eprint={2608.24150},
      archivePrefix={arXiv},
}

@article{preconditioning2012,
  title = {Preconditioning for {{Sparse Linear Systems}} at the {{Dawn}} of the 21st {{Century}}: {{History}}, {{Current Developments}}, and {{Future Perspectives}}},
  shorttitle = {Preconditioning for {{Sparse Linear Systems}} at the {{Dawn}} of the 21st {{Century}}},
  author = {Ferronato, Massimiliano},
  year = 2012,
  journal = {International Scholarly Research Notices},
  volume = {2012},
  number = {1},
  pages = {127647},
  issn = {2356-7872},
  doi = {10.5402/2012/127647},
  copyright = {Copyright \copyright{} 2012 Massimiliano Ferronato.},
  langid = {english},
}

@article{argyropoulosRecentAdvancesNumerical2015,
  title = {Recent Advances on the Numerical Modelling of Turbulent Flows},
  author = {Argyropoulos, C. D. and Markatos, N. C.},
  year = 2015,
  month = {jan},
  journal = {Applied Mathematical Modelling},
  volume = {39},
  number = {2},
  pages = {693--732},
  issn = {0307-904X},
  doi = {10.1016/j.apm.2014.07.001}
}

@book{Keiper2018-tf,
  editor    = {Keiper, Winfried and Milde, Anja and Volkwein, Stefan},
  title     = {Reduced-Order Modeling (ROM) for Simulation and Optimization: Powerful Algorithms as Key Enablers for Scientific Computing},
  publisher = {Springer International Publishing},
  address   = {Cham},
  year      = {2018},
  edition   = {1},
  doi       = {10.1007/978-3-319-75319-5},
  isbn      = {978-3-319-75319-5}
}

@article{reducedorder2004,
  title = {Reduced-Order Modeling: New Approaches for Computational Physics},
  shorttitle = {Reduced-Order Modeling},
  author = {Lucia, David J. and Beran, Philip S. and Silva, Walter A.},
  year = 2004,
  month = {feb},
  journal = {Progress in Aerospace Sciences},
  volume = {40},
  number = {1},
  pages = {51--117},
  issn = {0376-0421},
  doi = {10.1016/j.paerosci.2003.12.001},
}

@article{mccullochDensitymatrixRenormalizationGroup2007,
  title = {From Density-Matrix Renormalization Group to Matrix Product States},
  author = {McCulloch, Ian P.},
  year = 2007,
  month = {oct},
  journal = {Journal of Statistical Mechanics: Theory and Experiment},
  volume = {2007},
  number = {10},
  pages = {P10014--P10014},
  issn = {1742-5468},
  doi = {10.1088/1742-5468/2007/10/P10014},
}

@article{orusTensorNetworksComplex2019a,
  title = {Tensor Networks for Complex Quantum Systems},
  author = {Or{\'u}s, Rom{\'a}n},
  year = 2019,
  month = {sep},
  journal = {Nature Reviews Physics},
  volume = {1},
  number = {9},
  pages = {538--550},
  publisher = {Nature Publishing Group},
  issn = {2522-5820},
  doi = {10.1038/s42254-019-0086-7},
  copyright = {2019 Springer Nature Limited},
  langid = {english}
}

@article{durbinRecentDevelopmentsTurbulence2018,
  title = {Some {{Recent Developments}} in {{Turbulence Closure Modeling}}},
  author = {Durbin, Paul A.},
  year = 2018,
  month = {jan},
  journal = {Annual Review of Fluid Mechanics},
  volume = {50},
  number = {Volume 50, 2018},
  pages = {77--103},
  publisher = {Annual Reviews},
  issn = {0066-4189, 1545-4479},
  doi = {10.1146/annurev-fluid-122316-045020},
  langid = {english}
}

@incollection{LESWBForlandi2000,
  title = {Large Eddy Simulations of Wall-Bounded Flows},
  booktitle = {Fluid Flow Phenomena: A Numerical Toolkit},
  author = {Orlandi, Paolo},
  editor = {Orlandi, Paolo},
  year = 2000,
  pages = {312--325},
  publisher = {Springer Netherlands},
  address = {Dordrecht},
  doi = {10.1007/978-94-011-4281-6_14},
  isbn = {978-0-7923-6095-7}
}

@misc{Cain2026shorsalgo,
      title={Shor's algorithm is possible with as few as 10,000 reconfigurable atomic qubits}, 
      author={Madelyn Cain and Qian Xu and Robbie King and Lewis R. B. Picard and Harry Levine and Manuel Endres and John Preskill and Hsin-Yuan Huang and Dolev Bluvstein},
      year={2026},
      eprint={2603.28627},
      archivePrefix={arXiv},
      primaryClass={quant-ph},
      url={https://arxiv.org/abs/2603.28627}, 
}

@article{Bruzewicz2019,
    author = {Bruzewicz, Colin D. and Chiaverini, John and McConnell, Robert and Sage, Jeremy M.},
    title = {Trapped-ion quantum computing: Progress and challenges},
    journal = {Applied Physics Reviews},
    volume = {6},
    number = {2},
    pages = {021314},
    year = {2019},
    month = {05},
    issn = {1931-9401},
    doi = {10.1063/1.5088164},
    url = {https://doi.org/10.1063/1.5088164}
}

@article{Bluvstein2024,
  title = {Logical quantum processor based on reconfigurable atom arrays},
  author = {Bluvstein, Dolev and Evered, Simon J. and Geim, Alexandra A. and Li, Sophie H. and Zhou, Hengyun and Manovitz, Tom and Ebadi, Sepehr and Cain, Madelyn and Kalinowski, Marcin and Hangleiter, Dominik and Bonilla Ataides, J. Pablo and Maskara, Nishad and Cong, Iris and Gao, Xun and Sales Rodriguez, Pedro and Karolyshyn, Thomas and Semeghini, Giulia and Gullans, Michael J. and Greiner, Markus and Vuleti{\'c}, Vladan and Lukin, Mikhail D.},
  journal = {Nature},
  volume = {626},
  number = {7997},
  pages = {58--65},
  year = {2024},
  doi = {10.1038/s41586-023-06927-3},
  url = {https://doi.org/10.1038/s41586-023-06927-3}
}

@article{Bluvstein2026,
  title = {A fault-tolerant neutral-atom architecture for universal quantum computation},
  author = {Bluvstein, Dolev and Geim, Alexandra A. and Li, Sophie H. and Evered, Simon J. and Bonilla Ataides, J. Pablo and Baranes, Gefen and Gu, Andi and Manovitz, Tom and Xu, Muqing and Kalinowski, Marcin and Majidy, Shayan and Kokail, Christian and Maskara, Nishad and Trapp, Elias C. and Stewart, Luke M. and Hollerith, Simon and Zhou, Hengyun and Gullans, Michael J. and Yelin, Susanne F. and Greiner, Markus and Vuleti{\'c}, Vladan and Cain, Madelyn and Lukin, Mikhail D.},
  journal = {Nature},
  volume = {649},
  number = {8095},
  pages = {39--46},
  year = {2026},
  doi = {10.1038/s41586-025-09848-5},
  url = {https://doi.org/10.1038/s41586-025-09848-5}
}

@Article{Bravyi2024,
author={Bravyi, Sergey
and Cross, Andrew W.
and Gambetta, Jay M.
and Maslov, Dmitri
and Rall, Patrick
and Yoder, Theodore J.},
title={High-threshold and low-overhead fault-tolerant quantum memory},
journal={Nature},
year={2024},
month={Mar},
day={01},
volume={627},
number={8005},
pages={778-782},
issn={1476-4687},
doi={10.1038/s41586-024-07107-7},
url={https://doi.org/10.1038/s41586-024-07107-7}
}

@article{Gross_2026,
   title={Tensor network lattice Boltzmann method for data-compressed fluid simulations},
   volume={460},
   ISSN={0045-7825},
   url={http://dx.doi.org/10.1016/j.cma.2026.119088},
   DOI={10.1016/j.cma.2026.119088},
   journal={Computer Methods in Applied Mechanics and Engineering},
   publisher={Elsevier BV},
   author={Gross, Lukas and Mounzer, Elie and Wawrzyniak, David M. and Winter, Josef M. and Adams, Nikolaus A.},
   year={2026},
   month={oct}, 
   pages={119088} 
}

@article{Tran2026,
title = {A tensor train-based isogeometric solver for large-scale 3D poisson problems},
journal = {Computer Methods in Applied Mechanics and Engineering},
volume = {453},
pages = {118802},
year = {2026},
issn = {0045-7825},
doi = {10.1016/j.cma.2026.118802},
url = {https://www.sciencedirect.com/science/article/pii/S0045782526000769},
author = {Quoc {Thai Tran} and Duc P. Truong and  {Kim Ø Rasmussen} and Boian Alexandrov}
}

@article{Fang_2023,
   title={Time-marching based quantum solvers for time-dependent linear differential equations},
   volume={7},
   ISSN={2521-327X},
   url={http://dx.doi.org/10.22331/q-2023-03-20-955},
   DOI={10.22331/q-2023-03-20-955},
   journal={Quantum},
   publisher={Verein zur Forderung des Open Access Publizierens in den Quantenwissenschaften},
   author={Fang, Di and Lin, Lin and Tong, Yu},
   year={2023},
   month={Mar}, 
   pages={955} 
}

@misc{Danis2025b,
      title={Tensor-Train Operator Inference}, 
      author={Engin Danis and Duc Truong and Kim {\O}. Rasmussen§ and Boian S. Alexandrov},
      year={2025},
      eprint={2509.08071},
      archivePrefix={arXiv},
      primaryClass={math.NA},
      url={https://arxiv.org/abs/2509.08071}, 
}

@article{Ion2022,
title = {Tensor train based isogeometric analysis for PDE approximation on parameter dependent geometries},
journal = {Computer Methods in Applied Mechanics and Engineering},
volume = {401},
pages = {115593},
year = {2022},
issn = {0045-7825},
doi = {10.1016/j.cma.2022.115593},
url = {https://www.sciencedirect.com/science/article/pii/S0045782522005618},
author = {Ion Gabriel Ion and Dimitrios Loukrezis and Herbert {De Gersem}}
}

@article{Preskill2018,
  author  = {Preskill, John},
  title   = {Quantum Computing in the NISQ era and beyond},
  journal = {Quantum},
  volume  = {2},
  pages   = {79},
  year    = {2018},
  doi     = {10.22331/q-2018-08-06-79},
  url     = {https://doi.org/10.22331/q-2018-08-06-79}
}

@article{alfonsi2011,
  title = {On {{Direct Numerical Simulation}} of {{Turbulent Flows}}},
  author = {Alfonsi, Giancarlo},
  year = 2011,
  month = {dec},
  journal = {Applied Mechanics Reviews},
  volume = {64},
  number = {020802},
  issn = {0003-6900},
  doi = {10.1115/1.4005282},
}

@article{H_lscher_2025,
   title={Quantum-inspired fluid simulation of two-dimensional turbulence with GPU acceleration},
   volume={7},
   ISSN={2643-1564},
   url={http://dx.doi.org/10.1103/PhysRevResearch.7.013112},
   DOI={10.1103/physrevresearch.7.013112},
   number={1},
   journal={Physical Review Research},
   publisher={American Physical Society (APS)},
   author={Hölscher, Leonhard and Rao, Pooja and Müller, Lukas and Klepsch, Johannes and Luckow, Andre and Stollenwerk, Tobias and Wilhelm, Frank K.},
   year={2025},
   month={jan} 
}

@misc{Over2026,
      title={Operator Learning for efficient Quantum Computation}, 
      author={Paul Over and Sergio Bengoechea and Leonardo Borello Busilacchi and Martin Kiffner and Thomas Rung and Alexios A. Michailidis},
      year={2026},
      eprint={2606.20184},
      archivePrefix={arXiv},
      primaryClass={quant-ph},
      url={https://arxiv.org/abs/2606.20184}, 
}

@article{van_H_lst_2026,
   title={Quantum-inspired tensor-network fractional-step method for incompressible flow in curvilinear coordinates},
   volume={325},
   ISSN={0010-4655},
   url={http://dx.doi.org/10.1016/j.cpc.2026.110169},
   DOI={10.1016/j.cpc.2026.110169},
   journal={Computer Physics Communications},
   publisher={Elsevier BV},
   author={van H{\"u}lst, Nis-Luca and Siegl, Pia and Over, Paul and Bengoechea, Sergio and Hashizume, Tomohiro and Cecile, Mario Guillaume and Rung, Thomas and Jaksch, Dieter},
   year={2026},
   month={aug}, pages={110169} }

@book{Pope2000, 
title={Turbulent Flows}, 
publisher={Cambridge University Press}, 
author={Pope, Stephen B.}, 
year={2000},
doi={ 10.1017/CBO9780511840531 }
}

@book{Anderson1995,
  title={Computational Fluid Dynamics: The Basics with Applications},
  author={Anderson, J.D.},
  isbn={9780070016859},
  lccn={94021237},
  year={1995},
  publisher={McGraw-Hill Education}
}

@article{Over2024b,
  title = {Quantum algorithm for the advection-diffusion equation by direct block encoding of the time-marching operator},
  author = {Over, Paul and Bengoechea, Sergio and Brearley, Peter and Laizet, Sylvain and Rung, Thomas},
  journal = {Phys. Rev. A},
  volume = {112},
  issue = {1},
  pages = {L010401},
  numpages = {6},
  year = {2025},
  month = {Jul},
  publisher = {American Physical Society},
  doi = {10.1103/d8hb-fv93},
  url = {https://link.aps.org/doi/10.1103/d8hb-fv93}
}

@article{Brearley2024,
  title = {Quantum algorithm for solving the advection equation using Hamiltonian simulation},
  author = {Brearley, Peter and Laizet, Sylvain},
  journal = {Phys. Rev. A},
  volume = {110},
  issue = {1},
  pages = {012430},
  numpages = {12},
  year = {2024},
  month = {Jul},
  publisher = {American Physical Society},
  doi = {10.1103/PhysRevA.110.012430},
  url = {https://link.aps.org/doi/10.1103/PhysRevA.110.012430}
}

@article{Over2024a,
    title = {Boundary treatment for variational quantum simulations of partial differential equations on quantum computers},
    journal = {Computers \& Fluids},
    volume = {288},
    pages = {106508},
    year = {2025},
    issn = {0045-7930},
    doi = {10.1016/j.compfluid.2024.106508},
    url = {https://www.sciencedirect.com/science/article/pii/S0045793024003396},
    author = {Paul Over and Sergio Bengoechea and Thomas Rung and Francesco Clerici and Leonardo Scandurra and Eugene {de Villiers} and Dieter Jaksch},
}

@article{Bengoechea2026,
   title={Quantum Time‐Marching Algorithms for Solving Linear Transport Problems Including Boundary Conditions},
   volume={127},
   ISSN={1097-0207},
   url={http://dx.doi.org/10.1002/nme.70326},
   DOI={10.1002/nme.70326},
   number={8},
   journal={International Journal for Numerical Methods in Engineering},
   publisher={Wiley},
   author={Bengoechea, Sergio and Over, Paul and Rung, Thomas},
   year={2026},
   month={apr} 
}

@article{Harrow2009,
  title = {Quantum Algorithm for Linear Systems of Equations},
  author = {Harrow, Aram W. and Hassidim, Avinatan and Lloyd, Seth},
  journal = {Phys. Rev. Lett.},
  volume = {103},
  issue = {15},
  pages = {150502},
  numpages = {4},
  year = {2009},
  month = {Oct},
  publisher = {American Physical Society},
  doi = {10.1103/PhysRevLett.103.150502},
}

@article{Giovannetti2008,
  title = {Quantum Random Access Memory},
  author = {Giovannetti, Vittorio and Lloyd, Seth and Maccone, Lorenzo},
  journal = {Phys. Rev. Lett.},
  volume = {100},
  issue = {16},
  pages = {160501},
  numpages = {4},
  year = {2008},
  month = {Apr},
  publisher = {American Physical Society},
  doi = {10.1103/PhysRevLett.100.160501},
  url = {https://link.aps.org/doi/10.1103/PhysRevLett.100.160501}
}

@article{Pan2014,
  title = {Experimental realization of quantum algorithm for solving linear systems of equations},
  author = {Pan, Jian and Cao, Yudong and Yao, Xiwei and Li, Zhaokai and Ju, Chenyong and Chen, Hongwei and Peng, Xinhua and Kais, Sabre and Du, Jiangfeng},
  journal = {Phys. Rev. A},
  volume = {89},
  issue = {2},
  pages = {022313},
  numpages = {5},
  year = {2014},
  month = {Feb},
  publisher = {American Physical Society},
  doi = {10.1103/PhysRevA.89.022313},
  url = {https://link.aps.org/doi/10.1103/PhysRevA.89.022313}
}

@article{Setty2025,
doi = {10.1088/2632-2153/ada3ab},
url = {https://doi.org/10.1088/2632-2153/ada3ab},
year = {2025},
month = {jan},
publisher = {IOP Publishing},
volume = {6},
number = {1},
pages = {015002},
author = {Setty, Abhishek and Abdusalamov, Rasul and Motzoi, Felix},
title = {Self-adaptive physics-informed quantum machine learning for solving differential equations},
journal = {Machine Learning: Science and Technology}
}

@article{1nts-v6y9,
  title = {Block encoding of sparse matrices via coherent permutation},
  author = {Setty, Abhishek},
  journal = {Phys. Rev. Res.},
  volume = {8},
  issue = {3},
  pages = {033282},
  numpages = {16},
  year = {2026},
  month = {Sep},
  publisher = {American Physical Society},
  doi = {10.1103/1nts-v6y9},
  url = {https://link.aps.org/doi/10.1103/1nts-v6y9}
}

@article{Kyriienko2021,
  title = {Solving nonlinear differential equations with differentiable quantum circuits},
  author = {Kyriienko, Oleksandr and Paine, Annie E. and Elfving, Vincent E.},
  journal = {Phys. Rev. A},
  volume = {103},
  issue = {5},
  pages = {052416},
  numpages = {22},
  year = {2021},
  month = {May},
  publisher = {American Physical Society},
  doi = {10.1103/PhysRevA.103.052416},
  url = {https://link.aps.org/doi/10.1103/PhysRevA.103.052416}
}

@misc{Umer2026Adaptive,
      title={Circuit Design Informed Adaptive Variational Quantum Algorithms}, 
      author={Muhammad Umer and Dimitris G. Angelakis},
      year={2026},
      eprint={2607.04110},
      archivePrefix={arXiv},
      primaryClass={quant-ph},
      url={https://arxiv.org/abs/2607.04110}, 
}

@article{Cai2013,
  title = {Experimental Quantum Computing to Solve Systems of Linear Equations},
  author = {Cai, X.-D. and Weedbrook, C. and Su, Z.-E. and Chen, M.-C. and Gu, Mile and Zhu, M.-J. and Li, Li and Liu, Nai-Le and Lu, Chao-Yang and Pan, Jian-Wei},
  journal = {Phys. Rev. Lett.},
  volume = {110},
  issue = {23},
  pages = {230501},
  numpages = {5},
  year = {2013},
  month = {Jun},
  publisher = {American Physical Society},
  doi = {10.1103/PhysRevLett.110.230501},
  url = {https://link.aps.org/doi/10.1103/PhysRevLett.110.230501}
}

@article{Childs2017,
  author   = {Childs, Andrew M. and Kothari, Robin and Somma, Rolando D.},
  journal  = {SIAM J. Comput.},
  title    = {Quantum Algorithm for Systems of Linear Equations with Exponentially Improved Dependence on Precision},
  year     = {2017},
  number   = {6},
  pages    = {1920-1950},
  volume   = {46},
  url      = {https://doi.org/10.1137/16M1087072},
}

@article{Childs_2020,
   title={Quantum Spectral Methods for Differential Equations},
   volume={375},
   ISSN={1432-0916},
   url={http://dx.doi.org/10.1007/s00220-020-03699-z},
   DOI={10.1007/s00220-020-03699-z},
   number={2},
   journal={Communications in Mathematical Physics},
   publisher={Springer Science and Business Media LLC},
   author={Childs, Andrew M. and Liu, Jin-Peng},
   year={2020},
   month={feb}, pages={1427–1457} 
}

@article{Umer_2025a,
   title={Probing the limits of variational quantum algorithms for nonlinear ground states on real quantum hardware: The effects of noise},
   volume={111},
   ISSN={2469-9934},
   url={http://dx.doi.org/10.1103/PhysRevA.111.012626},
   DOI={10.1103/physreva.111.012626},
   number={1},
   journal={Physical Review A},
   publisher={American Physical Society (APS)},
   author={Umer, Muhammad and Mastorakis, Eleftherios and Evangelou, Sofia and Angelakis, Dimitris G.},
   year={2025},
   month={jan} }

@article{Umer_2025b,
   title={Efficient estimation and sequential optimization of cost functions in variational quantum algorithms},
   volume={10},
   ISSN={2058-9565},
   url={http://dx.doi.org/10.1088/2058-9565/add55e},
   DOI={10.1088/2058-9565/add55e},
   number={3},
   journal={Quantum Science and Technology},
   publisher={IOP Publishing},
   author={Umer, Muhammad and Mastorakis, Eleftherios and Angelakis, Dimitris G},
   year={2025},
   month={may}, pages={035022} }

@article{Itani2024,
    author = {Itani, Wael and Sreenivasan, Katepalli R. and Succi, Sauro},
    title = {Quantum algorithm for lattice Boltzmann (QALB) simulation of incompressible fluids with a nonlinear collision term},
    journal = {Physics of Fluids},
    volume = {36},
    number = {1},
    pages = {017112},
    year = {2024},
    month = {01},
    issn = {1070-6631},
    doi = {10.1063/5.0176569},
    url = {https://doi.org/10.1063/5.0176569},
}

@inproceedings{Gily_n_2019,
   title={Quantum singular value transformation and beyond: exponential improvements for quantum matrix arithmetics},
   url={http://dx.doi.org/10.1145/3313276.3316366},
   DOI={10.1145/3313276.3316366},
   booktitle={Proceedings of the 51st Annual ACM SIGACT Symposium on Theory of Computing},
   publisher={ACM},
   author={Gilyén, András and Su, Yuan and Low, Guang Hao and Wiebe, Nathan},
   year={2019},
   month={jun}, pages={193–204},
   collection={STOC ’19} 
}

@article{Montanaro_2016,
   title={Quantum algorithms: an overview},
   volume={2},
   ISSN={2056-6387},
   url={http://dx.doi.org/10.1038/npjqi.2015.23},
   DOI={10.1038/npjqi.2015.23},
   number={1},
   journal={npj Quantum Information},
   publisher={Springer Science and Business Media LLC},
   author={Montanaro, Ashley},
   year={2016},
   month={jan} }

@article{Low2017,
  title = {Optimal Hamiltonian Simulation by Quantum Signal Processing},
  author = {Low, Guang Hao and Chuang, Isaac L.},
  journal = {Phys. Rev. Lett.},
  volume = {118},
  issue = {1},
  pages = {010501},
  numpages = {5},
  year = {2017},
  month = {Jan},
  publisher = {American Physical Society},
  doi = {10.1103/PhysRevLett.118.010501},
  url = {https://link.aps.org/doi/10.1103/PhysRevLett.118.010501}
}

@misc{Brassard_2002,
   title={Quantum amplitude amplification and estimation},
   ISBN={9780821878958},
   ISSN={0271-4132},
   url={http://dx.doi.org/10.1090/conm/305/05215},
   DOI={10.1090/conm/305/05215},
   journal={Quantum Computation and Information},
   publisher={American Mathematical Society},
   author={Brassard, Gilles and Høyer, Peter and Mosca, Michele and Tapp, Alain},
   year={2002},
   pages={53–74} 
}

@article{Lapworth2025,
   title={Preconditioned block encodings for quantum linear systems},
   volume={10},
   ISSN={2058-9565},
   url={http://dx.doi.org/10.1088/2058-9565/ae0f4b},
   DOI={10.1088/2058-9565/ae0f4b},
   number={4},
   journal={Quantum Science and Technology},
   publisher={IOP Publishing},
   author={Lapworth, Leigh and Sünderhauf, Christoph},
   year={2025},
   month={oct}, pages={045064} 
}

@article{Shao2018,
  title = {Quantum circulant preconditioner for a linear system of equations},
  author = {Shao, Changpeng and Xiang, Hua},
  journal = {Phys. Rev. A},
  volume = {98},
  issue = {6},
  pages = {062321},
  numpages = {9},
  year = {2018},
  month = {Dec},
  publisher = {American Physical Society},
  doi = {10.1103/PhysRevA.98.062321},
  url = {https://link.aps.org/doi/10.1103/PhysRevA.98.062321}
}

@InProceedings{Ambainis2012,
  author =	{Ambainis, Andris},
  title =	{{Variable time amplitude amplification and quantum algorithms for linear algebra problems}},
  booktitle =	{29th International Symposium on Theoretical Aspects of Computer Science (STACS 2012)},
  pages =	{636--647},
  series =	{Leibniz International Proceedings in Informatics (LIPIcs)},
  ISBN =	{978-3-939897-35-4},
  ISSN =	{1868-8969},
  year =	{2012},
  volume =	{14},
  editor =	{D\"{u}rr, Christoph and Wilke, Thomas},
  publisher =	{Schloss Dagstuhl -- Leibniz-Zentrum f{\"u}r Informatik},
  address =	{Dagstuhl, Germany},
  URL =		{https://drops.dagstuhl.de/entities/document/10.4230/LIPIcs.STACS.2012.636},
  URN =		{urn:nbn:de:0030-drops-34261},
  doi =		{10.4230/LIPIcs.STACS.2012.636}
}

@misc{meng2026geometricencodingturbulenceendtoend,
      title={Geometric encoding of turbulence for end-to-end quantum simulation}, 
      author={Zhaoyuan Meng and Xiao-Ming Zhang and Xiao Yuan and Yue Yang},
      year={2026},
      eprint={2508.05346},
      archivePrefix={arXiv},
      primaryClass={quant-ph},
      url={https://arxiv.org/abs/2508.05346}, 
}

@misc{zhuang2025pathwaypracticalquantumadvantage,
      title={A Pathway to Practical Quantum Advantage in Solving Navier-Stokes Equations}, 
      author={Xi-Ning Zhuang and Zhao-Yun Chen and Ming-Yang Tan and Jiaxuan Zhang and Chuang-Chao Ye and Tian-Hao Wei and Teng-Yang Ma and Cheng Xue and Huan-Yu Liu and Qing-Song Li and Tai-Ping Sun and Xiao-Fan Xu and Yun-Jie Wang and Yu-Chun Wu and Guo-Ping Guo},
      year={2025},
      eprint={2509.08807},
      archivePrefix={arXiv},
      primaryClass={quant-ph},
      url={https://arxiv.org/abs/2509.08807}, 
}

@book{Elman2014,
    author = {Elman, Howard and Silvester, David and Wathen, Andy},
    title = {Finite Elements and Fast Iterative Solvers: with Applications in Incompressible Fluid Dynamics},
    publisher = {Oxford University Press},
    year = {2014},
    month = {06},
    isbn = {9780199678792},
    doi = {10.1093/acprof:oso/9780199678792.001.0001},
    url = {https://doi.org/10.1093/acprof:oso/9780199678792.001.0001},
}

@article{Budinski2021,
  author   = {Budinski, Ljubomir},
  journal  = {Quantum Information Processing},
  title    = {Quantum algorithm for the advection-diffusion equation simulated with the lattice Boltzmann method},
  year     = {2021},
  issn     = {1573-1332},
  number   = {2},
  pages    = {57},
  volume   = {20},
  doi      = {10.1007/s11128-021-02996-3},
  refid    = {Budinski2021},
  url      = {https://doi.org/10.1007/s11128-021-02996-3},
}

@article{Budinski2022,
author = {Ljubomir, Budinski},
title = {Quantum algorithm for the Navier–Stokes equations by using the streamfunction-vorticity formulation and the lattice Boltzmann method},
journal = {International Journal of Quantum Information},
volume = {20},
number = {02},
pages = {2150039},
year = {2022},
doi = {10.1142/S0219749921500398},
URL = {https://doi.org/10.1142/S0219749921500398},
}

@article{zamanStepbyStepHHLAlgorithm2023,
  title = {A {{Step-by-Step HHL Algorithm Walkthrough}} to {{Enhance Understanding}} of {{Critical Quantum Computing Concepts}}},
  author = {Zaman, Anika and Morrell, Hector Jose and Wong, Hiu Yung},
  year = 2023,
  journal = {IEEE Access},
  volume = {11},
  pages = {77117--77131},
  issn = {2169-3536},
  doi = {10.1109/ACCESS.2023.3297658},
}

@article{White1992,
  title = {Density matrix formulation for quantum renormalization groups},
  author = {White, Steven R.},
  journal = {Phys. Rev. Lett.},
  volume = {69},
  issue = {19},
  pages = {2863--2866},
  numpages = {0},
  year = {1992},
  month = {Nov},
  publisher = {American Physical Society},
  doi = {10.1103/PhysRevLett.69.2863},
  url = {https://link.aps.org/doi/10.1103/PhysRevLett.69.2863}
}

@article{Peddinti2024,
  title={Quantum-inspired framework for computational fluid dynamics},
  author={Peddinti, Raghavendra Dheeraj and Pisoni, Stefano and Marini, Alessandro and Lott, Philippe and Argentieri, Henrique and Tiunov, Egor and Aolita, Leandro},
  journal={Communications Physics},
  volume={7},
  number={1},
  pages={135},
  year={2024},
  publisher={Nature Publishing Group UK London},
  doi       = {10.1038/s42005-024-01623-8},
  url = {https://doi.org/10.1038/s42005-024-01623-8},
}

@Article{Gourianov2022,
  author    = {Nikita Gourianov and Michael Lubasch and Sergey Dolgov and Quincy Y. van den Berg and Hessam Babaee and Peyman Givi and Martin Kiffner and Dieter Jaksch},
  journal   = {Nature Computational Science},
  title     = {A quantum-inspired approach to exploit turbulence structures},
  year      = {2022},
  month     = {jan},
  number    = {1},
  pages     = {30--37},
  volume    = {2},
  doi       = {10.1038/s43588-021-00181-1},
  publisher = {Springer Science and Business Media {LLC}},
}

@article{liuSurveyImprovementApplication2022,
  title = {Survey on the {{Improvement}} and {{Application}} of {{HHL Algorithm}}},
  author = {Liu, Xiaonan and Xie, Haoshan and Liu, Zhengyu and Zhao, Chenyan},
  year = 2022,
  month = {aug},
  journal = {Journal of Physics: Conference Series},
  volume = {2333},
  number = {1},
  pages = {012023},
  publisher = {IOP Publishing},
  issn = {1742-6596},
  doi = {10.1088/1742-6596/2333/1/012023},
  langid = {english},
}

@article{Orus2013,
author = {Orus, Roman},
year = {2013},
month = {06},
pages = {},
title = {A Practical Introduction to Tensor Networks: Matrix Product States and Projected Entangled Pair States},
volume = {349},
journal = {Annals of Physics},
doi = {10.1016/j.aop.2014.06.013}
}

@article{Oseledets2011,
author = {Oseledets, I. V.},
title = {{Tensor-Train} Decomposition},
journal = {SIAM Journal on Scientific Computing},
volume = {33},
number = {5},
pages = {2295--2317},
year = {2011},
doi = {10.1137/090752286},
URL = {https://doi.org/10.1137/090752286},
}

@article{Olsacher2025,
doi = {10.1088/2058-9565/ad9ed5},
url = {https://doi.org/10.1088/2058-9565/ad9ed5},
year = {2025},
month = {jan},
publisher = {IOP Publishing},
volume = {10},
number = {1},
pages = {015065},
author = {Olsacher, Tobias and Kraft, Tristan and Kokail, Christian and Kraus, Barbara and Zoller, Peter},
title = {Hamiltonian and Liouvillian learning in weakly-dissipative quantum many-body systems},
journal = {Quantum Science and Technology}
}

@article{Pastori2022,
  title = {Characterization and Verification of Trotterized Digital Quantum Simulation Via Hamiltonian and Liouvillian Learning},
  author = {Pastori, Lorenzo and Olsacher, Tobias and Kokail, Christian and Zoller, Peter},
  journal = {PRX Quantum},
  volume = {3},
  issue = {3},
  pages = {030324},
  numpages = {22},
  year = {2022},
  month = {Aug},
  publisher = {American Physical Society},
  doi = {10.1103/PRXQuantum.3.030324},
  url = {https://link.aps.org/doi/10.1103/PRXQuantum.3.030324}
}

@article{Kraft2026,
   title={Bounded-Error Quantum Simulation via Hamiltonian and Lindbladian Learning},
   volume={16},
   ISSN={2160-3308},
   url={http://dx.doi.org/10.1103/s96t-n8tx},
   DOI={10.1103/s96t-n8tx},
   number={3},
   journal={Physical Review X},
   publisher={American Physical Society (APS)},
   author={Kraft, Tristan and Joshi, Manoj K. and Lam, William T. and Olsacher, Tobias and Kranzl, Florian and Franke, Johannes and Joshi, Lata Kh and Blatt, Rainer and Smerzi, Augusto and França, Daniel Stilck and Vermersch, Benoît and Kraus, Barbara and Roos, Christian F. and Zoller, Peter},
   year={2026},
   month={aug} 
}

@article{Gunn2025,
  title = {Phases of matrix product states with symmetric quantum circuits and symmetric measurements with feedforward},
  author = {Gunn, David and Styliaris, Georgios and Kraft, Tristan and Kraus, Barbara},
  journal = {Phys. Rev. B},
  volume = {111},
  issue = {11},
  pages = {115110},
  numpages = {26},
  year = {2025},
  month = {Mar},
  publisher = {American Physical Society},
  doi = {10.1103/PhysRevB.111.115110},
  url = {https://link.aps.org/doi/10.1103/PhysRevB.111.115110}
}

@misc{connor2025tensornetworkmethodsgrosspitaevskii,
      title={Tensor network methods for the Gross-Pitaevskii equation on fine grids}, 
      author={Ryan J. J. Connor and Callum W. Duncan and Andrew J. Daley},
      year={2025},
      eprint={2507.01149},
      archivePrefix={arXiv},
      primaryClass={cond-mat.quant-gas},
      url={https://arxiv.org/abs/2507.01149}, 
}

@article{Ye2024, 
    title={Quantized tensor networks for solving the Vlasov–Maxwell equations}, 
    volume={90}, 
    DOI={10.1017/S0022377824000503}, 
    number={3}, 
    journal={Journal of Plasma Physics}, 
    author={Ye, Erika and Loureiro, Nuno F.}, year={2024}, 
    pages={805900301}
}

@article{Clader2013,
  title = {Preconditioned Quantum Linear System Algorithm},
  author = {Clader, B. D. and Jacobs, B. C. and Sprouse, C. R.},
  journal = {Phys. Rev. Lett.},
  volume = {110},
  issue = {25},
  pages = {250504},
  numpages = {5},
  year = {2013},
  month = {Jun},
  publisher = {American Physical Society},
  doi = {10.1103/PhysRevLett.110.250504},
  url = {https://link.aps.org/doi/10.1103/PhysRevLett.110.250504}
}

@incollection{Sanavio_2024,
    author = {Claudio Sanavio and Sauro Succi},
    title = {Quantum Computing for Simulation of Fluid Dynamics},
    booktitle = {Quantum Information Science - Recent Advances and Computational Science Applications},
    publisher = {IntechOpen},
    address = {London},
    year = {2024},
    editor = {René Steijl},
    chapter = {1},
    doi = {10.5772/intechopen.1005242},
    url = {https://doi.org/10.5772/intechopen.1005242}
}

@article{Burkard2023,
  title = {Semiconductor spin qubits},
  author = {Burkard, Guido and Ladd, Thaddeus D. and Pan, Andrew and Nichol, John M. and Petta, Jason R.},
  journal = {Rev. Mod. Phys.},
  volume = {95},
  issue = {2},
  pages = {025003},
  numpages = {58},
  year = {2023},
  month = {Jun},
  publisher = {American Physical Society},
  doi = {10.1103/RevModPhys.95.025003},
  url = {https://link.aps.org/doi/10.1103/RevModPhys.95.025003}
}

@article{GonzalezZalba2021,
author={Gonzalez-Zalba, M. F.
and de Franceschi, S.
and Charbon, E.
and Meunier, T.
and Vinet, M.
and Dzurak, A. S.},
title={Scaling silicon-based quantum computing using CMOS technology},
journal={Nature Electronics},
year={2021},
month={Dec},
day={01},
volume={4},
number={12},
pages={872-884},
issn={2520-1131},
doi={10.1038/s41928-021-00681-y},
url={https://doi.org/10.1038/s41928-021-00681-y}
}

@article{Pankovich2024,
  title = {High-Photon-Loss Threshold Quantum Computing Using GHZ-State Measurements},
  author = {Pankovich, Brendan and Kan, Angus and Wan, Kwok Ho and Ostmann, Maike and Neville, Alex and Omkar, Srikrishna and Sohbi, Adel and Br\'adler, Kamil},
  journal = {Phys. Rev. Lett.},
  volume = {133},
  issue = {5},
  pages = {050604},
  numpages = {6},
  year = {2024},
  month = {Aug},
  publisher = {American Physical Society},
  doi = {10.1103/PhysRevLett.133.050604},
  url = {https://link.aps.org/doi/10.1103/PhysRevLett.133.050604}
}

@article{AghaeeRad2025,
author={Aghaee Rad, H.
and Ainsworth, T.
and Alexander, R. N.
and Altieri, B.
and Askarani, M. F.
and Baby, R.
and Banchi, L.
and Baragiola, B. Q.
and Bourassa, J. E.
and Chadwick, R. S.
and Charania, I.
and Chen, H.
and Collins, M. J.
and Contu, P.
and D'Arcy, N.
and Dauphinais, G.
and De Prins, R.
and Deschenes, D.
and Di Luch, I.
and Duque, S.
and Edke, P.
and Fayer, S. E.
and Ferracin, S.
and Ferretti, H.
and Gefaell, J.
and Glancy, S.
and Gonz{\'a}lez-Arciniegas, C.
and Grainge, T.
and Han, Z.
and Hastrup, J.
and Helt, L. G.
and Hillmann, T.
and Hundal, J.
and Izumi, S.
and Jaeken, T.
and Jonas, M.
and Kocsis, S.
and Krasnokutska, I.
and Larsen, M. V.
and Laskowski, P.
and Laudenbach, F.
and Lavoie, J.
and Li, M.
and Lomonte, E.
and Lopetegui, C. E.
and Luey, B.
and Lund, A. P.
and Ma, C.
and Madsen, L. S.
and Mahler, D. H.
and Mantilla Calder{\'o}n, L.
and Menotti, M.
and Miatto, F. M.
and Morrison, B.
and Nadkarni, P. J.
and Nakamura, T.
and Neuhaus, L.
and Niu, Z.
and Noro, R.
and Papirov, K.
and Pesah, A.
and Phillips, D. S.
and Plick, W. N.
and Rogalsky, T.
and Rortais, F.
and Sabines-Chesterking, J.
and Safavi-Bayat, S.
and Sazhaev, E.
and Seymour, M.
and Rezaei Shad, K.
and Silverman, M.
and Srinivasan, S. A.
and Stephan, M.
and Tang, Q. Y.
and Tasker, J. F.
and Teo, Y. S.
and Then, R. B.
and Tremblay, J. E.
and Tzitrin, I.
and Vaidya, V. D.
and Vasmer, M.
and Vernon, Z.
and Villalobos, L. F. S. S. M.
and Walshe, B. W.
and Weil, R.
and Xin, X.
and Yan, X.
and Yao, Y.
and Zamani Abnili, M.
and Zhang, Y.},
title={Scaling and networking a modular photonic quantum computer},
journal={Nature},
year={2025},
month={Feb},
day={01},
volume={638},
number={8052},
pages={912-919},
issn={1476-4687},
doi={10.1038/s41586-024-08406-9},
url={https://doi.org/10.1038/s41586-024-08406-9}
}

@article{Slussarenko_2019,
   title={Photonic quantum information processing: A concise review},
   volume={6},
   ISSN={1931-9401},
   url={http://dx.doi.org/10.1063/1.5115814},
   DOI={10.1063/1.5115814},
   number={4},
   journal={Applied Physics Reviews},
   publisher={AIP Publishing},
   author={Slussarenko, Sergei and Pryde, Geoff J.},
   year={2019},
   month={oct} 
}

@article{Akhtar2023,
author={Akhtar, M.
and Bonus, F.
and Lebrun-Gallagher, F. R.
and Johnson, N. I.
and Siegele-Brown, M.
and Hong, S.
and Hile, S. J.
and Kulmiya, S. A.
and Weidt, S.
and Hensinger, W. K.},
title={A high-fidelity quantum matter-link between ion-trap microchip modules},
journal={Nature Communications},
year={2023},
month={Feb},
day={08},
volume={14},
number={1},
pages={531},
issn={2041-1723},
doi={10.1038/s41467-022-35285-3},
url={https://doi.org/10.1038/s41467-022-35285-3}
}

@article{Xu2024,
    author={Xu, Qian
    and Bonilla Ataides, J. Pablo
    and Pattison, Christopher A.
    and Raveendran, Nithin
    and Bluvstein, Dolev
    and Wurtz, Jonathan
    and Vasi{\'{c}}, Bane
    and Lukin, Mikhail D.
    and Jiang, Liang
    and Zhou, Hengyun},
    title={Constant-overhead fault-tolerant quantum computation with reconfigurable atom arrays},
    journal={Nature Physics},
    year={2024},
    month={Jul},
    day={01},
    volume={20},
    number={7},
    pages={1084-1090},
    issn={1745-2481},
    doi={10.1038/s41567-024-02479-z},
    url={https://doi.org/10.1038/s41567-024-02479-z}
}

@misc{aasen2025roadmapfaulttolerantquantum,
      title={Roadmap to fault tolerant quantum computation using topological qubit arrays}, 
      author={David Aasen and Morteza Aghaee and Zulfi Alam and Mariusz Andrzejczuk and Andrey Antipov and Mikhail Astafev and Lukas Avilovas and Amin Barzegar and Bela Bauer and Jonathan Becker and Juan M. Bello-Rivas and Umesh Bhaskar and Alex Bocharov and Srini Boddapati and David Bohn and Jouri Bommer and Parsa Bonderson and Jan Borovsky and Leo Bourdet and Samuel Boutin and Tom Brown and Gary Campbell and Lucas Casparis and Srivatsa Chakravarthi and Rui Chao and Benjamin J. Chapman and Sohail Chatoor and Anna Wulff Christensen and Patrick Codd and William Cole and Paul Cooper and Fabiano Corsetti and Ajuan Cui and Wim van Dam and Tareq El Dandachi and Sahar Daraeizadeh and Adrian Dumitrascu and Andreas Ekefjärd and Saeed Fallahi and Luca Galletti and Geoff Gardner and Raghu Gatta and Haris Gavranovic and Michael Goulding and Deshan Govender and Flavio Griggio and Ruben Grigoryan and Sebastian Grijalva and Sergei Gronin and Jan Gukelberger and Jeongwan Haah and Marzie Hamdast and Esben Bork Hansen and Matthew Hastings and Sebastian Heedt and Samantha Ho and Justin Hogaboam and Laurens Holgaard and Kevin Van Hoogdalem and Jinnapat Indrapiromkul and Henrik Ingerslev and Lovro Ivancevic and Sarah Jablonski and Thomas Jensen and Jaspreet Jhoja and Jeffrey Jones and Kostya Kalashnikov and Ray Kallaher and Rachpon Kalra and Farhad Karimi and Torsten Karzig and Seth Kimes and Vadym Kliuchnikov and Maren Elisabeth Kloster and Christina Knapp and Derek Knee and Jonne Koski and Pasi Kostamo and Jamie Kuesel and Brad Lackey and Tom Laeven and Jeffrey Lai and Gijs de Lange and Thorvald Larsen and Jason Lee and Kyunghoon Lee and Grant Leum and Kongyi Li and Tyler Lindemann and Marijn Lucas and Roman Lutchyn and Morten Hannibal Madsen and Nash Madulid and Michael Manfra and Signe Brynold Markussen and Esteban Martinez and Marco Mattila and Jake Mattinson and Robert McNeil and Antonio Rodolph Mei and Ryan V. Mishmash and Gopakumar Mohandas and Christian Mollgaard and Michiel de Moor and Trevor Morgan and George Moussa and Anirudh Narla and Chetan Nayak and Jens Hedegaard Nielsen and William Hvidtfelt Padkær Nielsen and Frédéric Nolet and Mike Nystrom and Eoin O'Farrell and Keita Otani and Adam Paetznick and Camille Papon and Andres Paz and Karl Petersson and Luca Petit and Dima Pikulin and Diego Olivier Fernandez Pons and Sam Quinn and Mohana Rajpalke and Alejandro Alcaraz Ramirez and Katrine Rasmussen and David Razmadze and Ben Reichardt and Yuan Ren and Ken Reneris and Roy Riccomini and Ivan Sadovskyy and Lauri Sainiemi and Juan Carlos Estrada Saldaña and Irene Sanlorenzo and Simon Schaal and Emma Schmidgall and Cristina Sfiligoj and Marcus P. da Silva and Shilpi Singh and Sarat Sinha and Mathias Soeken and Patrick Sohr and Tomas Stankevic and Lieuwe Stek and Patrick Strøm-Hansen and Eric Stuppard and Aarthi Sundaram and Henri Suominen and Judith Suter and Satoshi Suzuki and Krysta Svore and Sam Teicher and Nivetha Thiyagarajah and Raj Tholapi and Mason Thomas and Dennis Tom and Emily Toomey and Josh Tracy and Matthias Troyer and Michelle Turley and Matthew D. Turner and Shivendra Upadhyay and Ivan Urban and Alexander Vaschillo and Dmitrii Viazmitinov and Dominik Vogel and Zhenghan Wang and John Watson and Alex Webster and Joseph Weston and Timothy Williamson and Georg W. Winkler and David J. van Woerkom and Brian Paquelet Wütz and Chung Kai Yang and Richard Yu and Emrah Yucelen and Jesús Herranz Zamorano and Roland Zeisel and Guoji Zheng and Justin Zilke and Andrew Zimmerman},
      year={2025},
      eprint={2502.12252},
      archivePrefix={arXiv},
      primaryClass={quant-ph},
      url={https://arxiv.org/abs/2502.12252}, 
}

@article{Wintersperger2023,
    author={Wintersperger, Karen
    and Dommert, Florian
    and Ehmer, Thomas
    and Hoursanov, Andrey
    and Klepsch, Johannes
    and Mauerer, Wolfgang
    and Reuber, Georg
    and Strohm, Thomas
    and Yin, Ming
    and Luber, Sebastian},
    title={Neutral atom quantum computing hardware: performance and end-user perspective},
    journal={EPJ Quantum Technology},
    year={2023},
    month={Aug},
    day={28},
    volume={10},
    number={1},
    pages={32},
    issn={2196-0763},
    doi={10.1140/epjqt/s40507-023-00190-1},
    url={https://doi.org/10.1140/epjqt/s40507-023-00190-1}
}

@article{Henriet2020quantumcomputing,
  doi = {10.22331/q-2020-09-21-327},
  url = {https://doi.org/10.22331/q-2020-09-21-327},
  title = {Quantum computing with neutral atoms},
  author = {Henriet, Lo{\"{i}}c and Beguin, Lucas and Signoles, Adrien and Lahaye, Thierry and Browaeys, Antoine and Reymond, Georges-Olivier and Jurczak, Christophe},
  journal = {{Quantum}},
  issn = {2521-327X},
  publisher = {{Verein zur F{\"{o}}rderung des Open Access Publizierens in den Quantenwissenschaften}},
  volume = {4},
  pages = {327},
  month = {sep},
  year = {2020}
}

@article{Michailidis2025,
author = {Michailidis, Alexios A. and Fenton, Christian and Kiffner, Martin},
title = {Element-wise Multiplication of Tensor Trains},
journal = {SIAM Journal on Scientific Computing},
volume = {47},
number = {5},
pages = {B1158-B1174},
year = {2025},
doi = {10.1137/24M1714149},
URL = {https://doi.org/10.1137/24M1714149},
eprint = {https://doi.org/10.1137/24M1714149}
}

@misc{Michailidis2024,
      title={Tensor Train Multiplication}, 
      author={Alexios A Michailidis and Christian Fenton and Martin Kiffner},
      year={2024},
      eprint={2410.19747},
      archivePrefix={arXiv},
      journal = {arXiv (pre-print)},
      primaryClass={physics.comp-ph},
      url={https://arxiv.org/abs/2410.19747}, 
}

@article{Termanova2024,
doi = {10.1088/1367-2630/ad985b},
url = {https://dx.doi.org/10.1088/1367-2630/ad985b},
year = {2024},
month = {dec},
publisher = {IOP Publishing},
volume = {26},
number = {12},
pages = {123019},
author = {Termanova, A and Melnikov, Ar and Mamenchikov, E and Belokonev, N and Dolgov, S and Berezutskii, A and Ellerbrock, R and Mansell, C and Perelshtein, M R},
title = {Tensor quantum programming},
journal = {New Journal of Physics},
}

@article{Terraneo2005,
  title = {Quantum computation and analysis of Wigner and Husimi functions: Toward a quantum image treatment},
  author = {Terraneo, M. and Georgeot, B. and Shepelyansky, D. L.},
  journal = {Phys. Rev. E},
  volume = {71},
  issue = {6},
  pages = {066215},
  numpages = {14},
  year = {2005},
  month = {Jun},
  publisher = {American Physical Society},
  doi = {10.1103/PhysRevE.71.066215},
  url = {https://link.aps.org/doi/10.1103/PhysRevE.71.066215}
}

@misc{li2026efficientendtoendquantumelliptic,
      title={Toward Efficient End-to-End Quantum Elliptic PDE Solvers: a Multilevel Correction Algorithm for Direct Observable Estimation}, 
      author={Xiantao Li},
      year={2026},
      eprint={2606.01270},
      archivePrefix={arXiv},
      primaryClass={quant-ph},
      url={https://arxiv.org/abs/2606.01270}, 
}

@misc{li2026exponentialreductionmeshdependence,
      title={Exponential Reduction of Mesh Dependence in Quantum Estimation of Parabolic PDE Observables}, 
      author={Xiantao Li},
      year={2026},
      eprint={2607.18113},
      archivePrefix={arXiv},
      primaryClass={quant-ph},
      url={https://arxiv.org/abs/2607.18113}, 
}

@Article{Chew2022,
author={Chew, Y.
and Tomita, T.
and Mahesh, T. P.
and Sugawa, S.
and de L{\'e}s{\'e}leuc, S.
and Ohmori, K.},
title={Ultrafast energy exchange between two single Rydberg atoms on a nanosecond timescale},
journal={Nature Photonics},
year={2022},
month={Oct},
day={01},
volume={16},
number={10},
pages={724-729},
issn={1749-4893},
doi={10.1038/s41566-022-01047-2},
url={https://doi.org/10.1038/s41566-022-01047-2}
}

@article{Bojovic2026,
    author={Bojovi{\'{c}}, Petar
    and Hilker, Timon
    and Wang, Si
    and Obermeyer, Johannes
    and Barendregt, Marnix
    and Tell, Dorothee
    and Chalopin, Thomas
    and Preiss, Philipp M.
    and Bloch, Immanuel
    and Franz, Titus},
    title={High-fidelity collisional quantum gates with fermionic atoms},
    journal={Nature},
    year={2026},
    month={Apr},
    day={01},
    volume={652},
    number={8110},
    pages={602-608},
    issn={1476-4687},
    doi={10.1038/s41586-026-10356-3},
    url={https://doi.org/10.1038/s41586-026-10356-3}
}

@article{Gyger2024,
  title = {Continuous operation of large-scale atom arrays in optical lattices},
  author = {Gyger, Flavien and Ammenwerth, Maximilian and Tao, Renhao and Timme, Hendrik and Snigirev, Stepan and Bloch, Immanuel and Zeiher, Johannes},
  journal = {Phys. Rev. Res.},
  volume = {6},
  issue = {3},
  pages = {033104},
  numpages = {9},
  year = {2024},
  month = {Jul},
  publisher = {American Physical Society},
  doi = {10.1103/PhysRevResearch.6.033104},
  url = {https://link.aps.org/doi/10.1103/PhysRevResearch.6.033104}
}

@article{Bagherimehrab_2024,
   title={Efficient quantum algorithm for all quantum wavelet transforms},
   volume={9},
   ISSN={2058-9565},
   url={http://dx.doi.org/10.1088/2058-9565/ad3d7f},
   DOI={10.1088/2058-9565/ad3d7f},
   number={3},
   journal={Quantum Science and Technology},
   publisher={IOP Publishing},
   author={Bagherimehrab, Mohsen and Aspuru-Guzik, Alán},
   year={2024},
   month={apr}, pages={035010} 
}

@misc{Horner2026,
      title={Tensor network compression using fluid dynamics as a testbed: Analytical foundations in one dimension}, 
      author={Matthew D. Horner and Callum W. Duncan and Oliver T. Brown and Stephen M. de Bruyn Kops and Muralikrishnan Gopalakrishnan Meena},
      year={2026},
      eprint={2606.17064},
      archivePrefix={arXiv},
      primaryClass={physics.comp-ph},
      url={https://arxiv.org/abs/2606.17064}, 
}

@article{NumezFernandez2025,
   title={Learning tensor networks with tensor cross interpolation: New algorithms and libraries},
   volume={18},
   ISSN={2542-4653},
   url={http://dx.doi.org/10.21468/SciPostPhys.18.3.104},
   DOI={10.21468/scipostphys.18.3.104},
   number={3},
   journal={SciPost Physics},
   publisher={Stichting SciPost},
   author={Núñez Fernández, Yuriel and Ritter, Marc K and Jeannin, Matthieu and Li, Jheng-Wei and Kloss, Thomas and Louvet, Thibaud and Terasaki, Satoshi and Parcollet, Olivier and von Delft, Jan and Shinaoka, Hiroshi and Waintal, Xavier},
   year={2025},
   month={Mar} }

@article{Ritter24,
  title = {Quantics Tensor Cross Interpolation for High-Resolution Parsimonious Representations of Multivariate Functions},
  author = {Ritter, Marc K. and N\'u\~nez Fern\'andez, Yuriel and Wallerberger, Markus and von Delft, Jan and Shinaoka, Hiroshi and Waintal, Xavier},
  journal = {Phys. Rev. Lett.},
  volume = {132},
  issue = {5},
  pages = {056501},
  numpages = {6},
  year = {2024},
  month = {Jan},
  publisher = {American Physical Society},
  doi = {10.1103/PhysRevLett.132.056501},
  url = {https://link.aps.org/doi/10.1103/PhysRevLett.132.056501}
}

@article{OSELEDETS2010TTCROSS,
title = {{TT}-cross approximation for multidimensional arrays},
journal = {Linear Algebra and its Applications},
volume = {432},
number = {1},
pages = {70-88},
year = {2010},
issn = {0024-3795},
doi = {10.1016/j.laa.2009.07.024},
url = {https://www.sciencedirect.com/science/article/pii/S0024379509003747},
author = {Ivan Oseledets and Eugene Tyrtyshnikov},
}

@misc{Amaral2026,
      title={A review of quantum machine learning and quantum-inspired applied methods to computational fluid dynamics}, 
      author={Cesar A. Amaral and Vinícius L. Oliveira and Juan P. L. C. Salazar and Eduardo I. Duzzioni},
      year={2026},
      eprint={2510.14099},
      archivePrefix={arXiv},
      primaryClass={quant-ph},
      url={https://arxiv.org/abs/2510.14099}, 
}

@article{CourantFriedrichsLewy1928,
  author = {Courant, R. and Friedrichs, K. and Lewy, H.},
  title = {{\"U}ber die partiellen Differenzengleichungen der mathematischen Physik},
  journal = {Mathematische Annalen},
  volume = {100},
  number = {1},
  pages = {32--74},
  year = {1928},
  publisher = {Springer},
  doi = {10.1007/BF01448839}
}

@book{Saad2003,
author = {Saad, Yousef},
title = {Iterative Methods for Sparse Linear Systems},
publisher = {Society for Industrial and Applied Mathematics},
year = {2003},
doi = {10.1137/1.9780898718003},
address = {},
edition   = {Second},
URL = {https://epubs.siam.org/doi/abs/10.1137/1.9780898718003},
eprint = {https://epubs.siam.org/doi/pdf/10.1137/1.9780898718003}
}

@article{Choi2012,
    author = {Choi, Haecheon and Moin, Parviz},
    title = {Grid-point requirements for large eddy simulation: Chapman’s estimates revisited},
    journal = {Physics of Fluids},
    volume = {24},
    number = {1},
    pages = {011702},
    year = {2012},
    month = {01},
    issn = {1070-6631},
    doi = {10.1063/1.3676783},
    url = {https://doi.org/10.1063/1.3676783},
}

@book{WILCOX06,
    author      = {Wilcox, David C.},
    title       = {Turbulence {Modeling} for {CFD}},
    address     = {La C{\~a}nada, CA},
    edition     = {3rd},
    isbn        = {978-1-928729-08-2},
    publisher   = {DCW Industries, Inc.},
    year        = {2006},
}

@article{Cai2023,
  title = {Quantum error mitigation},
  author = {Cai, Zhenyu and Babbush, Ryan and Benjamin, Simon C. and Endo, Suguru and Huggins, William J. and Li, Ying and McClean, Jarrod R. and O'Brien, Thomas E.},
  journal = {Rev. Mod. Phys.},
  volume = {95},
  issue = {4},
  pages = {045005},
  numpages = {37},
  year = {2023},
  month = {Dec},
  publisher = {American Physical Society},
  doi = {10.1103/RevModPhys.95.045005},
  url = {https://link.aps.org/doi/10.1103/RevModPhys.95.045005}
}

@article{Meng2024,
    author={Meng, Zhaoyuan
    and Zhong, Jiarun
    and Xu, Shibo
    and Wang, Ke
    and Chen, Jiachen
    and Jin, Feitong
    and Zhu, Xuhao
    and Gao, Yu
    and Wu, Yaozu
    and Zhang, Chuanyu
    and Wang, Ning
    and Zou, Yiren
    and Zhang, Aosai
    and Cui, Zhengyi
    and Shen, Fanhao
    and Bao, Zehang
    and Zhu, Zitian
    and Tan, Ziqi
    and Li, Tingting
    and Zhang, Pengfei
    and Xiong, Shiying
    and Li, Hekang
    and Guo, Qiujiang
    and Wang, Zhen
    and Song, Chao
    and Wang, H.
    and Yang, Yue},
    title={Simulating unsteady flows on a superconducting quantum processor},
    journal={Communications Physics},
    year={2024},
    month={Oct},
    day={25},
    volume={7},
    number={1},
    pages={349},
    issn={2399-3650},
    doi={10.1038/s42005-024-01845-w},
    url={https://doi.org/10.1038/s42005-024-01845-w}
}

@article{Chen2024,
title = {Enabling large-scale and high-precision fluid simulations on near-term quantum computers},
journal = {Computer Methods in Applied Mechanics and Engineering},
volume = {432},
pages = {117428},
year = {2024},
issn = {0045-7825},
doi = {10.1016/j.cma.2024.117428},
url = {https://www.sciencedirect.com/science/article/pii/S0045782524006832},
author = {Zhao-Yun Chen and Teng-Yang Ma and Chuang-Chao Ye and Liang Xu and Wen Bai and Lei Zhou and Ming-Yang Tan and Xi-Ning Zhuang and Xiao-Fan Xu and Yun-Jie Wang and Tai-Ping Sun and Yong Chen and Lei Du and Liang-Liang Guo and Hai-Feng Zhang and Hao-Ran Tao and Tian-Le Wang and Xiao-Yan Yang and Ze-An Zhao and Peng Wang and Sheng Zhang and Ren-Ze Zhao and Chi Zhang and Zhi-Long Jia and Wei-Cheng Kong and Meng-Han Dou and Jun-Chao Wang and Huan-Yu Liu and Cheng Xue and Peng-Jun-Yi Zhang and Sheng-Hong Huang and Peng Duan and Yu-Chun Wu and Guo-Ping Guo}
}

@article{GonzalezConde2025,
  title = {Quantum Carleman linearization efficiency in nonlinear fluid dynamics},
  author = {Gonzalez-Conde, Javier and Lewis, Dylan and Bharadwaj, Sachin S. and Sanz, Mikel},
  journal = {Phys. Rev. Res.},
  volume = {7},
  issue = {2},
  pages = {023254},
  numpages = {16},
  year = {2025},
  month = {Jun},
  publisher = {American Physical Society},
  doi = {10.1103/PhysRevResearch.7.023254},
  url = {https://link.aps.org/doi/10.1103/PhysRevResearch.7.023254}
}

@misc{Jesus2026,
  title={Time evolution of nonlinear dynamics on a quantum processor}, 
  author={José Diogo da Costa Jesus and Abhishek Setty and Tommaso Calarco and Dieter Jaksch and Francisco Cárdenas López and Felix Motzoi},
  year={2026},
  eprint={2608.13041},
  archivePrefix={arXiv},
  primaryClass={quant-ph},
  url={https://arxiv.org/abs/2608.13041}, 
}

@article{Loio2025,
  title = {Correlations, spectra, and entanglement transitions in ensembles of matrix product states},
  author = {L\'oio, Hugo and Cecile, Guillaume and Gopalakrishnan, Sarang and Lami, Guglielmo and De Nardis, Jacopo},
  journal = {Phys. Rev. B},
  volume = {112},
  issue = {3},
  pages = {035127},
  numpages = {9},
  year = {2025},
  month = {Jul},
  publisher = {American Physical Society},
  doi = {10.1103/ymzz-923j},
  url = {https://link.aps.org/doi/10.1103/ymzz-923j}
}

@phdthesis{Gourianov2022b,
  author       = {Gourianov, Nikita},
  title        = {Exploiting the Structure of Turbulence with Tensor Networks},
  school       = {University of Oxford},
  year         = {2022},
  type         = {PhD thesis},
  address      = {Oxford, UK},
  url          = {https://ora.ox.ac.uk/objects/uuid:9e3f4786-ad68-4913-9a0d-e9b1e108128f}
}

@article{Krovi_2023,
   title={Improved quantum algorithms for linear and nonlinear differential equations},
   volume={7},
   ISSN={2521-327X},
   url={http://dx.doi.org/10.22331/q-2023-02-02-913},
   DOI={10.22331/q-2023-02-02-913},
   journal={Quantum},
   publisher={Verein zur Forderung des Open Access Publizierens in den Quantenwissenschaften},
   author={Krovi, Hari},
   year={2023},
   month={feb}, pages={913} }

@article{Liu_2021,
   title={Efficient quantum algorithm for dissipative nonlinear differential equations},
   volume={118},
   ISSN={1091-6490},
   url={http://dx.doi.org/10.1073/pnas.2026805118},
   DOI={10.1073/pnas.2026805118},
   number={35},
   journal={Proceedings of the National Academy of Sciences},
   publisher={Proceedings of the National Academy of Sciences},
   author={Liu, Jin-Peng and Kolden, Herman {\O}ie and Krovi, Hari K. and Loureiro, Nuno F. and Trivisa, Konstantina and Childs, Andrew M.},
   year={2021},
   month={aug} 
}

@article{Esmaeilifar2024,
    author = {Esmaeilifar, Esmaeil and Ahn, Doyeol and Myong, Rho Shin},
    title = {Quantum algorithm for nonlinear Burgers' equation for high-speed compressible flows},
    journal = {Physics of Fluids},
    volume = {36},
    number = {10},
    pages = {106110},
    year = {2024},
    month = {10},
    issn = {1070-6631},
    doi = {10.1063/5.0231994},
    url = {https://doi.org/10.1063/5.0231994},
}

@misc{Uchida_2026,
      title={Quantum simulation of Burgers turbulence: Nonlinear transformation and direct evaluation of statistical quantities}, 
      author={Fumio Uchida and Koichi Miyamoto and Soichiro Yamazaki and Kotaro Fujisawa and Naoki Yoshida},
      year={2026},
      eprint={2412.17206},
      archivePrefix={arXiv},
      primaryClass={quant-ph},
      url={https://arxiv.org/abs/2412.17206}, 
}

@misc{Wu_2025,
      title={Quantum Algorithms for Nonlinear Dynamics: Revisiting Carleman Linearization with No Dissipative Conditions}, 
      author={Hsuan-Cheng Wu and Jingyao Wang and Xiantao Li},
      year={2025},
      eprint={2405.12714},
      archivePrefix={arXiv},
      primaryClass={quant-ph},
      url={https://arxiv.org/abs/2405.12714}, 
}

@article{Carleman1932,
author = {Torsten Carleman},
title = {{Application de la théorie des équations intégrales linéaires aux systèmes d'équations différentielles non linéaires}},
volume = {59},
journal = {Acta Mathematica},
number = {none},
publisher = {Institut Mittag-Leffler},
pages = {63 -- 87},
year = {1932},
doi = {10.1007/BF02546499},
URL = {https://doi.org/10.1007/BF02546499}
}

@misc{vanhülst2025quantumsolverspredictiveaeroacoustic,
      title={Quantum Solvers: Predictive Aeroacoustic \& Aerodynamic modeling}, 
      author={Nis-Luca van H{\"u}lst and Theofanis Panagos and Greta Sophie Reese and Shahram Panahiyan and Tomohiro Hashizume},
      year={2025},
      eprint={2507.21683},
      archivePrefix={arXiv},
      primaryClass={quant-ph},
      url={https://arxiv.org/abs/2507.21683}, 
}

@article{McClean2016,
  author  = {McClean, Jarrod R. and Romero, Jonathan and Babbush, Ryan and Aspuru-Guzik, Al{\'a}n},
  title   = {The theory of variational hybrid quantum-classical algorithms},
  journal = {New Journal of Physics},
  volume  = {18},
  number  = {2},
  pages   = {023023},
  year    = {2016},
  doi     = {10.1088/1367-2630/18/2/023023},
  url     = {https://doi.org/10.1088/1367-2630/18/2/023023}
}

@article{Larocca2023,
  author  = {Larocca, Martín and Ju, Nathan and García-Martín, Diego and Coles, Patrick J. and Cerezo, Marco},
  title   = {Theory of overparametrization in quantum neural networks},
  journal = {Nature Computational Science},
  volume  = {3},
  number  = {6},
  pages   = {542--551},
  year    = {2023},
  doi     = {10.1038/s43588-023-00467-6},
  url     = {https://doi.org/10.1038/s43588-023-00467-6}
}

@misc{hashizume2026,
      title={Quantum computation at the edge of chaos}, 
      author={Tomohiro Hashizume and Zhengjun Wang and Frank Schlawin and Dieter Jaksch},
      year={2026},
      eprint={2604.15441},
      archivePrefix={arXiv},
      primaryClass={quant-ph},
      url={https://arxiv.org/abs/2604.15441}, 
}

@book{Watrous_2018, 
place={Cambridge}, 
title={The Theory of Quantum Information}, 
publisher={Cambridge University Press}, 
author={Watrous, John}, 
year={2018}
}

@article{Raissi2019,
title = {Physics-informed neural networks: A deep learning framework for solving forward and inverse problems involving nonlinear partial differential equations},
journal = {Journal of Computational Physics},
volume = {378},
pages = {686-707},
year = {2019},
issn = {0021-9991},
doi = {10.1016/j.jcp.2018.10.045},
url = {https://www.sciencedirect.com/science/article/pii/S0021999118307125},
author = {M. Raissi and P. Perdikaris and G.E. Karniadakis}
}

@article{Stinespring1955,
  title={Positive functions on {$C^*$}-algebras},
  author={Stinespring, W. Forrest},
  journal={Proceedings of the American Mathematical Society},
  volume={6},
  number={2},
  pages={211--216},
  year={1955},
  publisher={JSTOR}
}

@misc{Ambainis2014,
      title={Quantum walk algorithm for element distinctness}, 
      author={Andris Ambainis},
      year={2014},
      eprint={quant-ph/0311001},
      archivePrefix={arXiv},
      primaryClass={quant-ph},
      url={https://arxiv.org/abs/quant-ph/0311001}, 
}

@misc{Aharonov2002,
      title={Quantum Walks On Graphs}, 
      author={Dorit Aharonov and Andris Ambainis and Julia Kempe and Umesh Vazirani},
      year={2002},
      eprint={quant-ph/0012090},
      archivePrefix={arXiv},
      primaryClass={quant-ph},
      url={https://arxiv.org/abs/quant-ph/0012090}, 
}

@article{Wawrzyniak2025,
title = {A quantum algorithm for the lattice-Boltzmann method advection-diffusion equation},
journal = {Computer Physics Communications},
volume = {306},
pages = {109373},
year = {2025},
issn = {0010-4655},
doi = {10.1016/j.cpc.2024.109373},
url = {https://www.sciencedirect.com/science/article/pii/S0010465524002960},
author = {David Wawrzyniak and Josef Winter and Steffen Schmidt and Thomas Indinger and Christian F. Janßen and Uwe Schramm and Nikolaus A. Adams}
}

@article{McClean2018,
  author  = {McClean, Jarrod R. and Boixo, Sergio and Smelyanskiy, Vadim N. and Babbush, Ryan and Neven, Hartmut},
  title   = {Barren plateaus in quantum neural network training landscapes},
  journal = {Nature Communications},
  year    = {2018},
  volume  = {9},
  number  = {1},
  pages   = {4812},
  doi     = {10.1038/s41467-018-07090-4},
  url     = {https://doi.org/10.1038/s41467-018-07090-4},
  issn    = {2041-1723}
}

@book{Paris2004,
  editor    = {Paris, Matteo G. A. and {\v R}eh{\'a}{\v c}ek, Jaroslav},
  title     = {Quantum State Estimation},
  series    = {Lecture Notes in Physics},
  volume    = {649},
  publisher = {Springer},
  address   = {Berlin, Heidelberg},
  year      = {2004},
  doi       = {10.1007/b98673},
  isbn      = {978-3-540-44481-7}
}

@article{Aaronson2018,
author = {Aaronson, Scott},
title = {Shadow Tomography of Quantum States},
journal = {SIAM Journal on Computing},
volume = {49},
number = {5},
pages = {STOC18-368-STOC18-394},
year = {2020},
doi = {10.1137/18M120275X}
}

@article{Preskill_2025,
   title={Beyond NISQ: The Megaquop Machine},
   volume={6},
   ISSN={2643-6817},
   url={http://dx.doi.org/10.1145/3723153},
   DOI={10.1145/3723153},
   number={3},
   journal={ACM Transactions on Quantum Computing},
   publisher={Association for Computing Machinery (ACM)},
   author={Preskill, John},
   year={2025},
   month={apr}, pages={1–7} 
}

@article{Acharya2025,
  author = {Acharya, Rajeev and
            Abanin, Dmitry A. and
            Aghababaie-Beni, Laleh and
            Aleiner, Igor and
            Andersen, Trond I. and
            Ansmann, Markus and
            Arute, Frank and
            Arya, Kunal and
            Asfaw, Abraham and
            Astrakhantsev, Nikita and
            Atalaya, Juan and
            Babbush, Ryan and
            Bacon, Dave and
            Ballard, Brian and
            Bardin, Joseph C. and
            Bausch, Johannes and
            Bengtsson, Andreas and
            Bilmes, Alexander and
            Blackwell, Sam and
            Boixo, Sergio and
            Bortoli, Gina and
            Bourassa, Alexandre and
            Bovaird, Jenna and
            Brill, Leon and
            Broughton, Michael and
            Browne, David A. and
            Buchea, Brett and
            Buckley, Bob B. and
            Buell, David A. and
            Burger, Tim and
            Burkett, Brian and
            Bushnell, Nicholas and
            Cabrera, Anthony and
            Campero, Juan and
            Chang, Hung-Shen and
            Chen, Yu and
            Chen, Zijun and
            Chiaro, Ben and
            Chik, Desmond and
            Chou, Charina and
            Claes, Jahan and
            Cleland, Agnetta Y. and
            Cogan, Josh and
            Collins, Roberto and
            Conner, Paul and
            Courtney, William and
            Crook, Alexander L. and
            Curtin, Ben and
            Das, Sayan and
            Davies, Alex and
            De Lorenzo, Laura and
            Debroy, Dripto M. and
            Demura, Sean and
            Devoret, Michel and
            Di Paolo, Agustin and
            Donohoe, Paul and
            Drozdov, Ilya and
            Dunsworth, Andrew and
            Earle, Clint and
            Edlich, Thomas and
            Eickbusch, Alec and
            Elbag, Aviv Moshe and
            Elzouka, Mahmoud and
            Erickson, Catherine and
            Faoro, Lara and
            Farhi, Edward and
            Ferreira, Vinicius S. and
            Burgos, Leslie Flores and
            Forati, Ebrahim and
            Fowler, Austin G. and
            Foxen, Brooks and
            Ganjam, Suhas and
            Garcia, Gonzalo and
            Gasca, Robert and
            Genois, {\'E}lie and
            Giang, William and
            Gidney, Craig and
            Gilboa, Dar and
            Gosula, Raja and
            Dau, Alejandro Grajales and
            Graumann, Dietrich and
            Greene, Alex and
            Gross, Jonathan A. and
            Habegger, Steve and
            Hall, John and
            Hamilton, Michael C. and
            Hansen, Monica and
            Harrigan, Matthew P. and
            Harrington, Sean D. and
            Heras, Francisco J. H. and
            Heslin, Stephen and
            Heu, Paula and
            Higgott, Oscar and
            Hill, Gordon and
            Hilton, Jeremy and
            Holland, George and
            Hong, Sabrina and
            Huang, Hsin-Yuan and
            Huff, Ashley and
            Huggins, William J. and
            Ioffe, Lev B. and
            Isakov, Sergei V. and
            Iveland, Justin and
            Jeffrey, Evan and
            Jiang, Zhang and
            Jones, Cody and
            Jordan, Stephen and
            Joshi, Chaitali and
            Juhas, Pavol and
            Kafri, Dvir and
            Kang, Hui and
            Karamlou, Amir H. and
            Kechedzhi, Kostyantyn and
            Kelly, Julian and
            Khaire, Trupti and
            Khattar, Tanuj and
            Khezri, Mostafa and
            Kim, Seon and
            Klimov, Paul V. and
            Klots, Andrey R. and
            Kobrin, Bryce and
            Kohli, Pushmeet and
            Korotkov, Alexander N. and
            Kostritsa, Fedor and
            Kothari, Robin and
            Kozlovskii, Borislav and
            Kreikebaum, John Mark and
            Kurilovich, Vladislav D. and
            Lacroix, Nathan and
            Landhuis, David and
            Lange-Dei, Tiano and
            Langley, Brandon W. and
            Laptev, Pavel and
            Lau, Kim-Ming and
            Le Guevel, Lo{\"i}ck and
            Ledford, Justin and
            Lee, Joonho and
            Lee, Kenny and
            Lensky, Yuri D. and
            Leon, Shannon and
            Lester, Brian J. and
            Li, Wing Yan and
            Li, Yin and
            Lill, Alexander T. and
            Liu, Wayne and
            Livingston, William P. and
            Locharla, Aditya and
            Lucero, Erik and
            Lundahl, Daniel and
            Lunt, Aaron and
            Madhuk, Sid and
            Malone, Fionn D. and
            Maloney, Ashley and
            Mandr{\`a}, Salvatore and
            Manyika, James and
            Martin, Leigh S. and
            Martin, Orion and
            Martin, Steven and
            Maxfield, Cameron and
            McClean, Jarrod R. and
            McEwen, Matt and
            Meeks, Seneca and
            Megrant, Anthony and
            Mi, Xiao and
            Miao, Kevin C. and
            Mieszala, Amanda and
            Molavi, Reza and
            Molina, Sebastian and
            Montazeri, Shirin and
            Morvan, Alexis and
            Movassagh, Ramis and
            Mruczkiewicz, Wojciech and
            Naaman, Ofer and
            Neeley, Matthew and
            Neill, Charles and
            Nersisyan, Ani and
            Neven, Hartmut and
            Newman, Michael and
            Ng, Jiun How and
            Nguyen, Anthony and
            Nguyen, Murray and
            Ni, Chia-Hung and
            Niu, Murphy Yuezhen and
            O'Brien, Thomas E. and
            Oliver, William D. and
            Opremcak, Alex and
            Ottosson, Kristoffer and
            Petukhov, Andre and
            Pizzuto, Alex and
            Platt, John and
            Potter, Rebecca and
            Pritchard, Orion and
            Pryadko, Leonid P. and
            Quintana, Chris and
            Ramachandran, Ganesh and
            Reagor, Matthew J. and
            Redding, John and
            Rhodes, David M. and
            Roberts, Gabrielle and
            Rosenberg, Eliott and
            Rosenfeld, Emma and
            Roushan, Pedram and
            Rubin, Nicholas C. and
            Saei, Negar and
            Sank, Daniel and
            Sankaragomathi, Kannan and
            Satzinger, Kevin J. and
            Schurkus, Henry F. and
            Schuster, Christopher and
            Senior, Andrew W. and
            Shearn, Michael J. and
            Shorter, Aaron and
            Shutty, Noah and
            Shvarts, Vladimir and
            Singh, Shraddha and
            Sivak, Volodymyr and
            Skruzny, Jindra and
            Small, Spencer and
            Smelyanskiy, Vadim and
            Smith, W. Clarke and
            Somma, Rolando D. and
            Springer, Sofia and
            Sterling, George and
            Strain, Doug and
            Suchard, Jordan and
            Szasz, Aaron and
            Sztein, Alex and
            Thor, Douglas and
            Torres, Alfredo and
            Torunbalci, M. Mert and
            Vaishnav, Abeer and
            Vargas, Justin and
            Vdovichev, Sergey and
            Vidal, Guifre and
            Villalonga, Benjamin and
            Heidweiller, Catherine Vollgraff and
            Waltman, Steven and
            Wang, Shannon X. and
            Ware, Brayden and
            Weber, Kate and
            Weidel, Travis and
            White, Theodore and
            Wong, Kristi and
            Woo, Bryan W. K. and
            Xing, Cheng and
            Yao, Z. Jamie and
            Yeh, Ping and
            Ying, Bicheng and
            Yoo, Juhwan and
            Yosri, Noureldin and
            Young, Grayson and
            Zalcman, Adam and
            Zhang, Yaxing and
            Zhu, Ningfeng and
            Zobrist, Nicholas and
            {Google Quantum AI and Collaborators}},
  title   = {Quantum Error Correction Below the Surface Code Threshold},
  journal = {Nature},
  volume  = {638},
  number  = {8052},
  pages   = {920--926},
  year    = {2025},
  doi     = {10.1038/s41586-024-08449-y},
  url     = {https://doi.org/10.1038/s41586-024-08449-y}
}

@article{Aaronson2015,
  author    = {Scott Aaronson},
  title     = {Read the Fine Print},
  journal   = {Nature Physics},
  volume    = {11},
  number    = {4},
  pages     = {291--293},
  year      = {2015},
  doi       = {10.1038/nphys3272},
  url       = {https://doi.org/10.1038/nphys3272}
}

@article{Boschung2016, 
title={Generalised higher-order Kolmogorov scales}, 
volume={794}, 
DOI={10.1017/jfm.2016.172}, 
journal={Journal of Fluid Mechanics}, 
author={Boschung, Jonas and Hennig, Fabian and Gauding, Michael and Pitsch, Heinz and Peters, Norbert}, 
year={2016}, 
pages={233–251}
}

@book{Dalzell_2025,
   title={Quantum Algorithms: A Survey of Applications and End-to-end Complexities},
   ISBN={9781009639668},
   url={http://dx.doi.org/10.1017/9781009639651},
   DOI={10.1017/9781009639651},
   publisher={Cambridge University Press},
   author={Dalzell, Alexander M. and McArdle, Sam and Berta, Mario and Bienias, Przemyslaw and Chen, Chi-Fang and Gilyén, András and Hann, Connor T. and Kastoryano, Michael J. and Khabiboulline, Emil T. and Kubica, Aleksander and Salton, Grant and Wang, Samson and Brandão, Fernando G. S. L.},
   year={2025},
   month={apr} 
}

@article{Malinverno2026,
       author = {{Malinverno}, Giulio and {Blasco Alberto}, Javier},
        title = "{A Review of the Current State-of-the-Art of Quantum Computing for CFD: Approaches, Advantages, and Limitations}",
      journal = {Aerotecnica Missili \& Spazio},
         year = 2026,
        month = {apr},
       volume = {105},
       number = {2},
        pages = {251-268},
          doi = {10.1007/s42496-025-00269-1},
       adsurl = {https://ui.adsabs.harvard.edu/abs/2026AeMiS.105..251M}
}

@book{Tennekes1972,
    author = {Tennekes, Henk and Lumley, John L.},
    title = {A First Course in Turbulence},
    publisher = {The MIT Press},
    year = {1972},
    month = {03},
    isbn = {9780262310901},
    doi = {10.7551/mitpress/3014.001.0001},
    url = {https://doi.org/10.7551/mitpress/3014.001.0001},
}

@article{Verma2018,
   title={Energy Spectra and Fluxes in Dissipation Range of Turbulent and Laminar Flows},
   volume={53},
   ISSN={1573-8507},
   url={http://dx.doi.org/10.1134/S0015462818050166},
   DOI={10.1134/s0015462818050166},
   number={6},
   journal={Fluid Dynamics},
   publisher={Pleiades Publishing Ltd},
   author={Verma, M. K. and Kumar, A. and Kumar, P. and Barman, S. and Chatterjee, A. G. and Samtaney, R. and Stepanov, R. A.},
   year={2018},
   month={nov}, pages={862–873} }

@article{Wang2025,
  author  = {Wang, Boyuan and Meng, Zhaoyuan and Zhao, Yaomin and Yang, Yue},
  title   = {Quantum lattice Boltzmann method for simulating nonlinear fluid dynamics},
  journal = {npj Quantum Information},
  year    = {2025},
  volume  = {11},
  number  = {1},
  pages   = {196},
  doi     = {10.1038/s41534-025-01142-6},
  url     = {https://doi.org/10.1038/s41534-025-01142-6}
}

@misc{Molpeceres2026,
      title={Benchmark of quantum algorithms for ground state preparation in the presence of noise}, 
      author={Daniel Molpeceres and Sirui Lu and J. Ignacio Cirac and Barbara Kraus},
      year={2026},
      eprint={2606.20551},
      archivePrefix={arXiv},
      primaryClass={quant-ph},
      url={https://arxiv.org/abs/2606.20551}, 
}

@article{Molpeceres2025,
  title = {Quantum algorithms for cooling: A simple case study},
  author = {Molpeceres, Daniel and Lu, Sirui and Cirac, J. Ignacio and Kraus, Barbara},
  journal = {Phys. Rev. Res.},
  volume = {7},
  issue = {3},
  pages = {033162},
  numpages = {36},
  year = {2025},
  month = {Aug},
  publisher = {American Physical Society},
  doi = {10.1103/4hx7-xnhw},
  url = {https://link.aps.org/doi/10.1103/4hx7-xnhw}
}

@misc{meng2026recursivesketchedinterpolationefficient,
      title={Recursive Sketched Interpolation: Efficient Hadamard Products of Tensor Trains}, 
      author={Zhaonan Meng and Yuehaw Khoo and Jiajia Li and E. Miles Stoudenmire},
      year={2026},
      eprint={2602.17974},
      archivePrefix={arXiv},
      primaryClass={quant-ph},
      url={https://arxiv.org/abs/2602.17974}, 
}

@article{Chen1998,
   author = "Chen, Shiyi and Doolen, Gary D.",
   title = "LATTICE BOLTZMANN METHOD FOR FLUID FLOWS", 
   journal= "Annual Review of Fluid Mechanics",
   year = "1998",
   volume = "30",
   number = "Volume 30, 1998",
   pages = "329-364",
   doi = "10.1146/annurev.fluid.30.1.329",
   url = "https://www.annualreviews.org/content/journals/10.1146/annurev.fluid.30.1.329",
   publisher = "Annual Reviews",
   issn = "1545-4479",
   type = "Journal Article",
}

@article{Campbell2017,
   title={Roads towards fault-tolerant universal quantum computation},
   volume={549},
   ISSN={1476-4687},
   url={http://dx.doi.org/10.1038/nature23460},
   DOI={10.1038/nature23460},
   number={7671},
   journal={Nature},
   publisher={Springer Science and Business Media LLC},
   author={Campbell, Earl T. and Terhal, Barbara M. and Vuillot, Christophe},
   year={2017},
   month={sept}, pages={172–179} 
}

@article{Bravyi2005,
  title = {Universal quantum computation with ideal Clifford gates and noisy ancillas},
  author = {Bravyi, Sergey and Kitaev, Alexei},
  journal = {Phys. Rev. A},
  volume = {71},
  issue = {2},
  pages = {022316},
  numpages = {14},
  year = {2005},
  month = {Feb},
  publisher = {American Physical Society},
  doi = {10.1103/PhysRevA.71.022316},
  url = {https://link.aps.org/doi/10.1103/PhysRevA.71.022316}
}

@article{Y.H.Qian_1992,
doi = {10.1209/0295-5075/17/6/001},
url = {https://doi.org/10.1209/0295-5075/17/6/001},
year = {1992},
month = {feb},
publisher = {},
volume = {17},
number = {6},
pages = {479},
author = {Y. H. Qian and D. D'Humières and P. Lallemand},
title = {Lattice BGK Models for Navier-Stokes Equation},
journal = {Europhysics Letters}
}

@article{Matthies_2024,
   title={Programmable adiabatic demagnetization for systems with trivial and topological excitations},
   volume={8},
   ISSN={2521-327X},
   url={http://dx.doi.org/10.22331/q-2024-10-23-1505},
   DOI={10.22331/q-2024-10-23-1505},
   journal={Quantum},
   publisher={Verein zur Forderung des Open Access Publizierens in den Quantenwissenschaften},
   author={Matthies, Anne and Rudner, Mark and Rosch, Achim and Berg, Erez},
   year={2024},
   month={oct}, pages={1505} 
}

@misc{Tiwari2025,
      title={Algorithmic Advances Towards a Realizable Quantum Lattice Boltzmann Method}, 
      author={Apurva Tiwari and Jason Iaconis and Jezer Jojo and Sayonee Ray and Martin Roetteler and Chris Hill and Jay Pathak},
      year={2025},
      eprint={2504.10870},
      archivePrefix={arXiv},
      primaryClass={quant-ph},
      url={https://arxiv.org/abs/2504.10870}, 
}

@article{Gidney_2021,
   title={How to factor 2048 bit RSA integers in 8 hours using 20 million noisy qubits},
   volume={5},
   ISSN={2521-327X},
   url={http://dx.doi.org/10.22331/q-2021-04-15-433},
   DOI={10.22331/q-2021-04-15-433},
   journal={Quantum},
   publisher={Verein zur Forderung des Open Access Publizierens in den Quantenwissenschaften},
   author={Gidney, Craig and Ekerå, Martin},
   year={2021},
   month={apr}, 
   pages={433} 
}

@misc{Beverland2022,
      title={Assessing requirements to scale to practical quantum advantage}, 
      author={Michael E. Beverland and Prakash Murali and Matthias Troyer and Krysta M. Svore and Torsten Hoefler and Vadym Kliuchnikov and Guang Hao Low and Mathias Soeken and Aarthi Sundaram and Alexander Vaschillo},
      year={2022},
      eprint={2211.07629},
      archivePrefix={arXiv},
      primaryClass={quant-ph},
      url={https://arxiv.org/abs/2211.07629}, 
}

@misc{menssen2026strategicplanneutralatom,
      title={Strategic Plan for Neutral Atom Quantum Computation}, 
      author={Adrian J. Menssen and Tout Wang and Michael Gullans and Tom Manovitz and Jacob M. Taylor and Jason Cong and Josiah Sinclair and Ziv Aqua and Daniel J. Blumenthal and J. Pablo Bonilla Ataides and Johannes Borregaard and Antoine Browaeys and Paola Cappellaro and Soonwon Choi and Alexandre Cooper and Robin Côté and Jacob P. Covey and Alexandre Dauphin and Ivana Dimitrova and Matt Eichenfield and Dirk Englund and Jacob Freedman and Akihisa Goban and Brandon Grinkemeyer and Andi Gu and Ruonan Han and Dominik Hangleiter and Aram W. Harrow and Liang Jiang and Eun-ah Kim and Felix W. Knollmann and Aleksander Kubica and Thierry Lahaye and Lucas Lassabliere and Joonho Lee and Bingzhao Li and Mo Li and Wan-Hsuan Lin and Mikhail D. Lukin and Varun Menon and Thomas Propson and Akbar Safari and Mark Saffman and Pascal Scholl and Alexander Schuckert and Giulia Semeghini and Jonathan Simon and David Spierings and Daniel Bochen Tan and Shai Tsesses and Vladan Vuletic and Hanrui Wang and Hanyu Wang and Susanne Yelin and Johannes Zeiher and Hengyun Zhou},
      year={2026},
      eprint={2607.21554},
      archivePrefix={arXiv},
      primaryClass={quant-ph},
      url={https://arxiv.org/abs/2607.21554}, 
}

@article{Malinverno2025,
  author  = {Malinverno, Giulio and Blasco Alberto, Javier},
  title   = {A Review of the Current State-of-the-Art of Quantum Computing for CFD: Approaches, Advantages, and Limitations},
  journal = {Aerotecnica Missili \& Spazio},
  volume  = {105},
  pages   = {251--268},
  year    = {2025},
  doi     = {10.1007/s42496-025-00269-1}
}

@book{Kruger2017,
  author    = {Kr\"uger, Timm and Kusumaatmaja, Halim and Kuzmin, Alexandr and Shardt, Orest and Silva, Goncalo and Viggen, Erlend Magnus},
  title     = {The Lattice Boltzmann Method: Principles and Practice},
  publisher = {Springer},
  address   = {Cham},
  year      = {2017},
  edition   = {1},
  doi       = {10.1007/978-3-319-44649-3}
}

@book{Succi2001,
    author = {Succi, Sauro},
    title = {The Lattice Boltzmann Equation for Fluid Dynamics and Beyond},
    publisher = {Oxford University Press},
    year = {2001},
    month = {06},
    isbn = {9780198503989},
    doi = {10.1093/oso/9780198503989.001.0001},
    url = {https://doi.org/10.1093/oso/9780198503989.001.0001},
}

@article{Lloyd2025,
  title = {Quasiparticle Cooling Algorithms for Quantum Many-Body State Preparation},
  author = {Lloyd, Jerome and Michailidis, Alexios A. and Mi, Xiao and Smelyanskiy, Vadim and Abanin, Dmitry A.},
  journal = {PRX Quantum},
  volume = {6},
  issue = {1},
  pages = {010361},
  numpages = {30},
  year = {2025},
  month = {Mar},
  publisher = {American Physical Society},
  doi = {10.1103/PRXQuantum.6.010361},
  url = {https://link.aps.org/doi/10.1103/PRXQuantum.6.010361}
}

@article{
Mi2024,
author = {X. Mi  and A. A. Michailidis  and S. Shabani  and K. C. Miao  and P. V. Klimov  and J. Lloyd  and E. Rosenberg  and R. Acharya  and I. Aleiner  and T. I. Andersen  and M. Ansmann  and F. Arute  and K. Arya  and A. Asfaw  and J. Atalaya  and J. C. Bardin  and A. Bengtsson  and G. Bortoli  and A. Bourassa  and J. Bovaird  and L. Brill  and M. Broughton  and B. B. Buckley  and D. A. Buell  and T. Burger  and B. Burkett  and N. Bushnell  and Z. Chen  and B. Chiaro  and D. Chik  and C. Chou  and J. Cogan  and R. Collins  and P. Conner  and W. Courtney  and A. L. Crook  and B. Curtin  and A. G. Dau  and D. M. Debroy  and A. Del Toro Barba  and S. Demura  and A. Di Paolo  and I. K. Drozdov  and A. Dunsworth  and C. Erickson  and L. Faoro  and E. Farhi  and R. Fatemi  and V. S. Ferreira  and L. F. Burgos  and E. Forati  and A. G. Fowler  and B. Foxen  and {\'E}. Genois  and W. Giang  and C. Gidney  and D. Gilboa  and M. Giustina  and R. Gosula  and J. A. Gross  and S. Habegger  and M. C. Hamilton  and M. Hansen  and M. P. Harrigan  and S. D. Harrington  and P. Heu  and M. R. Hoffmann  and S. Hong  and T. Huang  and A. Huff  and W. J. Huggins  and L. B. Ioffe  and S. V. Isakov  and J. Iveland  and E. Jeffrey  and Z. Jiang  and C. Jones  and P. Juhas  and D. Kafri  and K. Kechedzhi  and T. Khattar  and M. Khezri  and M. Kieferov{\'a}  and S. Kim  and A. Kitaev  and A. R. Klots  and A. N. Korotkov  and F. Kostritsa  and J. M. Kreikebaum  and D. Landhuis  and P. Laptev  and K.-M. Lau  and L. Laws  and J. Lee  and K. W. Lee  and Y. D. Lensky  and B. J. Lester  and A. T. Lill  and W. Liu  and A. Locharla  and F. D. Malone  and O. Martin  and J. R. McClean  and M. McEwen  and A. Mieszala  and S. Montazeri  and A. Morvan  and R. Movassagh  and W. Mruczkiewicz  and M. Neeley  and C. Neill  and A. Nersisyan  and M. Newman  and J. H. Ng  and A. Nguyen  and M. Nguyen  and M. Y. Niu  and T. E. O'Brien  and A. Opremcak  and A. Petukhov  and R. Potter  and L. P. Pryadko  and C. Quintana  and C. Rocque  and N. C. Rubin  and N. Saei  and D. Sank  and K. Sankaragomathi  and K. J. Satzinger  and H. F. Schurkus  and C. Schuster  and M. J. Shearn  and A. Shorter  and N. Shutty  and V. Shvarts  and J. Skruzny  and W. C. Smith  and R. Somma  and G. Sterling  and D. Strain  and M. Szalay  and A. Torres  and G. Vidal  and B. Villalonga  and C. V. Heidweiller  and T. White  and B. W. K. Woo  and C. Xing  and Z. J. Yao  and P. Yeh  and J. Yoo  and G. Young  and A. Zalcman  and Y. Zhang  and N. Zhu  and N. Zobrist  and H. Neven  and R. Babbush  and D. Bacon  and S. Boixo  and J. Hilton  and E. Lucero  and A. Megrant  and J. Kelly  and Y. Chen  and P. Roushan  and V. Smelyanskiy  and D. A. Abanin },
title = {Stable quantum-correlated many-body states through engineered dissipation},
journal = {Science},
volume = {383},
number = {6689},
pages = {1332--1337},
year = {2024},
doi = {10.1126/science.adh9932},
url = {https://www.science.org/doi/abs/10.1126/science.adh9932},
eprint = {https://www.science.org/doi/pdf/10.1126/science.adh9932}
}

@article{Harrington_2022,
   title={Engineered dissipation for quantum information science},
   volume={4},
   ISSN={2522-5820},
   url={http://dx.doi.org/10.1038/s42254-022-00494-8},
   DOI={10.1038/s42254-022-00494-8},
   number={10},
   journal={Nature Reviews Physics},
   publisher={Springer Science and Business Media LLC},
   author={Harrington, Patrick M. and Mueller, Erich J. and Murch, Kater W.},
   year={2022},
   month={aug}, pages={660–671} 
}

@article{Low_2019,
   title={Hamiltonian Simulation by Qubitization},
   volume={3},
   ISSN={2521-327X},
   url={http://dx.doi.org/10.22331/q-2019-07-12-163},
   DOI={10.22331/q-2019-07-12-163},
   journal={Quantum},
   publisher={Verein zur Forderung des Open Access Publizierens in den Quantenwissenschaften},
   author={Low, Guang Hao and Chuang, Isaac L.},
   year={2019},
   month={jul}, pages={163} 
}

@article{Cerezo2021,
  author  = {Cerezo, M. and Arrasmith, A. and Babbush, R. and Benjamin, S. C. and Endo, S. and Fujii, K. and McClean, J. R. and Mitarai, K. and Yuan, X. and Cincio, L. and Coles, P. J.},
  title   = {Variational Quantum Algorithms},
  journal = {Nature Reviews Physics},
  volume  = {3},
  pages   = {625--644},
  year    = {2021},
  doi     = {10.1038/s42254-021-00348-9},
  url     = {https://doi.org/10.1038/s42254-021-00348-9}
}

@misc{georgescu2026efficientexpressiveboundaryconditions,
      title={Efficient and Expressive Boundary Conditions in Quantum Lattice Boltzmann Methods}, 
      author={Călin A. Georgescu and Matthias Möller},
      year={2026},
      eprint={2606.01426},
      archivePrefix={arXiv},
      primaryClass={quant-ph},
      url={https://arxiv.org/abs/2606.01426}, 
}

@misc{duong2026quantumlatticeboltzmanndenoising,
      title={Quantum Lattice Boltzmann with Denoising Collision Operators}, 
      author={Trong Duong and Matthias Möller and Norbert Hosters},
      year={2026},
      eprint={2604.09997},
      archivePrefix={arXiv},
      primaryClass={quant-ph},
      url={https://arxiv.org/abs/2604.09997}, 
}

@misc{akshay2024tensornetworksbasedquantum,
      title={Tensor networks based quantum optimization algorithm}, 
      author={V. Akshay and Ar. Melnikov and A. Termanova and M. R. Perelshtein},
      year={2024},
      eprint={2404.15048},
      archivePrefix={arXiv},
      primaryClass={quant-ph},
      url={https://arxiv.org/abs/2404.15048}, 
}

@article{Spall1992,
  author={Spall, J.C.},
  journal={IEEE Transactions on Automatic Control}, 
  title={Multivariate stochastic approximation using a simultaneous perturbation gradient approximation}, 
  year={1992},
  volume={37},
  number={3},
  pages={332-341},
  doi={10.1109/9.119632}
}

@article{Martyn2021,
  title = {Grand Unification of Quantum Algorithms},
  author = {Martyn, John M. and Rossi, Zane M. and Tan, Andrew K. and Chuang, Isaac L.},
  journal = {PRX Quantum},
  volume = {2},
  issue = {4},
  pages = {040203},
  numpages = {40},
  year = {2021},
  month = {Dec},
  publisher = {American Physical Society},
  doi = {10.1103/PRXQuantum.2.040203},
  url = {https://link.aps.org/doi/10.1103/PRXQuantum.2.040203}
}

@article{Wawrzyniak2025b,
title = {Linearized quantum lattice-Boltzmann method for the advection-diffusion equation using dynamic circuits},
journal = {Computer Physics Communications},
volume = {317},
pages = {109856},
year = {2025},
issn = {0010-4655},
doi = {10.1016/j.cpc.2025.109856},
url = {https://www.sciencedirect.com/science/article/pii/S0010465525003583},
author = {David Wawrzyniak and Josef Winter and Steffen Schmidt and Thomas Indinger and Christian F. Janßen and Uwe Schramm and Nikolaus A. Adams}
}

@inproceedings{Chakraborty2019,
  doi = {10.4230/LIPICS.ICALP.2019.33},
  
  url = {https://drops.dagstuhl.de/entities/document/10.4230/LIPIcs.ICALP.2019.33},
  
  author = {Chakraborty, Shantanav and Gilyén, András and Jeffery, Stacey},
  
  language = {en},
  
  title = {The Power of Block-Encoded Matrix Powers: Improved Regression Techniques via Faster Hamiltonian Simulation},
  
  journal = {LIPIcs, Volume 132, ICALP 2019},
  
  volume = {132},
  
  pages = {33:1-33:14},
  
  publisher = {Schloss Dagstuhl – Leibniz-Zentrum für Informatik},
  
  year = {2019},
  
  copyright = {Creative Commons Attribution 3.0 Unported license}
}

@article{Spall1998ANOO,
  title={AN OVERVIEW OF THE SIMULTANEOUS PERTURBATION METHOD FOR EFFICIENT OPTIMIZATION},
  author={James C. Spall},
  journal={Johns Hopkins Apl Technical Digest},
  year={1998},
  volume={19},
  pages={482-492},
  url={https://api.semanticscholar.org/CorpusID:7988308}
}

@book{NocedalWright2006,
  title = {Numerical Optimization},
  author = {Nocedal, Jorge and Wright, Stephen J.},
  series = {Springer Series in Operations Research and Financial Engineering},
  edition = {2},
  publisher = {Springer},
  address = {New York, NY},
  year = {2006},
  doi = {10.1007/978-0-387-40065-5},
  isbn = {978-0-387-40065-5}
}

@article{Mastorakis_2026,
   title={Resource-efficient Hadamard test tailored variational framework for nonlinear dynamics on quantum computers},
   volume={11},
   ISSN={2058-9565},
   url={http://dx.doi.org/10.1088/2058-9565/ae3a13},
   DOI={10.1088/2058-9565/ae3a13},
   number={1},
   journal={Quantum Science and Technology},
   publisher={IOP Publishing},
   author={Mastorakis, Eleftherios and Umer, Muhammad and Guevara-Bertsch, Milena and Ulmanis, Juris and Rohde, Felix and Angelakis, Dimitris G},
   year={2026},
   month={feb}, pages={015061} }

@article{Lacatus2025,
author = {Lăcătuş, Monica and Möller, Matthias},
title = {Surrogate Quantum Circuit Design for the Lattice Boltzmann Collision Operator},
journal = {International Journal for Numerical Methods in Engineering},
volume = {127},
number = {4},
pages = {e70286},
doi = {10.1002/nme.70286},
url = {https://onlinelibrary.wiley.com/doi/abs/10.1002/nme.70286},
eprint = {https://onlinelibrary.wiley.com/doi/pdf/10.1002/nme.70286},
year = {2026}
}

@misc{zhang2025datadrivenquantumkoopmanmethod,
      title={Data-driven quantum Koopman method for simulating nonlinear dynamics}, 
      author={Baoyang Zhang and Zhen Lu and Yaomin Zhao and Yue Yang},
      year={2025},
      eprint={2507.21890},
      archivePrefix={arXiv},
      primaryClass={quant-ph},
      url={https://arxiv.org/abs/2507.21890}, 
}

@article{Setty_2026,
doi = {10.1088/1751-8121/ae5cef},
url = {https://doi.org/10.1088/1751-8121/ae5cef},
year = {2026},
month = {may},
publisher = {IOP Publishing},
volume = {59},
number = {18},
pages = {185303},
author = {Setty, Abhishek},
title = {A quantum linear systems pathway for solving differential equations},
journal = {Journal of Physics A: Mathematical and Theoretical}
}

@article{Lubasch2020,
  title = {Variational quantum algorithms for nonlinear problems},
  author = {Lubasch, Michael and Joo, Jaewoo and Moinier, Pierre and Kiffner, Martin and Jaksch, Dieter},
  journal = {Phys. Rev. A},
  volume = {101},
  issue = {1},
  pages = {010301},
  numpages = {7},
  year = {2020},
  month = {Jan},
  publisher = {American Physical Society},
  doi = {10.1103/PhysRevA.101.010301},
  url = {https://link.aps.org/doi/10.1103/PhysRevA.101.010301}
}

@article{Bharti2022,
  author  = {Bharti, Kishor and Cervera-Lierta, Alba and Kyaw, Thi Ha and Haug, Tobias and Alperin-Lea, Samuel and Anand, Abhinav and Degroote, Matthias and Heimonen, Henrik and Kottmann, Jakob S. and Menke, Tim and Mukherjee, Sudipto and Sim, Sukin and Singh, Harshad and Sun, Shouvanik and Tew, David P. and Wang, Liwei and Yuan, Xiao and Aspuru-Guzik, Al{\'a}n},
  title   = {Noisy intermediate-scale quantum algorithms},
  journal = {Reviews of Modern Physics},
  volume  = {94},
  number  = {1},
  pages   = {015004},
  year    = {2022},
  doi     = {10.1103/RevModPhys.94.015004},
  url     = {https://doi.org/10.1103/RevModPhys.94.015004}
}

@misc{Tserkis_2026,
      title={Depth optimization of CNOT ladder circuits}, 
      author={Spyros Tserkis and Muhammad Umer and Dimitris G. Angelakis},
      year={2026},
      eprint={2511.13256},
      archivePrefix={arXiv},
      primaryClass={quant-ph},
      url={https://arxiv.org/abs/2511.13256}, 
}

@misc{Tserkis_2026_MCT,
      title={Minimum Toffoli depth for the multi-controlled Toffoli gate via teleportation}, 
      author={Spyros Tserkis and Muhammad Umer and Eleftherios Mastorakis and Dimitris G. Angelakis},
      year={2026},
      eprint={2604.25861},
      archivePrefix={arXiv},
      primaryClass={quant-ph},
      url={https://arxiv.org/abs/2604.25861}, 
}

@article{Chapman1979,
author = {Chapman, Dean R.},
title = {Computational Aerodynamics Development and Outlook},
journal = {AIAA Journal},
volume = {17},
number = {12},
pages = {1293-1313},
year = {1979},
doi = {10.2514/3.61311},
URL = {https://doi.org/10.2514/3.61311},
eprint = {https://doi.org/10.2514/3.61311}
}

@inproceedings{Griffin2019,
  title={Investigation of quantum algorithms for direct numerical simulation of the Navier-Stokes equations},
  author={Kevin Patrick Griffin and Suhas S. Jain and Timothy Flint and W. H. R. Chan},
  year={2020},
  url={https://api.semanticscholar.org/CorpusID:231773866}
}

@inproceedings{Bharadwaj2020,
    author={Bharadwaj, Sachin S and Sreenivasan, Katepalli R},
    title={Quantum computation of fluid dynamics},
    booktitle = {Indian Academy of Scieinces Conference Series},
    year = {2020},
    volume = {3},
    doi = {10.29195/iascs.03.01.0015},
    pages = {77-96},
}

@article{MoinMahesh1998,
  author  = {Moin, Parviz and Mahesh, Krishnan},
  title   = {Direct Numerical Simulation: A Tool in Turbulence Research},
  journal = {Annual Review of Fluid Mechanics},
  year    = {1998},
  volume  = {30},
  number  = {1},
  pages   = {539--578},
  doi     = {10.1146/annurev.fluid.30.1.539},
  url     = {https://www.annualreviews.org/doi/10.1146/annurev.fluid.30.1.539}
}

@book{Sagaut2006,
  author    = {Sagaut, Pierre},
  title     = {Large Eddy Simulation for Incompressible Flows: An Introduction},
  edition   = {3},
  year      = {2006},
  publisher = {Springer},
  address   = {Berlin, Heidelberg},
  doi       = {10.1007/978-3-662-04416-2},
  url       = {https://link.springer.com/book/10.1007/978-3-662-04416-2}
}

@book{hirsch2007numerical,
  title     = {Numerical Computation of Internal and External Flows (Second Edition)},
  author    = {Hirsch, Charles},
  year      = {2007},
  publisher = {Butterworth-Heinemann},
  edition = {2},
  address   = {Oxford, UK},
  isbn      = {978-0-7506-6594-0},
  doi = {10.1016/B978-075066594-0/50037-0},
  url = {https://www.sciencedirect.com/science/article/pii/B9780750665940500370}
}

@article{Schollwock2011,
  author  = {Schollwöck, Ulrich},
  title   = {The Density-Matrix Renormalization Group in the Age of Matrix Product States},
  journal = {Annals of Physics},
  year    = {2011},
  volume  = {326},
  number  = {1},
  pages   = {96--192},
  doi     = {10.1016/j.aop.2010.09.012}
}

@article{Arenstein2026,
AUTHOR = {Arenstein, Lucas and Mikkelsen, Martin and Kastoryano, Michael},
TITLE = {Fast and Flexible Quantum-Inspired Differential Equation Solvers with Data Integration},
JOURNAL = {Mathematics},
VOLUME = {14},
YEAR = {2026},
NUMBER = {12},
ARTICLE-NUMBER = {2069},
URL = {https://www.mdpi.com/2227-7390/14/12/2069},
ISSN = {2227-7390},
DOI = {10.3390/math14122069}
}

@article{Kazeev2012,
author = {Kazeev, Vladimir A. and Khoromskij, Boris N.},
title = {Low-Rank Explicit QTT Representation of the Laplace Operator and Its Inverse},
journal = {SIAM Journal on Matrix Analysis and Applications},
volume = {33},
number = {3},
pages = {742-758},
year = {2012},
doi = {10.1137/100820479},

URL = {https://doi.org/10.1137/100820479},
eprint = {https://doi.org/10.1137/100820479}
}

@techreport{Slotnick2014CFDV2,
  title={CFD Vision 2030 Study: A Path to Revolutionary Computational Aerosciences},
  author={Jeffrey P. Slotnick and Abdollah Khodadoust and Juan J. Alonso and David L. Darmofal and William Gropp and Elizabeth A. Lurie and Dimitri J. Mavriplis},
  year={2014},
  url={https://api.semanticscholar.org/CorpusID:215813979}
}

@article{Syamlal2024VQCFD,
  title={Computational Fluid Dynamics on Quantum Computers},
  author={Madhava Syamlal and Carter Copen and Mariko Takahashi and Benjamin Hall QubitSolve Inc. and Infleqtion},
  journal={AIAA AVIATION FORUM AND ASCEND 2024},
  year={2024},
  url={https://api.semanticscholar.org/CorpusID:270765032}
}

@article{Yepez2001,
  title = {Quantum lattice-gas model for computational fluid dynamics},
  author = {Yepez, Jeffrey},
  journal = {Phys. Rev. E},
  volume = {63},
  issue = {4},
  pages = {046702},
  numpages = {18},
  year = {2001},
  month = {Mar},
  publisher = {American Physical Society},
  doi = {10.1103/PhysRevE.63.046702},
  url = {https://link.aps.org/doi/10.1103/PhysRevE.63.046702}
}

@article{Kitaev1995QuantumMA,
  title={Quantum measurements and the Abelian Stabilizer Problem},
  author={Alexei Y. Kitaev},
  journal={Electron. Colloquium Comput. Complex.},
  year={1995},
  volume={TR96},
  url={https://api.semanticscholar.org/CorpusID:17023060}
}

@article{Berry_2017,
   title={Quantum Algorithm for Linear Differential Equations with Exponentially Improved Dependence on Precision},
   volume={356},
   ISSN={1432-0916},
   url={http://dx.doi.org/10.1007/s00220-017-3002-y},
   DOI={10.1007/s00220-017-3002-y},
   number={3},
   journal={Communications in Mathematical Physics},
   publisher={Springer Science and Business Media LLC},
   author={Berry, Dominic W. and Childs, Andrew M. and Ostrander, Aaron and Wang, Guoming},
   year={2017},
   month={oct}, pages={1057–1081} }

@inproceedings{Berry_2015,
   title={Hamiltonian Simulation with Nearly Optimal Dependence on all Parameters},
   url={http://dx.doi.org/10.1109/FOCS.2015.54},
   DOI={10.1109/focs.2015.54},
   booktitle={2015 IEEE 56th Annual Symposium on Foundations of Computer Science},
   publisher={IEEE},
   author={Berry, Dominic W. and Childs, Andrew M. and Kothari, Robin},
   year={2015},
   month={oct}, pages={792–809} }

@article{PhysRevLett.114.090502,
  title = {Simulating Hamiltonian Dynamics with a Truncated Taylor Series},
  author = {Berry, Dominic W. and Childs, Andrew M. and Cleve, Richard and Kothari, Robin and Somma, Rolando D.},
  journal = {Phys. Rev. Lett.},
  volume = {114},
  issue = {9},
  pages = {090502},
  numpages = {5},
  year = {2015},
  month = {Mar},
  publisher = {American Physical Society},
  doi = {10.1103/PhysRevLett.114.090502},
  url = {https://link.aps.org/doi/10.1103/PhysRevLett.114.090502}
}

@book{LeVeque2007,
  title={Finite Difference Methods for Ordinary and Partial Differential Equations: Steady-State and Time-Dependent Problems},
  author={LeVeque, R.J.},
  isbn={9780898717839},
  lccn={2007061732}, 
  url={https://epubs.siam.org/doi/book/10.1137/1.9780898717839},
  year={2007},
  publisher={Society for Industrial and Applied Mathematics}
}

@Article{Cirac2000,
author={Cirac, J. I.
and Zoller, P.},
title={A scalable quantum computer with ions in an array of microtraps},
journal={Nature},
year={2000},
month={Apr},
day={01},
volume={404},
number={6778},
pages={579-581},
issn={1476-4687},
doi={10.1038/35007021},
url={https://doi.org/10.1038/35007021}
}

@article{Childs_2021,
   title={High-precision quantum algorithms for partial differential equations},
   volume={5},
   ISSN={2521-327X},
   url={http://dx.doi.org/10.22331/q-2021-11-10-574},
   DOI={10.22331/q-2021-11-10-574},
   journal={Quantum},
   publisher={Verein zur Forderung des Open Access Publizierens in den Quantenwissenschaften},
   author={Childs, Andrew M. and Liu, Jin-Peng and Ostrander, Aaron},
   year={2021},
   month={nov}, pages={574} }

@article{Jaksch2000,
  title = {Fast Quantum Gates for Neutral Atoms},
  author = {Jaksch, D. and Cirac, J. I. and Zoller, P. and Rolston, S. L. and C\^ot\'e, R. and Lukin, M. D.},
  journal = {Phys. Rev. Lett.},
  volume = {85},
  issue = {10},
  pages = {2208--2211},
  numpages = {0},
  year = {2000},
  month = {Sep},
  publisher = {American Physical Society},
  doi = {10.1103/PhysRevLett.85.2208},
  url = {https://link.aps.org/doi/10.1103/PhysRevLett.85.2208}
}

@article{Cirac1995,
  title = {Quantum Computations with Cold Trapped Ions},
  author = {Cirac, J. I. and Zoller, P.},
  journal = {Phys. Rev. Lett.},
  volume = {74},
  issue = {20},
  pages = {4091--4094},
  numpages = {0},
  year = {1995},
  month = {May},
  publisher = {American Physical Society},
  doi = {10.1103/PhysRevLett.74.4091},
  url = {https://link.aps.org/doi/10.1103/PhysRevLett.74.4091}
}

@Article{Jaksch2023,
  author    = {Jaksch, Dieter and Givi, Peyman and Daley, Andrew J and Rung, Thomas},
  journal   = {AIAA journal},
  title     = {Variational quantum algorithms for computational fluid dynamics},
  year      = {2023},
  number    = {5},
  pages     = {1885--1894},
  volume    = {61},
  doi       = {10.2514/1.J062426},
  publisher = {American Institute of Aeronautics and Astronautics},
  url       = {https://arc.aiaa.org/doi/abs/10.2514/1.J062426},
}

@article{Siegl_2026,
   title={Tensor-programmable quantum circuits for solving differential equations},
   volume={8},
   ISSN={2643-1564},
   url={http://dx.doi.org/10.1103/2qzh-yf49},
   DOI={10.1103/2qzh-yf49},
   number={1},
   journal={Physical Review Research},
   publisher={American Physical Society (APS)},
   author={Siegl, Pia and Reese, Greta Sophie and Hashizume, Tomohiro and van H{\"u}lst, Nis-Luca and Jaksch, Dieter},
   year={2026},
   month={jan} 
}

@article{Kazeev2014,
    doi = {10.1371/journal.pcbi.1003359},
    author = {Kazeev, Vladimir AND Khammash, Mustafa AND Nip, Michael AND Schwab, Christoph},
    journal = {PLOS Computational Biology},
    publisher = {Public Library of Science},
    title = {Direct Solution of the Chemical Master Equation Using Quantized Tensor Trains},
    year = {2014},
    month = {03},
    volume = {10},
    url = {https://doi.org/10.1371/journal.pcbi.1003359},
    pages = {1-19},
    number = {3},

}

@article{Grasedyck2013,
  author  = {Grasedyck, Lars and Kressner, Daniel and Tobler, Christine},
  title   = {A literature survey of low-rank tensor approximation techniques},
  journal = {GAMM-Mitteilungen},
  volume  = {36},
  number  = {1},
  pages   = {53--78},
  year    = {2013},
  doi     = {10.1002/gamm.201310004}
}

@article{Hauru2021,
   title={Riemannian optimization of isometric tensor networks},
   volume={10},
   ISSN={2542-4653},
   url={http://dx.doi.org/10.21468/SciPostPhys.10.2.040},
   DOI={10.21468/scipostphys.10.2.040},
   number={2},
   journal={SciPost Physics},
   publisher={Stichting SciPost},
   author={Hauru, Markus and Van Damme, Maarten and Haegeman, Jutho},
   year={2021},
   month={Feb} 
}

@article{Luchnikov2021,
   title={Riemannian geometry and automatic differentiation for optimization problems of quantum physics and quantum technologies},
   volume={23},
   ISSN={1367-2630},
   url={http://dx.doi.org/10.1088/1367-2630/ac0b02},
   DOI={10.1088/1367-2630/ac0b02},
   number={7},
   journal={New Journal of Physics},
   publisher={IOP Publishing},
   author={Luchnikov, Ilia A and Krechetov, Mikhail E and Filippov, Sergey N},
   year={2021},
   month={July}, 
   pages={073006} 
}

@book{Absil2007,
    author = {Absil, P.-A. and Mahony, R. and Sepulchre, R.},
    title = {Optimization Algorithms on Matrix Manifolds},
    year = {2007},
    isbn = {0691132984},
    publisher = {Princeton University Press},
    address = {USA}
}

@article{Schon2005,
  title = {Sequential Generation of Entangled Multiqubit States},
  author = {Sch\"on, C. and Solano, E. and Verstraete, F. and Cirac, J. I. and Wolf, M. M.},
  journal = {Phys. Rev. Lett.},
  volume = {95},
  issue = {11},
  pages = {110503},
  numpages = {4},
  year = {2005},
  month = {Sep},
  publisher = {American Physical Society},
  doi = {10.1103/PhysRevLett.95.110503},
  url = {https://link.aps.org/doi/10.1103/PhysRevLett.95.110503}
}

@article{Katabarwa2024,
  title = {Early Fault-Tolerant Quantum Computing},
  author = {Katabarwa, Amara and Gratsea, Katerina and Caesura, Athena and Johnson, Peter D.},
  journal = {PRX Quantum},
  volume = {5},
  issue = {2},
  pages = {020101},
  numpages = {20},
  year = {2024},
  month = {Jun},
  publisher = {American Physical Society},
  doi = {10.1103/PRXQuantum.5.020101},
  url = {https://link.aps.org/doi/10.1103/PRXQuantum.5.020101}
}

@inproceedings{Dangwal2025, 
author = {Dangwal, Siddharth and Vittal, Suhas and Seifert, Lennart Maximilian and Chong, Frederic T. and Ravi, Gokul Subramanian}, 
title = {Variational Quantum Algorithms in the era of Early Fault Tolerance}, 
year = {2025}, 
isbn = {9798400712616}, 
publisher = {Association for Computing Machinery}, 
address = {New York, NY, USA}, 
url = {https://doi.org/10.1145/3695053.3731112}, 
doi = {10.1145/3695053.3731112}, 
booktitle = {Proceedings of the 52nd Annual International Symposium on Computer Architecture}, pages = {1417–1431}, 
numpages = {15}, location = { }, 
series = {ISCA '25} 
}

@article{Sayginel2024,
doi = {10.1088/2058-9565/ad0571},
url = {https://doi.org/10.1088/2058-9565/ad0571},
year = {2023},
month = {nov},
publisher = {IOP Publishing},
volume = {9},
number = {1},
pages = {015015},
author = {Sayginel, Hasan and Jamet, Francois and Agarwal, Abhishek and Browne, Dan E and Rungger, Ivan},
title = {A fault-tolerant variational quantum algorithm with limited T-depth},
journal = {Quantum Science and Technology}
}

@article{Bairey2019,
  title = {Learning a Local Hamiltonian from Local Measurements},
  author = {Bairey, Eyal and Arad, Itai and Lindner, Netanel H.},
  journal = {Phys. Rev. Lett.},
  volume = {122},
  issue = {2},
  pages = {020504},
  numpages = {5},
  year = {2019},
  month = {Jan},
  publisher = {American Physical Society},
  doi = {10.1103/PhysRevLett.122.020504},
  url = {https://link.aps.org/doi/10.1103/PhysRevLett.122.020504}
}

@article{Ran2020,
  title = {Encoding of matrix product states into quantum circuits of one- and two-qubit gates},
  author = {Ran, Shi-Ju},
  journal = {Phys. Rev. A},
  volume = {101},
  issue = {3},
  pages = {032310},
  numpages = {7},
  year = {2020},
  month = {Mar},
  publisher = {American Physical Society},
  doi = {10.1103/PhysRevA.101.032310},
  url = {https://link.aps.org/doi/10.1103/PhysRevA.101.032310}
}

@misc{vanhülst2026quantuminspiredsimulation2dturbulent,
      title={Quantum-Inspired Simulation of 2D Turbulent Rayleigh-B\'enard Convection}, 
      author={Nis-Luca van H{\"u}lst and Mario Guillaume Cecile and Hai-Yen Van and Tomohiro Hashizume and Eugene de Villiers and Dieter Jaksch},
      year={2026},
      eprint={2604.16179},
      archivePrefix={arXiv},
      primaryClass={physics.flu-dyn},
      url={https://arxiv.org/abs/2604.16179}, 
}

@misc{arenstein2026gridsolutionmultiassetoptions,
      title={Full grid solution for multi-asset options pricing with tensor networks}, 
      author={Lucas Arenstein and Michael Kastoryano},
      year={2026},
      eprint={2601.00009},
      archivePrefix={arXiv},
      primaryClass={q-fin.CP},
      url={https://arxiv.org/abs/2601.00009}, 
}

@article{Kiffner2023,
  title = {Tensor network reduced order models for wall-bounded flows},
  author = {Kiffner, Martin and Jaksch, Dieter},
  journal = {Phys. Rev. Fluids},
  volume = {8},
  issue = {12},
  pages = {124101},
  numpages = {20},
  year = {2023},
  month = {Dec},
  publisher = {American Physical Society},
  doi = {10.1103/PhysRevFluids.8.124101},
  url = {https://link.aps.org/doi/10.1103/PhysRevFluids.8.124101}
}

@article{Khoromskij_2011,
author = {Khoromskij, Boris N. and Schwab, Christoph},
title = {Tensor-Structured Galerkin Approximation of Parametric and Stochastic Elliptic PDEs},
journal = {SIAM Journal on Scientific Computing},
volume = {33},
number = {1},
pages = {364-385},
year = {2011},
doi = {10.1137/100785715},
URL = {https://doi.org/10.1137/100785715},
eprint ={https://doi.org/10.1137/100785715}
}

@article{Bengoechea_2025,
   title={Toward Variational Quantum Algorithms for Generalized Linear and Nonlinear Transport Phenomena},
   ISSN={1533-385X},
   url={http://dx.doi.org/10.2514/1.J065582},
   DOI={10.2514/1.j065582},
   journal={AIAA Journal},
   publisher={American Institute of Aeronautics and Astronautics (AIAA)},
   author={Bengoechea, Sergio and Over, Paul and Jaksch, Dieter and Rung, Thomas},
   year={2025},
   month={nov}, 
   pages={1–20} 
}

@article{Dawson2005blj,
    author = "Dawson, Christopher M. and Nielsen, Michael A.",
    title = "{The Solovay-Kitaev algorithm}",
    eprint = "quant-ph/0505030",
    archivePrefix = "arXiv",
    doi = "10.26421/QIC6.1-6",
    journal = "Quant. Inf. Comput.",
    volume = "6",
    pages = "081--095",
    year = "2006"
}

@article{Ross2014okw,
    author = "Ross, Neil J. and Selinger, Peter",
    title = "{Optimal ancilla-free Clifford+$T$ approximation of $Z$-rotations}",
    eprint = "1403.2975",
    archivePrefix = "arXiv",
    primaryClass = "quant-ph",
    doi = "10.26421/QIC16.11-12-1",
    journal = "Quant. Inf. Comput.",
    volume = "16",
    number = "11-12",
    pages = "0901--0953",
    year = "2016"
}

@article{Vidal2003,
  title = {Efficient Classical Simulation of Slightly Entangled Quantum Computations},
  author = {Vidal, Guifr\'e},
  journal = {Phys. Rev. Lett.},
  volume = {91},
  issue = {14},
  pages = {147902},
  numpages = {4},
  year = {2003},
  month = {Oct},
  publisher = {American Physical Society},
  doi = {10.1103/PhysRevLett.91.147902},
  url = {https://link.aps.org/doi/10.1103/PhysRevLett.91.147902}
}

@article{Bridgeman_2017,
   title={Hand-waving and interpretive dance: an introductory course on tensor networks},
   volume={50},
   ISSN={1751-8121},
   url={http://dx.doi.org/10.1088/1751-8121/aa6dc3},
   DOI={10.1088/1751-8121/aa6dc3},
   number={22},
   journal={Journal of Physics A: Mathematical and Theoretical},
   publisher={IOP Publishing},
   author={Bridgeman, Jacob C and Chubb, Christopher T},
   year={2017},
   month={may}, 
   pages={223001} 
}

@article{Lanyon_2017,
   title={Efficient tomography of a quantum many-body system},
   volume={13},
   ISSN={1745-2481},
   url={http://dx.doi.org/10.1038/nphys4244},
   DOI={10.1038/nphys4244},
   number={12},
   journal={Nature Physics},
   publisher={Springer Science and Business Media LLC},
   author={Lanyon, B. P. and Maier, C. and Holzäpfel, M. and Baumgratz, T. and Hempel, C. and Jurcevic, P. and Dhand, I. and Buyskikh, A. S. and Daley, A. J. and Cramer, M. and Plenio, M. B. and Blatt, R. and Roos, C. F.},
   year={2017},
   month={sept}, pages={1158–1162} 
}

@article{Sato2021Poisson,
  title = {Variational quantum algorithm based on the minimum potential energy for solving the Poisson equation},
  author = {Sato, Yuki and Kondo, Ruho and Koide, Satoshi and Takamatsu, Hideki and Imoto, Nobuyuki},
  journal = {Phys. Rev. A},
  volume = {104},
  issue = {5},
  pages = {052409},
  numpages = {15},
  year = {2021},
  month = {Nov},
  publisher = {American Physical Society},
  doi = {10.1103/PhysRevA.104.052409},
  url = {https://link.aps.org/doi/10.1103/PhysRevA.104.052409}
}

@article{Pool2024,
  title = {Nonlinear dynamics as a ground-state solution on quantum computers},
  author = {Pool, Albert J. and Somoza, Alejandro D. and Mc Keever, Conor and Lubasch, Michael and Horstmann, Birger},
  journal = {Phys. Rev. Res.},
  volume = {6},
  issue = {3},
  pages = {033257},
  numpages = {19},
  year = {2024},
  month = {Sep},
  publisher = {American Physical Society},
  doi = {10.1103/PhysRevResearch.6.033257},
  url = {https://link.aps.org/doi/10.1103/PhysRevResearch.6.033257}
}

@article{Barison2022,
  title = {Variational dynamics as a ground-state problem on a quantum computer},
  author = {Barison, Stefano and Vicentini, Filippo and Cirac, Ignacio and Carleo, Giuseppe},
  journal = {Phys. Rev. Res.},
  volume = {4},
  issue = {4},
  pages = {043161},
  numpages = {12},
  year = {2022},
  month = {Dec},
  publisher = {American Physical Society},
  doi = {10.1103/PhysRevResearch.4.043161},
  url = {https://link.aps.org/doi/10.1103/PhysRevResearch.4.043161}
}

@article{Trahan2023,
AUTHOR = {Trahan, Corey Jason and Loveland, Mark and Davis, Noah and Ellison, Elizabeth},
TITLE = {A Variational Quantum Linear Solver Application to Discrete Finite-Element Methods},
JOURNAL = {Entropy},
VOLUME = {25},
YEAR = {2023},
NUMBER = {4},
ARTICLE-NUMBER = {580},
URL = {https://www.mdpi.com/1099-4300/25/4/580},
PubMedID = {37190367},
ISSN = {1099-4300},
DOI = {10.3390/e25040580}
}

@article{Li_2008,
   title={A public turbulence database cluster and applications to study Lagrangian evolution of velocity increments in turbulence},
   volume={9},
   ISSN={1468-5248},
   url={http://dx.doi.org/10.1080/14685240802376389},
   DOI={10.1080/14685240802376389},
   journal={Journal of Turbulence},
   publisher={Informa UK Limited},
   author={Li, Yi and Perlman, Eric and Wan, Minping and Yang, Yunke and Meneveau, Charles and Burns, Randal and Chen, Shiyi and Szalay, Alexander and Eyink, Gregory},
   year={2008},
   month={Jan}, 
   pages={N31} 
}

@article{Cramer_2010,
   title={Efficient quantum state tomography},
   volume={1},
   ISSN={2041-1723},
   url={http://dx.doi.org/10.1038/ncomms1147},
   DOI={10.1038/ncomms1147},
   number={1},
   journal={Nature Communications},
   publisher={Springer Science and Business Media LLC},
   author={Cramer, Marcus and Plenio, Martin B. and Flammia, Steven T. and Somma, Rolando and Gross, David and Bartlett, Stephen D. and Landon-Cardinal, Olivier and Poulin, David and Liu, Yi-Kai},
   year={2010},
   month={Dec} 
}

@article{Todorova2020,
    title = {Quantum algorithm for the collisionless Boltzmann equation},
    journal = {Journal of Computational Physics},
    volume = {409},
    pages = {109347},
    year = {2020},
    issn = {0021-9991},
    doi = {10.1016/j.jcp.2020.109347},
    url = {https://www.sciencedirect.com/science/article/pii/S0021999120301212},
    author = {Blaga N. Todorova and René Steijl}
}

@misc{Chen2021,
      title={Quantum Finite Volume Method for Computational Fluid Dynamics with Classical Input and Output}, 
      author={Zhao-Yun Chen and Cheng Xue and Si-Ming Chen and Bing-Han Lu and Yu-Chun Wu and Ju-Chun Ding and Sheng-Hong Huang and Guo-Ping Guo},
      year={2021},
      eprint={2102.03557},
      archivePrefix={arXiv},
      primaryClass={quant-ph},
      url={https://arxiv.org/abs/2102.03557} 
}

\end{document}